\documentclass[12pt]{article}
\usepackage{amsmath}
\usepackage{amssymb}
\usepackage{graphicx}
\usepackage{axodraw}
\usepackage{epsfig}
\usepackage[dvips,usenames]{color}

\begin{document}
\baselineskip 0.6cm

\def\simgt{\mathrel{\lower2.5pt\vbox{\lineskip=0pt\baselineskip=0pt
           \hbox{$>$}\hbox{$\sim$}}}}
\def\simlt{\mathrel{\lower2.5pt\vbox{\lineskip=0pt\baselineskip=0pt
           \hbox{$<$}\hbox{$\sim$}}}}
\def\simprop{\mathrel{\lower3.0pt\vbox{\lineskip=1.0pt\baselineskip=0pt
             \hbox{$\propto$}\hbox{$\sim$}}}}
\def\lg{\mathrel{\lower2.5pt\vbox{\lineskip=0pt\baselineskip=0pt
           \hbox{$<$}\hbox{$>$}}}}
\def\lrpartial{\stackrel{\leftrightarrow}{\partial}}
\def\lrD{\stackrel{\leftrightarrow}{\cal D}}
\def\chairup{\begin{picture}(15,15)
  \Line(4,10)(11,10) \Line(4,4)(11,4) \Line(4,4)(4,10) \Line(11,4)(11,10)
  \Line(4,4)(2.5,1) \Line(11,4)(12.5,1) \Line(2.5,1)(12.5,1)
  \Line(3,1)(3,-5) \Line(12,1)(12,-5)
  \Line(4.5,1)(4.5,-2.5) \Line(10.5,1)(10.5,-2.5)
\end{picture}}
\def\chairdown{\begin{picture}(15,15)
  \Line(4,-5)(11,-5) \Line(4,1)(11,1) \Line(4,1)(4,-5) \Line(11,1)(11,-5)
  \Line(4,1)(2.5,4) \Line(11,1)(12.5,4) \Line(2.5,4)(12.5,4)
  \Line(3,4)(3,10) \Line(12,4)(12,10)
  \Line(4.5,4)(4.5,7.5) \Line(10.5,4)(10.5,7.5)
\end{picture}}
\def\manup{\begin{picture}(15,15)
  \CArc(7.5,8)(2,0,360)
  \Line(7.5,5.5)(3,2.5) \Line(7.5,5.5)(12,2.5) \Line(7.5,5.5)(7.5,0.5)
  \Line(7.5,0.5)(5,-5) \Line(7.5,0.5)(10,-5)
\end{picture}}
\def\mandown{\begin{picture}(15,15)
  \CArc(7.5,-3)(2,0,360)
  \Line(7.5,-0.5)(3,2.5) \Line(7.5,-0.5)(12,2.5) \Line(7.5,-0.5)(7.5,4.5)
  \Line(7.5,4.5)(5,10) \Line(7.5,4.5)(10,10)
\end{picture}}
\def\ftchairup{\begin{picture}(12,12)
  \Line(3.2,8)(8.8,8.0) \Line(3.2,3.2)(8.8,3.2)
  \Line(3.2,3.2)(3.2,8.0) \Line(8.8,3.2)(8.8,8.0)
  \Line(3.2,3.2)(2.0,0.8) \Line(8.8,3.2)(10.0,0.8) \Line(2.0,0.8)(10.0,0.8)
  \Line(2.4,0.8)(2.4,-4.0) \Line(9.6,0.8)(9.6,-4.0)
  \Line(3.6,0.8)(3.6,-2.0) \Line(8.4,0.8)(8.4,-2.0)
\end{picture}}
\def\ftchairdown{\begin{picture}(12,12)
  \Line(3.2,-4.0)(8.8,-4.0) \Line(3.2,0.8)(8.8,0.8)
  \Line(3.2,0.8)(3.2,-4.0) \Line(8.8,0.8)(8.8,-4.0)
  \Line(3.2,0.8)(2.0,3.2) \Line(8.8,0.8)(10.0,3.2) \Line(2.0,3.2)(10.0,3.2)
  \Line(2.4,3.2)(2.4,8.0) \Line(9.6,3.2)(9.6,8.0)
  \Line(3.6,3.2)(3.6,6.0) \Line(8.4,3.2)(8.4,6.0)
\end{picture}}
\def\MARU#1{{\rm\ooalign{\hfil\lower.168ex\hbox{#1}\hfil
  \circ\mathhexbox20D}}}

\begin{titlepage}

\begin{flushright}
UCB-PTH-11/02 \\
\end{flushright}

\vskip 1.3cm

\begin{center}
{\Large \bf Physical Theories, Eternal Inflation, and Quantum Universe}

\vskip 0.7cm

{\large Yasunori Nomura}

\vskip 0.4cm

{\it Berkeley Center for Theoretical Physics, Department of Physics,\\
 University of California, Berkeley, CA 94720, USA}

\vskip 0.2cm

{\it Theoretical Physics Group, Lawrence Berkeley National Laboratory,
 CA 94720, USA}

\vskip 0.8cm

\abstract{Infinities in eternal inflation have long been plaguing 
cosmology, making any predictions highly sensitive to how they are 
regulated.  The problem exists already at the level of semi-classical 
general relativity, and has a priori nothing to do with quantum gravity. 
On the other hand, we know that certain problems in semi-classical 
gravity, for example physics of black holes and their evaporation, 
have led to understanding of surprising, quantum natures of spacetime 
and gravity, such as the holographic principle and horizon complementarity.

In this paper, we present a framework in which well-defined predictions 
are obtained in an eternally inflating multiverse, based on the principles 
of quantum mechanics.  We show that the {\it entire} multiverse is 
described {\it purely} from the viewpoint of a single ``observer,'' 
who describes the world as a quantum state defined on his/her past 
light cones bounded by the (stretched) apparent horizons.  We find 
that quantum mechanics plays an essential role in regulating infinities. 
The framework is ``gauge invariant,'' i.e.\ predictions do not depend 
on how spacetime is parametrized, as it should be in a theory of 
quantum gravity.

Our framework provides a fully unified treatment of quantum measurement 
processes and the multiverse.  We conclude that the eternally inflating 
multiverse and many worlds in quantum mechanics are the same.  Other 
important implications include: global spacetime can be viewed as 
a derived concept; the multiverse is a transient phenomenon during 
the world relaxing into a supersymmetric Minkowski state.  We also 
present a theory of ``initial conditions'' for the multiverse.  By 
extrapolating our framework to the extreme, we arrive at a picture 
that the entire multiverse is a fluctuation in the stationary, fractal 
``mega-multiverse,'' in which an infinite sequence of multiverse 
productions occurs.

The framework discussed here does not suffer from problems/paradoxes 
plaguing other measures proposed earlier, such as the youngness paradox, 
the Boltzmann brain problem, and a peculiar ``end'' of time.}

\newpage

\tableofcontents

\end{center}
\end{titlepage}

\section{Introduction and Summary}
\label{sec:intro}

A combination of eternal inflation~\cite{Guth:1982pn} and the string 
landscape~\cite{Bousso:2000xa} provides a theoretically well-motivated 
framework in which we might understand how nature selects our local 
universe to take the form as we see today.  On one hand, the string 
landscape provides an enormous number of quantum field theories, of 
$O(10^{500})$ or more, as consistent vacuum solutions to string theory. 
On the other hand, eternal inflation physically realizes all these 
vacua in spacetime, allowing us to have an anthropic understanding 
for the structure of physical theories~\cite{Barrow-Tipler}, including 
the value of the cosmological constant~\cite{Weinberg:1987dv}.

This elegant picture, however, suffers from the issue of predictivity 
arising from infinities associated with eternally inflating spacetime, 
known as the measure problem~\cite{Guth:2000ka}.  In fact, it is often 
said that {\it ``In an eternally inflating universe, anything that can 
happen will happen; in fact, it will happen an infinite number of 
times''}~\cite{Guth:2000ka-2}.  While this sentence captures well 
the origin of the issue, its precise meaning is not as clear as one 
might naively think.  Consider the part ``anything that can happen 
will happen.''  One might think it to mean that for a given initial 
state, anything that is allowed to happen in the landscape will happen. 
However, any single ``observer'' (geodesic) traveling in eternally 
inflating spacetime will see only a finite number of vacua before 
he/she hits a (big crunch or black hole) singularity or is dissipated 
into time-like infinity of Minkowski space.  Namely, for this observer, 
something that can happen may not happen.  We are thus led to interpret 
the phrase to mean either ``anything that can happen will happen with 
a nonzero probability'' or ``anything that can happen will happen at 
least to somebody.''  The former interpretation, however, leads to the 
question of what probability we are talking about, and the latter the 
question of whether considering the histories of {\it all} different 
observers physically makes sense despite the fact that they can never 
communicate with each other at late times.  Similar questions also 
apply to the part ``it will happen an infinite number of times.'' 
Indeed, any randomly selected observer will see a particular event 
only a finite number of times.

The first aim of this paper is to present a view on these and related 
issues in eternal inflation, by providing a well-defined framework for 
prediction in eternally inflating spacetime, i.e.\ in the eternally 
inflating multiverse.  Our framework is based on the following three 
basic observations:
\begin{itemize}
\item
In any physical theory, predicting (or postdicting) something means 
that, given what we ``already know,'' we obtain information about 
something we do {\it not} know.  In fact, there are two aspects in 
this:\ one is the issue of dynamical evolution and the other about 
probabilities.  Suppose one wants to predict the future (or explore the 
past) in a classical world, given perfect knowledge about the present. 
This requires one to know the underlying, dynamical evolution law, 
but only that.  On the other hand, if one's future is determined 
only probabilistically, or if one's current knowledge is imperfect, 
then making predictions/postdictions requires a definition of the 
probabilities, {\it in addition to} knowing the evolution law.  The 
measure problem in eternal inflation is of this second kind.
\item
Let us assume that the underlying evolution law is known (e.g.\ 
the landscape scalar potential in string theory).  Making 
predictions/postdictions is then {\it equivalent} to providing 
relative weights among all physical possibilities that are consistent 
with the information we already have.  In particular, since our 
knowledge is in principle limited to what occurred within our past 
light cone, having a framework for prediction/postdiction is the same 
as coming up with a prescription for finding appropriate {\it samples 
for past light cones} that are consistent with our prior knowledge.
\item
From the point of view of a single observer, infinities in eternal 
inflation arise only if one repeats ``experiments'' an infinite number 
of times, i.e.\ only if the observer travels through the multiverse 
an infinite number of times.  This implies that by considering a huge, 
but finite, number of observers emigrating from a fixed initial region 
at a very early moment, only a finite number of---indeed, only a very 
small fraction of---observers see past light cones that are consistent 
with the conditions we specify as our prior knowledge.  In particular, 
if our knowledge contains exact information about any non-static 
observable, then the number of past light cones satisfying the prior 
conditions {\it and} encountered by one of the observers is always 
finite, no matter how large the number of observers we consider.
\end{itemize}

These observations naturally lead us to introduce the following general 
framework for making predictions, which consists of two elements:
\begin{itemize}
\item[(i)]
We first phrase a physical question in the form:\ given a set of conditions 
$A$ imposed on a past light cone, what is the probability of this 
light cone to have a property $B$ or to evolve into another past light 
cone having a property $C$?  We will argue that any physical questions 
associated with prediction/postdiction can always be formulated in 
this manner.
\item[(ii)]
We then find an ensemble of past light cones satisfying conditions $A$ 
by ``scanning'' what observers {\it actually see}.  Specifically, we 
consider a set of geodesics emanating from randomly distributed spacetime 
points on a fixed, ``initial'' (space-like or null) hypersurface at a 
very early moment in the history of spacetime.  We then keep track of 
past light cones along each of these geodesics, and if we find a light 
cone that satisfies $A$, then we ``record'' it as an element of the 
ensemble.  Answers to any physical questions can then be obtained by 
simply counting the numbers of relevant light cones:
\begin{equation}
  \frac{{\cal N}_{A \cap B}}{{\cal N}_A} \rightarrow P(B|A),
\qquad\quad
  \frac{{\cal N}_{A \rightarrow C}}{{\cal N}_A} \rightarrow P(C|A),
\label{eq:intro-def}
\end{equation}
where ${\cal N}_A$, ${\cal N}_{A \cap B}$, and ${\cal N}_{A \rightarrow C}$ 
are the numbers of the recorded light cones that satisfy the specified 
conditions.  Note that these numbers are countable and finite if conditions 
$A$ involve a specification of any non-static observable, e.g.\ the 
value of a parameter that can play the role of time.  The predictions 
in Eq.~(\ref{eq:intro-def}) can be made arbitrary precise by making 
the number of the geodesics arbitrarily large.
\end{itemize}
\begin{figure}[t]
\begin{center}
\begin{picture}(300,190)(0,0)
  \GOval(150,30)(20,30)(0){0.8} \Text(184,21)[lt]{\large $\Sigma$}
  \Vertex(125.7,32.3){1.5} \LongArrowArc(-51,-32)(188,20,64)
  \DashLine(117.0,52.4)(117,39){2} \DashLine(117.0,52.4)(126,43){2}
  \Line(93.4,86.4)(94.7,87.7)
  \Line(93.8,83.8)(96.5,86.5)
  \Line(94.3,81.3)(98.4,85.4)
  \Line(94.7,78.7)(100.2,84.2)
  \Line(95.1,76.1)(102.0,83.0)
  \Line(95.6,73.6)(103.9,81.9)
  \Line(96.0,71.0)(105.7,80.7)
  \Line(96.4,68.4)(107.5,79.5)
  \Line(96.9,65.9)(109.3,78.3)
  \Line(103.4,69.4)(111.2,77.2)
  \Line(93.4,88.2)(97.4,64.4) \Line(93.4,88.2)(113.4,75.4)
  \Line(93.0,88.8)(97.0,65.0) \Line(93.0,88.8)(113.0,76.0)
  \Line(92.6,89.4)(96.6,65.6) \Line(92.6,89.4)(112.6,76.6)
  \DashLine(61.3,118.8)(70,95){2}  \DashLine(61.3,118.8)(88,111){2}
  \Vertex(138.1,25.1){1.5} \LongArrowArc(-118,-20)(260,10,45)
  \DashLine(126.3,68.9)(124,53){2} \DashLine(126.3,68.9)(137,56){2}
  \DashLine(95.0,129.1)(97,106){2} \DashLine(95.0,129.1)(117,116){2}
  \Vertex(145.7,36.7){1.5} \LongArrowArc(-274,22)(420,2,22)
  \Line(142.6,55.6)(145.3,58.3)
  \Line(141.4,51.4)(146.2,56.2)
  \Line(143.8,50.8)(147.1,54.1)
  \Line(147.3,51.3)(148.0,52.0)
  \Line(144.2,60.4)(141,50.4) \Line(144.2,60.4)(148,51.4)
  \Line(144.2,61.0)(141,51.0) \Line(144.2,61.0)(148,52.0)
  \Line(144.2,61.6)(141,51.6) \Line(144.2,61.6)(148,52.6)
  \DashLine(136.8,109.3)(133,91){2}  \DashLine(136.8,109.3)(147,94){2}
  \Line(122.7,151.7)(125.6,154.6)
  \Line(121.7,147.7)(127.1,153.1)
  \Line(120.7,143.7)(128.6,151.6)
  \Line(119.7,139.7)(130.1,150.1)
  \Line(118.7,135.7)(131.6,148.6)
  \Line(119.3,133.3)(133.1,147.1)
  \Line(123.1,134.1)(134.6,145.6)
  \Line(126.8,134.8)(136.1,144.1)
  \Line(130.6,135.6)(137.5,142.5)
  \Line(134.4,136.4)(139.0,141.0)
  \Line(138.2,137.2)(140.5,139.5)
  \Line(124.1,155.8)(118.2,132.4) \Line(123.9,155.8)(142,137.4)
  \Line(123.9,156.4)(118.0,133.0) \Line(123.9,156.4)(142,138.0)
  \Line(123.7,157.0)(117.8,133.6) \Line(123.9,157.0)(142,138.6)
  \Vertex(150.0,18.0){1.5} \LongArrowArcn(670,18)(520,180,162)
  \DashLine(150.7,45.2)(147,35){2}   \DashLine(150.7,45.2)(155,36){2}
  \DashLine(156.4,99.3)(146,81){2}   \DashLine(156.4,99.3)(162,80){2}
  \DashLine(167.7,152.6)(148,131){2} \DashLine(167.7,152.6)(175,126){2}
  \Vertex(158.9,39.3){1.5} \LongArrowArcn(410,-5)(255,170,138)
  \DashLine(169.1,78.6)(159,66){2}   \DashLine(169.1,78.6)(171,63){2}
  \DashLine(195.8,133.3)(179,121){2} \DashLine(195.8,133.3)(194,113){2}
  \Vertex(167.7,22.4){1.5} \LongArrowArcn(354,-24)(192,166,120)
  \DashLine(175.6,46.9)(167,37){2}   \DashLine(175.6,46.9)(177,34){2}
  \Line(197.7,89.3)(199.8,87.2)
  \Line(195.8,88.2)(199.5,84.5)
  \Line(194.0,87.0)(199.2,81.8)
  \Line(192.2,85.8)(198.8,79.2)
  \Line(190.3,84.7)(198.5,76.5)
  \Line(188.5,83.5)(198.1,73.9)
  \Line(186.7,82.3)(197.8,71.2)
  \Line(184.8,81.2)(195.3,70.7)
  \Line(183.0,80.0)(188.9,74.1)
  \Line(181.2,78.8)(182.5,77.5)
  \Line(200.7,91.5)(181,79)     \Line(200.7,91.5)(198,70)
  \Line(200.3,91.0)(180.6,78.5) \Line(200.3,91.0)(197.6,69.5)
  \Line(199.9,90.5)(180.2,78.0) \Line(199.9,90.5)(197.2,69.0)
  \DashLine(236.7,128.0)(212,120){2} \DashLine(236.7,128.0)(228,104){2}
  \Text(165.2,22.4)[r]{$p_i$} \Text(262,141)[l]{$g_i$}
\end{picture}
\caption{A schematic picture for obtaining samples of past light cones 
 that satisfy specified conditions $A$.  The light cones to be selected 
 are depicted by shaded triangles (with the figure showing ${\cal N}_A 
 = 4$).  Note that a single geodesic may encounter relevant light 
 cones multiple times in its history.}
\label{fig:presc}
\end{center}
\end{figure}
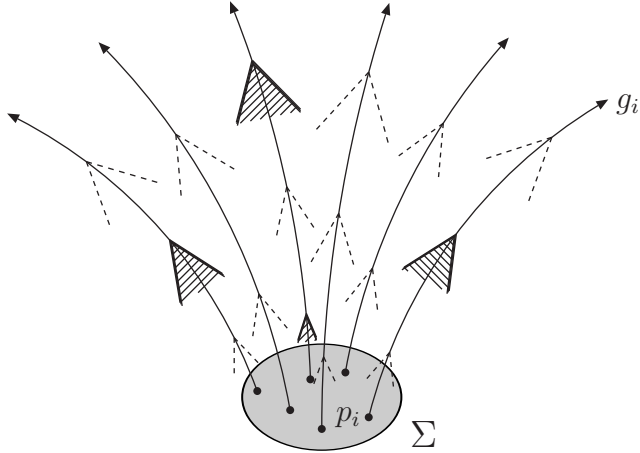
A schematic depiction of the procedure of sampling past light cones 
is given in Fig.~\ref{fig:presc}.  Note that this procedure corresponds 
precisely to simulating the entire multiverse many times as viewed from 
a single observer (geodesic).  Also, in this method, any time coordinate 
becomes simply a spurious parameter, exactly like a variable $t$ used 
in a parametric representation of a curve on a plane, $(x(t), y(t))$. 
It plays {\it no role} in defining probabilities, and time evolution of 
a physical quantity $X$ is nothing more than a correlation between $X$ 
and a quantity that can play the role of time, such as the average 
temperature of the cosmic microwave background (CMB) in our universe.

It is illuminating to highlight a close similarity between the prescription 
here and the formulation of usual quantum mechanics.  Consider a quantum 
mechanical system in a state $\left| \psi \right> = \sum_i a_i \left| 
\phi_i \right>$ with $a_i \neq 0$.  If we perform an experiment many 
times on this system, we would find that anything that can happen 
(represented by $\left| \phi_i \right>$) will happen (with a nonzero 
probability $|a_i|^2/\sum_j |a_j|^2$).  Moreover, by performing it 
an infinite number of times, all these results occur an infinite number 
of times.  In our prescription, $\left| \psi \right>$ corresponds to 
the multiverse, while $\left| \phi_i \right>$ to a collection of past 
light cones along a geodesic in this spacetime.  Performing an 
experiment then corresponds to sending an observer (geodesic) from 
an initial hypersurface located at the ``beginning'' of spacetime.

The second aim of this paper is to formulate the above framework, 
described using the semi-classical picture, in the context of quantum 
mechanics, where we will see that the similarity just illustrated is 
more than an analogy.  The necessity of a quantum mechanical treatment 
may sound obvious since we live in a quantum mechanical world, but it 
is much more fundamental than one might naively think.  Consider a set 
of geodesics that scan past light cones dense enough at late times, 
i.e., when conditions $A$ can be satisfied.  Because of the rapid 
exponential expansion of spacetime, these geodesics must have been 
much closer at early times---in fact, there are generically a huge 
number of geodesics emanating from an early space-like hypersurface 
of the Planck size.  On the other hand, theories of quantum gravity 
suggest that the Planck size region can contain only $O(1)$ (or smaller) 
information.  How can such a state evolve into many different universes 
in the future?  Is it reasonable to expect that the procedure~(ii) 
above, based on the classical spacetime picture, can select the 
appropriate ensemble of past light cones, despite the fact that 
it involves scales shorter than the Planck length at early times?

We find that quantum mechanics plays a crucial role in answering 
these questions---in fact, the ultimate resolution of the measure 
problem {\it requires} quantum mechanical interpretation of the 
multiverse.  Our basic assumption here is that the laws of quantum 
mechanics---deterministic, unitary evolution of quantum states and the 
superposition principle---are not violated when physics is described 
from an observer's point of view.  As we will argue, this implies 
that the {\it complete} description of the multiverse can be obtained 
{\it purely} from the viewpoint of a single observer traveling 
the multiverse.  All the information on the multiverse is contained 
in the (stretched) apparent horizon and spacetime therein, as 
seen from that observer.  This situation is similar to describing 
a black hole from the viewpoint of a distant observer using 
``complementarity''~\cite{Susskind:1993if}---indeed the situation 
for a black hole arises as a special case of our general description. 
Our construction is strongly motivated by such complementarity view of 
spacetime, as well as the holographic principle~\cite{'tHooft:1993gx}.

To exemplify this picture further, let us consider that spacetime was 
initially in a highly symmetric configuration, e.g.\ four dimensional 
de~Sitter spacetime, with the fields sitting in a local minimum of 
the potential.  From the semi-classical analysis, we know that even a 
tiny region of this initial configuration evolves into infinitely many 
different universes.  On the other hand, the holographic principle, 
or de~Sitter entropy, says that such a small region can have only 
finite degrees of freedom.  This implies that in the quantum picture, 
the origin of various semi-classical universes cannot be attributed to 
the difference of initial conditions---they must arise as different 
``outcomes'' of a quantum state $\Psi$, which is uniquely determined 
once an initial condition is given.  This clearly answers one question 
raised above:\ how can an initial state having only $O(1)$ information 
evolve into different universes?  In fact, the initial state evolves into 
the {\it unique} future state, which however is a probabilistic mixture 
of {\it different} (semi-classical) universes.  The meaning of the 
sampling procedure in (ii) also becomes clear---sending a geodesic 
corresponds to making a ``measurement'' on $\Psi$, and the sampling 
corresponds to {\it defining} probabilities through ``repeated 
measurements.''  In particular, we need not take too seriously the 
sub-Planckian distances appearing in the procedure---the semi-classical 
picture of the multiverse is simply a pictorial way of representing 
probabilistic processes, e.g.\ bubble nucleation processes, occurring 
in the quantum universe.%
\footnote{While completing this paper, I learned that Raphael Bousso 
 also arrived at a similar picture in the context of geometric 
 cutoff measures~\cite{Bousso}.}

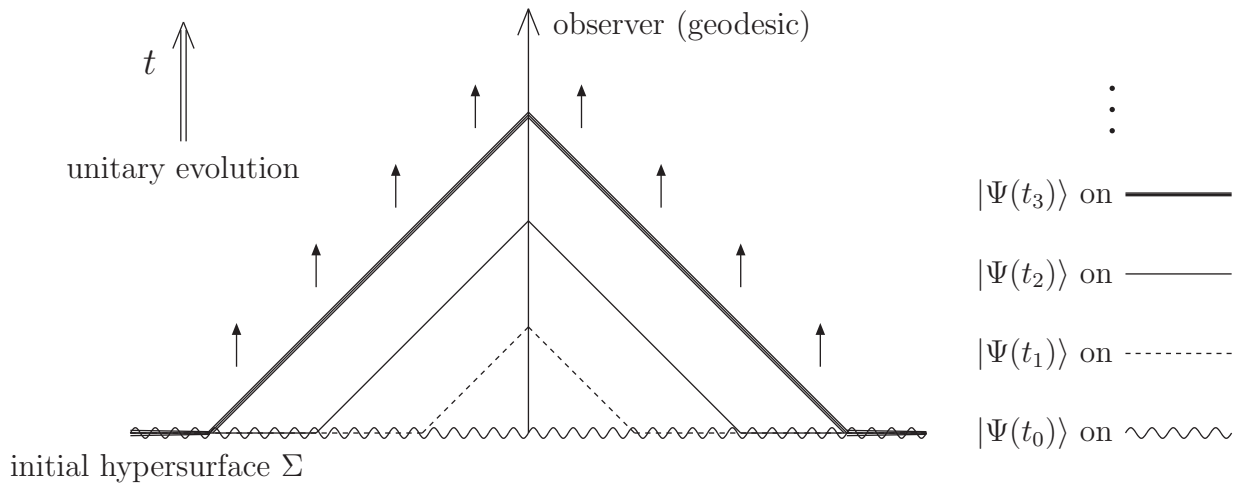
\begin{figure}[t]
\begin{center}
\begin{picture}(460,200)(-45,-5)
  \Photon(0,20)(300,20){2}{40}
  \Text(10,12)[t]{initial hypersurface $\Sigma$}
  \Line(150,20)(150,180) \Line(150,180)(146,170) \Line(150,180)(154,170)
  \Text(160,180)[lt]{observer (geodesic)}
  \Line(19,130)(19,172) \Line(21,130)(21,172)
  \Line(20,175)(16,165) \Line(20,175)(24,165)
  \Text(10,165)[rt]{\large $t$}
  \Text(20,125)[t]{unitary evolution}
  \DashLine(110,20)(150,60){2} \DashLine(190,20)(150,60){2}
  \DashLine(0,20)(110,20){2} \DashLine(190,20)(300,20){2}
  \Line(70,20)(150,100) \Line(230,20)(150,100)
  \Line(0,20)(70,20) \Line(230,20)(300,20)
  \Line(30,20)(150,140) \Line(31,20)(150,139) \Line(29,20)(150,141)
  \Line(270,20)(150,140) \Line(269,20)(150,139) \Line(271,20)(150,141)
  \Line(0,20)(30,20) \Line(0,21)(30,20.5) \Line(0,19)(30,19.5)
  \Line(270,20)(300,20) \Line(270,21)(300,20.5) \Line(270,19)(300,19.5)
  \LongArrow(40,45)(40,60) \LongArrow(260,45)(260,60)
  \LongArrow(70,75)(70,90) \LongArrow(230,75)(230,90)
  \LongArrow(100,105)(100,120) \LongArrow(200,105)(200,120)
  \LongArrow(130,135)(130,150) \LongArrow(170,135)(170,150)
  \Text(320,20)[l]{$\left| \Psi(t_0) \right>$ on}
  \Photon(375,20)(415,20){2}{5.33}
  \Text(320,50)[l]{$\left| \Psi(t_1) \right>$ on}
  \DashLine(375,50)(415,50){2}
  \Text(320,80)[l]{$\left| \Psi(t_2) \right>$ on}
  \Line(375,80)(415,80)
  \Text(320,110)[l]{$\left| \Psi(t_3) \right>$ on}
  \Line(375,110)(415,110) \Line(375,110.5)(415,110.5) 
  \Line(375,109.5)(415,109.5)
  \Vertex(370,134){1}
  \Vertex(370,142){1}
  \Vertex(370,150){1}
\end{picture}
\caption{The entire multiverse can be described purely from the viewpoint 
 of a single ``observer'' in terms of a quantum state $\left| \Psi(t) 
 \right>$, which is defined on the observer's past light cones (and 
 on the initial hypersurface $\Sigma$ at early times; later, the light 
 cones are bounded by the apparent horizons).  Once the initial condition 
 is given, the state is uniquely determined according to unitary, quantum 
 mechanical evolution.}
\label{fig:states}
\end{center}
\end{figure}

The remaining task to define a complete, quantum mechanical framework 
for prediction is to come up with the explicit formalism of implementing 
procedure~(ii), given originally within the semi-classical picture. 
To do so, we first define $\Psi$ more carefully.  For simplicity, we 
take it to be a pure state $\left| \Psi \right>$ (although an extension 
to the mixed state case is straightforward).  A crucial point is that 
we describe the system {\it from a single observer's point of view}. 
This allows us to consider the state $\left| \Psi(t) \right>$, where the 
time {\it parameter} $t$ is taken as a proper time along the observer, 
which can be assigned a simple, invariant meaning.  (We take the 
Schr\"{o}dinger picture throughout.)  We consider that $\left| \Psi(t) 
\right>$ is defined on observer's past light cones bounded by the 
(stretched) apparent horizons, as viewed from the observer.  This 
restriction on spacetime regions comes from consistency of quantum 
descriptions for systems with gravity.  If we provide an initial condition 
on a space-like hypersurface $\Sigma$, then the state is given on past 
light cones and $\Sigma$ at early times (as in Fig.~\ref{fig:states}), 
and later, on past light cones inside and at the horizons (see 
Figs.~\ref{fig:stretched}, \ref{fig:apparent} and Eq.~(\ref{eq:H-decomp}) 
in Section~\ref{subsec:single}).

In general, the state $\left| \Psi(t) \right>$ can be expanded as
\begin{equation}
  \left| \Psi(t) \right> = \sum_i c_i(t) \left| \alpha_i \right>,
\label{eq:Psi-t}
\end{equation}
where $\left| \alpha_i \right>$ span a complete set of orthonormal states 
in the landscape (i.e.\ all possible past light cones within the horizons), 
and the index $i$ may take continuous values (in which case the sum over 
$i$ should be interpreted as an integral).  Here, we have taken the basis 
states $\left| \alpha_i \right>$ to be independent of $t$, so that all 
the $t$ dependence of $\left| \Psi(t) \right>$ come from those of the 
coefficients $c_i(t)$.  Once the initial condition is given, the state 
$\left| \Psi(t) \right>$ is determined uniquely.  (Our framework does 
not provide the initial condition; rather, it is modular with respect 
to initial conditions.)

We next introduce an operator ${\cal O}_A$ which projects onto the states 
consistent with conditions A imposed on the past light cone:
\begin{equation}
  {\cal O}_A = \sum_i \left| \alpha_{A,i} \right> \left< \alpha_{A,i} \right|,
\label{eq:O_A}
\end{equation}
where $\left| \alpha_{A,i} \right>$ represent the complete set of states 
that satisfy conditions $A$.  The conditions can be imposed either within 
or on the past light cone, since they can be transformed to each other 
using (supposedly known) deterministic, quantum evolution.  Note that, 
as long as conditions $A$ are physical (i.e.\ do not depend on arbitrary 
parametrization $t$), the operator ${\cal O}_A$ does not depend on $t$, 
so that it can be written in the form of Eq.~(\ref{eq:O_A}).

The probability that a past light cone satisfying $A$ also has a property 
$B$ is then given by
\begin{equation}
  P_\Sigma(B|A) = \frac{\int\!dt \left< \Psi(t) \right| 
    {\cal O}_{A \cap B} \left| \Psi(t) \right>}
    {\int\!dt \left< \Psi(t) \right| {\cal O}_A \left| \Psi(t) \right>} 
  \rightarrow
    \frac{\bigl[ \sum_{i,j} |c_i(t_j)|^2 \bigr]_{A \cap B}}
    {\bigl[ \sum_{i,j} |c_i(t_j)|^2 \bigr]_A},
\label{eq:probability}
\end{equation}
which is the quantum version of the probability $P(B|A)$ in 
Eq.~(\ref{eq:intro-def}).  Here, we have put the subscript $\Sigma$ 
to remind us that the probability depends on initial conditions. 
The rightmost expression arises when conditions $A$ specify an exact 
``time,'' e.g.\ by requiring a particular value for a parameter that 
has smooth $t$ dependence; in such a case, the conditions select definite 
(discrete) times $t_j$.  Note that specifying physical ``time'' is 
different from specifying time {\it parameter} $t$; in fact, the 
same ``time'' can occur multiple times in the history at $t = t_j$ 
($j = 1,\cdots$).

The probability defined in Eq.~(\ref{eq:probability}) inherits the 
following crucial properties from the semi-classical definition:
\begin{itemize}
\item
{\bf well-defined} --- While there are cosmic histories which encounter 
$A$-satisfying past light cones an infinite number of times, they compose 
only a measure zero set.  In particular, histories encountering $n$ 
such light cones occur only with $e^{-c n}$ probabilities with $c > 0$, 
because, starting from any generic state, the probability of encountering 
$A$-satisfying light cones is exponentially small.  This makes the sums 
(time integrals) in Eq.~(\ref{eq:probability}) converge, and so makes 
$P_\Sigma(B|A)$ well-defined.
\item
{\bf ``gauge invariant''} --- The probability $P_\Sigma(B|A)$ does not 
depend on parametrization of time $t$, as is clear from the expressions 
in Eq.~(\ref{eq:probability}).  The role of time in making predictions 
can then be played by a physical quantity that has smooth dependence 
on $t$~\cite{DeWitt:1967yk}, which need not be defined globally in 
the entire spacetime.  In fact, it can be arbitrary information on any 
physical system, as long as it is not about a static property of the 
system.
\end{itemize}
Since quantum evolution of the state is deterministic, the probability 
for a future event $C$ to happen, $P_\Sigma(C|A)$, is also calculated 
once the values of $P_\Sigma(B|A)$ are known for all $B$'s that could 
affect the likelihood for $C$ to occur (assuming, of course, that the 
evolution law is known).

In principle, the definition of Eq.~(\ref{eq:probability}) can be used 
to answer any physical questions associated with prediction and postdiction 
in the multiverse.  However, since $\left| \Psi(t) \right>$ involves 
degrees of freedom on horizons, as well as those in the bulk of spacetime, 
its complete evolution can be obtained only with the knowledge of quantum 
gravity.  This difficulty is avoided if we focus on the bulk degrees of 
freedom, by considering a {\it bulk density matrix}
\begin{equation}
  \rho_{\rm bulk}(t) = {\rm Tr}_{\rm horizon} 
    \left| \Psi(t) \right> \left< \Psi(t) \right|,
\label{eq:rho_bulk-intro}
\end{equation}
where ${\rm Tr}_{\rm horizon}$ represents the partial trace over 
the horizon degrees of freedom.  This description allows us to make 
predictions/postdictions without knowing quantum gravity, since the 
evolution of $\rho_{\rm bulk}(t)$ can be determined by semi-classical 
calculations in low energy quantum field theories.  The cost is that 
(apparent) unitarity violation is induced in processes involving 
horizons, such as black hole evaporation.

Our framework has a number of implications.  Major ones are:
\begin{itemize}
\item
{\bf The measure problem in eternal inflation is solved.}  The principle 
behind the solution is quantum mechanics, where quantum states are 
defined in the spacetime region bounded by (stretched) apparent horizons 
{\it as viewed from a single ``observer'' (geodesic)}.  No ad~hoc cutoff 
needs to be introduced, and problems/paradoxes plaguing other measures 
are absent.  Transformations between descriptions based on different 
observers are possible, but they are in general highly nonlocal when 
the observers are not in causal contact at late times.
\item
{\bf The multiverse and many worlds in quantum mechanics are the same.} 
Our framework provides a {\it unified treatment} of quantum measurement 
processes and the eternally inflating multiverse, usually associated 
with vastly different scales---smaller than atomic and much larger than 
the universe.  The probabilities in microscopic quantum processes and 
in the multiverse are both understood as ``branching'' in the entire 
multiverse state $\left| \Psi(t) \right>$ (or equivalently, as a 
``misalignment'' in Hilbert space between $\left| \Psi(t) \right>$ 
and the basis of local observables).%
\footnote{The conjecture that the multiverse and many worlds may be the 
 same thing appeared earlier in Ref.~\cite{Susskind}.  (I thank Leonard 
 Susskind for bringing this to my attention.)  Here we show that they are, 
 in fact, identical, following the same formula for the probabilities, 
 Eqs.~(\ref{eq:probability-AB}) and (\ref{eq:probability-AC}).  This 
 provides complete unification of the two concepts.}
\item
{\bf Global spacetime can be viewed as a derived concept.}  By rearranging 
various terms in the multiverse state $\left| \Psi(t) \right>$, the 
global spacetime picture can be ``reconstructed.''  In particular, this 
allows us to connect the quantum probabilities with the semi-classical 
probabilities, defined using geodesics traveling through global 
spacetime of the multiverse.
\item
{\bf The multiverse is a transient phenomenon while relaxing into 
a supersymmetric Minkowski state.}  Our framework suggests that the 
ultimate fate of the multiverse is a supersymmetric Minkowski state, 
which is free from further decay.  The other components in $\left| \Psi(t) 
\right>$, e.g.\ ones hitting big crunch or black hole singularities, 
simply ``disappear.''  This leads to the picture that our entire 
multiverse arises as a transient phenomenon, during the process of 
some initial state (either determined by quantum gravity or generated 
as a ``fluctuation'' in a larger structure) relaxing into a stable, 
supersymmetric final state.
\end{itemize}
Other implications are also discussed throughout the paper.

We emphasize that the multiverse state $\left| \Psi(t) \right>$ is 
literally ``everything'' in terms of making predictions---even we 
ourselves appear as {\it a part of} $\left| \Psi(t) \right>$ at some 
time(s) $t_j$.  Once the state is given, any physical predictions 
can be obtained using Eq.~(\ref{eq:probability}), which is nothing 
but the standard Born rule.  There is no need to introduce wavefunction 
collapse, environmental decoherence, or anything like those---indeed, 
there is no ``external observer'' that performs measurements on 
$\left| \Psi(t) \right>$, and there is no ``environment'' with which 
$\left| \Psi(t) \right>$ interacts.  From the viewpoint of a single 
observer (geodesic), probabilities keep being ``diluted'' because 
of continuous branching of the state into different semi-classical 
possibilities, caused by the fact that the evolution of $\left| 
\Psi(t) \right>$ is not along an axis in Hilbert space determined 
by operators local in spacetime.  Indeed, the probabilities are 
constantly being dissipated into (fractal) Minkowski space, which 
acts as an infinite reservoir of (coarse-grained) entropy---this is 
the ultimate reason behind the well-definedness of the probabilities. 
It is quite remarkable that this simple and satisfactory picture comes 
with a drastic change of our view on spacetime:\ {\it the entire world 
can be described within the spacetime region inside the causal patch 
of a single geodesic, bounded by its (stretched) apparent horizons}. 
All the information about the world is contained in a {\it single} 
quantum state $\left| \Psi(t) \right>$ defined in this region---we do 
not even need to consider an ensemble of quantum states.  In this sense, 
we may even say that spacetime exists only to the extent that it can 
be accessed directly by {\it the} single ``observer'' (geodesic).

The organization of this paper is the following.  In 
Section~\ref{sec:prediction}, we provide detailed discussions 
on what prediction and postdiction mean in physical theories, 
especially in the context of the multiverse.  We present precise 
definitions of prediction/postdiction, as well as their practical 
implementations in realistic contexts.  In Section~\ref{sec:framework}, 
we present our explicit framework for prediction, using the 
semi-classical picture; but we will also see that the framework 
cries for a quantum mechanical treatment of the multiverse.  Based 
on these results, in Section~\ref{sec:quantum} we introduce a fully, 
quantum mechanical framework for prediction.  We carefully consider 
quantum states, and argue that they must be defined in the region 
within (stretched) apparent horizons, viewed from a single ``observer'' 
traveling through the multiverse.  We also introduce an approximation 
scheme in which horizon degrees of freedom are integrated out, so that 
the probabilities can be calculated without knowing quantum gravity.

The issue of initial conditions is discussed in Section~\ref{sec:initial}, 
both from the perspective of making predictions in eternally inflating 
universe and addressing fundamental questions regarding spacetime.  In 
Section~\ref{sec:measure-spacetime}, we present a unified picture of the 
eternally inflating multiverse and quantum measurement processes---in 
fact, this unification is an automatic consequence of our treatment 
of the multiverse.  We also discuss a connection of our picture, based 
on a single ``observer,'' with the global spacetime picture.  In 
Section~\ref{sec:no-problem}, we show that the framework does not suffer 
from problems plaguing other measures proposed so far~\cite{Guth:2000ka} 
based on geometric cutoffs; in particular, it avoids a peculiar 
conclusion that time should ``end''~\cite{Bousso:2010yn}.  Finally, 
in Section~\ref{sec:discuss} we summarize what we learned about eternal 
inflation, and discuss further issues such as the beginning/fate of the 
multiverse, the possibility of a fully holographic description, and a 
structure even larger than the entire multiverse (``mega-multiverse'') 
which is suggested by extrapolating our framework to the extreme.

Appendix~\ref{app:fuzzy} discusses an interpretation of the framework 
in terms of ``fuzzy'' time cutoff.  Appendix~\ref{app:calc} provides 
sample calculations of probabilities in toy landscapes, both in the 
semi-classical and quantum pictures.  Appendix~\ref{app:no-cloning} 
presents an analysis of a gedanken experiment which provides a nontrivial 
consistency check of our framework.

\section{What Questions Should We Ask?}
\label{sec:prediction}

In this section, we discuss in detail what are ``legitimate'' physical 
questions we would ask in the context of the eternally inflating 
multiverse.  For example, we may want to ask what are the results of 
future measurements, what is the probability of certain Lagrangian 
describing our universe, or if the observations we have already made 
are consistent with the assumption of typicality in the multiverse. 
We will see that answering these questions is boiled down to coming 
up with a well-defined prescription for selecting appropriate samples 
of past light cones (either finite or infinite numbers) that are 
consistent with our ``prior knowledge'' about our past light cone. 
The actual prescription will be given in the next section, where we 
will see that it naturally calls for a quantum mechanical treatment 
of the multiverse.

\subsection{Predictions in a strict sense}
\label{subsec:strict}

Suppose you want to ``predict'' what physics you will see at the LHC 
experiments, assuming you have a perfect knowledge about the underlying 
theory governing evolution of systems, including the multi-dimensional 
scalar potential in the string landscape (but, of course, {\it not} 
about which vacuum you live).  Specifically, you want to know what is the 
probability for you to find that weak scale supersymmetry, technicolor, 
just the standard model, or something else, is the Lagrangian describing 
your local universe, i.e.\ physics within your Hubble horizon.%
\footnote{If the LHC has answers to these questions by the time you read 
 this paper, then you should appropriately replace statements below with 
 those regarding some other experiment whose results are not yet available 
 to you.}

In the multiverse context, what this ``really'' means is the following. 
First, you specify all you know about your past light cone.%
\footnote{Here, I consider ``you'' as a point in spacetime, for simplicity. 
 To be precise, ``you'' must (at least) be some pattern of electromagnetic 
 activities (neural signals) corresponding to a brain in your ``current'' 
 moment, which occur inside but around the tip of the light cone.  In 
 fact, if you are not a Boltzmann brain, which we assume throughout this 
 section, your entire body (indeed, the entire you from the birth to the 
 present) must exist inside the light cone.  For more discussions on these 
 and related issues, see Section~\ref{subsec:paradoxes}. \label{ft:you}}
This includes the existence of you, the room around you (if you are reading 
this in a room), the earth, and (the observable part of) the universe. 
You then ask what is the probability of this ``initial situation''---i.e.\ 
the particular configuration within your past light cone---to evolve into 
a certain ``final situation''---a past light cone lying in your future 
in which you will somehow learn that physics at the TeV scale is, e.g., 
weak scale supersymmetry (either from a friend, paper, ...).  In fact, 
possible past light cones in your future must include ones in which the 
LHC fails, e.g.\ due to some accidents, so that no relevant measurement 
will be made.  After including all these possible (mutually exclusive) 
futures, the probabilities should add up to one.

Of course, you practically do not know everything you need to specify 
{\it uniquely} your past light cone.  (You even do not know the location 
of a coffee cup next to you with infinite precision.)  This implies that, 
even at the classical level, the ``initial situation'' you prepare must 
be an ensemble of ``situations''---i.e.\ a collection of past light 
cones---that are consistent with the knowledge you have about your 
past light cone.%
\footnote{In fact, in a classical world, if you have a {\it complete} 
 knowledge about your past light cone, you would not even ask about 
 future probabilities, since the physics is fully deterministic 
 (assuming the initial conditions at the earliest moment are also 
 known).  For example, you would be able to ``calculate'' if the LHC 
 fails or not, and you would know physics at very high energies through 
 tiny effects encoded in higher dimension operators in the effective 
 theory.  This apparently does not sound to be the case in a quantum 
 world, since even if you completely specify your past light cone, 
 outcome of future experiments can only be determined probabilistically. 
 However, for a given initial quantum state, its future sates {\it are} 
 uniquely determined.  In fact, the concept of making ``predictions'' 
 in the sense considered here is relevant only if our knowledge about 
 the current state of the system is incomplete.}
Therefore, a specification of the ``initial situation'' requires some 
method that assigns relative weights among these ``initial past light 
cones''---namely, you need to find samples of past light cones that 
``faithfully'' represent incompleteness of your knowledge.  In fact, 
in the global spacetime picture of the multiverse, the number of past 
light cones that satisfy any input knowledge will be infinite, so some 
regularization prescription will be needed to define the samples.  (This 
is the measure problem in the eternally inflating multiverse.)  On the 
other hand, once such samples are given, it is {\it conceptually} a 
straightforward matter to work out predictions, following the procedure 
described above.

\subsection{Predictions in practice}
\label{subsec:practice}

While the definition of predictions described in the previous subsection 
is ``rigorous,'' it is not very practical.  In practice, when we ask 
questions, e.g., about physics at the TeV scale, we only want to keep 
track of global quantities which are (reasonably) uniform in our horizon, 
such as the CMB temperature, the size of its statistical fluctuation, 
masses of ``elementary'' particles, the gauge group, and so on.  In 
this context, how can we make predictions, e.g., about the probability 
of our local universe being described by a certain Lagrangian?  We 
emphasize that this issue is logically separate from that of defining 
probabilities; rather, the issue here is to come up with reasonable 
approximation schemes which we may hope to be tractable.

Suppose you only specify that your past light cone---more precisely the 
intersection of your past light cone and your own bubble universe---is 
described by the standard models of particle physics and cosmology at 
energies below e.g.\ a TeV, and that the CMB temperature at the tip of 
the cone is $T_{\rm CMB} = 2.725~{\rm K}$.  Here, the latter condition 
is imposed to specify a physical ``time,'' which in the previous treatment 
was done by giving a particular configuration in the past light cone 
(e.g.\ the ``current'' state of your brain; see footnote~\ref{ft:you}). 
Then the ensemble you need to prepare as an ``initial situation'' 
includes past light cones in which physics at the TeV scale is the 
minimal standard model, weak scale supersymmetry, technicolor, and 
so on (as long as these theories arise in consistent vacua in the 
landscape).

Now, in selecting these light cones, you did {\it not} impose any 
condition about yourself.  However, as long as the chance that you 
are born does not depend much on TeV-scale physics, you can expect
\begin{equation}
  \frac{{\cal N}_{\mbox{\footnotesize ``you'' in SM}}}
    {{\cal N}_{\mbox{\footnotesize SM}}}
  \sim \frac{{\cal N}_{\mbox{\footnotesize ``you'' in SUSY}}}
    {{\cal N}_{\mbox{\footnotesize SUSY}}}
  \sim \frac{{\cal N}_{\mbox{\footnotesize ``you'' in technicolor}}}
    {{\cal N}_{\mbox{\footnotesize technicolor}}}
  \sim \cdots,
\label{eq:approx-1}
\end{equation}
where ${\cal N}_{\mbox{\footnotesize ``you'' in SM}}$ and 
${\cal N}_{\mbox{\footnotesize SM}}$ (and similarly for others) are, 
respectively, the numbers of past light cones selected {\it with} and 
{\it without} the condition that you are in the light cones (around 
the tips).  In this case, you can use ${\cal N}_{\mbox{\footnotesize SM}}$, 
${\cal N}_{\mbox{\footnotesize SUSY}}$, and ${\cal N}_{\mbox{\footnotesize 
technicolor}}$, instead of ${\cal N}_{\mbox{\footnotesize  ``you'' 
in SM}}$, ${\cal N}_{\mbox{\footnotesize  ``you'' in SUSY}}$, and 
${\cal N}_{\mbox{\footnotesize  ``you'' in technicolor}}$, to compute 
relative probabilities for each TeV-scale physics describing your own 
universe.%
\footnote{To compute absolute probabilities, these numbers must be divided 
 by the number of past light cones that are selected only with the condition 
 that they are consistent with your current knowledge about the physical 
 laws in your own universe, e.g., constraints from collider and rare decay 
 experiments.}
Of course, in the context of predicting results at the LHC, ``SUSY'' 
should mean the existence of superpartners within the reach of the LHC.

Note that when selecting samples for ${\cal N}_{\mbox{\footnotesize SM}}$, 
${\cal N}_{\mbox{\footnotesize SUSY}}$, ${\cal N}_{\mbox{\footnotesize 
technicolor}}$, ..., the condition was imposed that physics below 
a TeV scale is described by the standard models of particle physics 
and cosmology, with various parameters---such as the electron mass, 
fine structure constant, dark matter abundance, and baryon-to-photon 
ratio---taking the observed values (within experimental errors).  This 
is important especially because your own existence may be strongly 
affected by the values of these parameters.  Since some of the 
parameters, e.g.\ cosmological ones, will depend highly on physics 
at high energies, the ratios between ${\cal N}_{\mbox{\footnotesize SM}}$, 
${\cal N}_{\mbox{\footnotesize SUSY}}$, ${\cal N}_{\mbox{\footnotesize 
technicolor}}$, ..., are in general {\it not} the same as the 
corresponding ratios obtained without conditioning the values of low 
energy parameters.  From the practical point of view, this makes the 
problem hard(er); but again, once appropriate samples of past light 
cones are obtained, it is straightforward to make probabilistic 
predictions for physics at the TeV scale.

\subsection{Predictions and postdictions}
\label{subsec:pred-postd}

At first sight, it might seem completely straightforward to apply the 
method described so far to ``postdict'' some of the physical parameters 
we already know---we simply need to select samples of past light cones 
without imposing any condition on those parameters.  This, however, needs 
to be done with care, because our own existence may be affected by the 
values of these parameters.  In the context of an approximation scheme 
discussed in the previous subsection, appropriate postdictions are 
obtained after specifying an ``anthropic factor'' $n(x_i)$---the probability 
of ourselves developing in the universe with parameters taking values 
$x_i$.  With this factor, the probability density for us to observe 
$x_i$ is given by
\begin{equation}
  P(x_i)\, dx_i \propto {\cal N}(x_i)\, n(x_i)\, dx_i,
\label{eq:prob-xi}
\end{equation}
where ${\cal N}(x_i)$ is the number of past light cones in which 
the parameters take values between $x_i$ and $x_i + dx_i$.  Note 
that, when selecting samples for ${\cal N}(x_i)$, we still need 
to specify the ``time'' at which postdictions are made (e.g.\ through 
$T_{\rm CMB,\, tip} = 2.725~{\rm K}$).  In fact, the level of the 
validity of Eq.~(\ref{eq:prob-xi}) is determined by how well $n(x_i)$ 
and the specification of ``time'' can mimic ourselves.%
\footnote{A simple approximation, appropriate in some cases, is obtained 
 by assuming that $n(x_i) = 1$ for habitable regions while $n(x_i) = 0$ 
 for regions hostile for life.  See e.g.\ Ref.~\cite{Hall:2007ja} for 
 more details about this approximation.}
Postdictions obtained in this way can then be used to test the multiverse 
hypothesis, by comparing them with the observed data.

A treatment of postdictions analogous to that of predictions in 
Section~\ref{subsec:strict} would require a subtle choice for the 
condition of selecting light cone samples.  Suppose we want to postdict 
the electron mass, $m_e$, and compare it with the experimental value, 
$m_{e,{\rm exp}} = 510.998910 \pm 0.000013~{\rm keV}$.  We would 
then need to prepare the ensemble of past light cones that have some 
``consciousnesses'' around the tip perceiving similar worlds as we 
see now {\it except} that the electron mass may not take the value 
$m_{e,{\rm exp}}$.  However, since worlds with different electron 
masses will not be identical, we will not be able to require these 
consciousnesses to have {\it identical} thoughts/memories.  In this 
sense, the concept of postdiction will have intrinsic ambiguities 
associated with the specification of reference observers, and 
its treatment can only be ``approximate,'' e.g., in the sense of 
Eq.~(\ref{eq:prob-xi}).

\section{A Framework for Prediction---Selecting the Ensemble of Past 
 Light Cones}
\label{sec:framework}

We have argued that, assuming the underlying theory is known, answering 
any physical questions in the multiverse is boiled down to finding 
appropriate samples for past light cones that satisfy your input 
information.  This is equivalent to specifying relative weights for 
various ``microscopic'' possibilities that are consistent with your input. 
In this section we provide an explicit prescription for obtaining an 
appropriate ensemble of past light cone samples.  We will focus on the 
implementation of the idea in a semi-classical world, deferring the 
complete, quantum mechanical formulation to Section~\ref{sec:quantum}. 
We will, however, see that our prescription naturally demands a quantum 
mechanical treatment of the multiverse.

\subsection{Semi-classical picture}
\label{subsec:classical}

A simple prescription for finding the required ensemble is to 
define it as the result of ``repeated experiments.''  In the context 
of the eternally inflating multiverse, this means that we need to 
``simulate'' many times the entire multiverse as viewed from a single 
``observer.''  Defining this way, the result will depend on the initial 
setup for the ``experiments,'' i.e.\ the initial condition for the 
evolution of the multiverse.  We will discuss this issue in more detail 
in Section~\ref{sec:initial}, but for now we simply assume that the 
multiverse starts from some initial state that is highly symmetric. 
In particular, we assume that we can choose a time variable in such 
a way that, at sufficiently early times, constant-time space-like 
hypersurfaces are homogeneous and isotropic.

We now focus on a generic, finite region $\Sigma$ on an initial 
space-like hypersurface which contains at least one (measure zero) 
eternally inflating point.  Because of the homogeneity of the 
hypersurface, the choice of the region can be arbitrary.  (If the 
volume of the initial hypersurface is finite, as in the case where 
the spacetime is born as a closed universe, then $\Sigma$ may be chosen 
to be the entire hypersurface.)  We then consider a set of randomly 
distributed spacetime points $p_i$ on $\Sigma$, and a future geodesic 
$g_i$ emanating from each $p_i$.  Here we choose all the $g_i$'s to be 
normal to $\Sigma$; with this choice, the symmetry of the spacetime 
is recovered in the limit of a large number of $p_i$, $N_p \rightarrow 
\infty$.%
\footnote{The condition that the symmetry recovers in the large $N_p$ 
 limit does not uniquely fix the initial condition; in particular, we can 
 have an arbitrary distribution for the magnitude of the $g_i$ velocities, 
 $f(|\vec{v}|)$, as long as their direction $\vec{v}/|\vec{v}|$ is random. 
 While $|\vec{v}|$ decays exponentially with time during inflation, this 
 can still affect predictions~\cite{Garriga:2006hw} albeit to a small 
 extent~\cite{Freivogel:2009it}.  Here we have chosen $f(|\vec{v}|) = 
 \delta(|\vec{v}|)$ simply to illustrate our method in a ``minimal'' setup. 
 The issue of initial conditions in eternally inflating universe will 
 be discussed in Section~\ref{subsec:eternal-init}.  Ultimately, when 
 we apply our framework to the entire multiverse, the initial condition 
 needs to be specified by the ``theory of the beginning.''  Candidates 
 for such a theory will be discussed in Sections~\ref{subsec:initial} 
 and \ref{subsec:mega-multiverse}.}

Despite the fact that the spacetime is eternally inflating, any of 
the geodesics $g_i$ (except for a measure zero subset) experiences 
only a finite number of cosmic phase transitions and ends up with one 
of the terminal vacua---i.e.\ vacua with absolutely zero or negative 
cosmological constants---or a black hole singularity.%
\footnote{We assume that the landscape has (at least one) terminal vacua, 
 as suggested by string theory.  This assumption is also needed to avoid 
 the Boltzmann brain problem; see Section~\ref{subsec:paradoxes}.}
This implies that if we follow the ``history'' of one of these $g_i$'s 
by keeping track of the evolution of its past light cone, the possibility 
of finding a light cone that is consistent with any input conditions 
is extremely small.  However, since the input conditions necessarily 
have some ``range,'' as discussed in Section~\ref{subsec:strict}, the 
probability of finding such a light cone is nonzero, $P > 0$.%
\footnote{If the landscape is decomposed into several disconnected 
 pieces, some choice for the initial state may lead to $P = 0$.  This 
 implies that our vacuum does not belong to the same landscape component 
 as the initial state, and we need to (re)choose another initial state 
 to make predictions/postdictions.  In order for the entire framework 
 to make sense (i.e.\ our own existence to be compatible with the 
 multiverse picture), the correct choice for the initial state must 
 lead to $P > 0$.}
Therefore, if we prepare a large number of $p_i$'s:
\begin{equation}
  N_p \gg \frac{1}{P},
\label{eq:N_p}
\end{equation}
then we can find many light cone samples that satisfy the conditions 
provided.  A schematic depiction of this procedure was given in 
Fig.~\ref{fig:presc} in Section~\ref{sec:intro}.  Note that, if the 
input involves a specification of an exact physical ``time,'' e.g.\ 
through the precise state of our own brains (Section~\ref{subsec:strict}) 
or through $T_{\rm CMB,\, tip}$ (Section~\ref{subsec:practice}), 
then the number of light cones selected is countable and finite for 
each geodesic.%
\footnote{The specification of an exact time is {\it not} a necessity. 
 Indeed, in a real world we never have an exact knowledge about the 
 current time, so that the ``initial situation'' light cones have 
 some spread in the time direction, requiring us to count the passing 
 of $g_i$ through tips of these light cones as a ``single event.'' 
 In many cases, however, our knowledge about the current time is 
 sufficiently precise for the present purposes, so that the error 
 arising from treating it as an exact time is negligible in the limit 
 where $N_p$ is large.}

Our proposal is to use the ensemble of past light cones obtained in 
this way to make predictions/postdictions according to the procedures 
described in Section~\ref{sec:prediction}.  Suppose, for example, that 
we obtain ${\cal N}_A$ past light cone samples satisfying prior conditions 
$A$, and that among these samples ${\cal N}_{A \cap B}$ past light 
cones also satisfy conditions $B$.  Then, the conditional probability 
of $B$ under $A$, $P(B|A)$, is given by
\begin{equation}
  \frac{{\cal N}_{A \cap B}}{{\cal N}_A} \,\,
    \stackrel{N_p \rightarrow \infty}{\longrightarrow} \,\,
    P(B|A).
\label{eq:final-1}
\end{equation}
Since the distribution of $p_i$'s is not correlated with the histories 
afterwards, the resulting probability does not depend on this distribution 
in the limit of very large $N_p$.  Also, since ${\cal N}_A$ and 
${\cal N}_{A \cap B}$ are both finite at arbitrarily large $N_p$, 
the probability is well-defined.

Similarly, in order to predict the future in the sense of 
Section~\ref{subsec:strict}, we can count the number of past 
light cones in the sample that evolve into a particular future 
situation $C$: ${\cal N}_{A \rightarrow C}$.  The probability that 
$C$ occurs under conditions $A$ is then given by
\begin{equation}
  \frac{{\cal N}_{A \rightarrow C}}{{\cal N}_A} \,\,
    \stackrel{N_p \rightarrow \infty}{\longrightarrow} \,\,
    P(C|A).
\label{eq:final-2}
\end{equation}
If we select a set of future situations $C_i$ which are exhaustive 
and mutually exclusive, then we have
\begin{equation}
  \sum_i P(C_i | A) = 1.
\label{eq:future}
\end{equation}
In the example of predicting the result of the LHC experiments (as in 
Sections~\ref{subsec:strict} and \ref{subsec:practice}), we may choose, 
e.g., $\{ C_i \} = $ the LHC will \{find just the Higgs boson, find 
supersymmetry, find technicolor, find any other theory or the result 
will be inconclusive, fail\}.%
\footnote{In fact, the two definitions of Eqs.~(\ref{eq:final-1}) and 
 (\ref{eq:final-2}) are not independent, since we can calculate the 
 latter, $P(C|A)$, from the former, $P(B|A)$, once we know $P(B|A)$ 
 for all $B$'s that could affect $C$'s. \label{ft:AB-AC}}

We note that the validity of our method is not limited to vacua 
with four spacetime dimensions.  Our prior conditions $A$ can be 
formulated in any numbers of non-compact and compact spatial dimensions. 
While it is appropriate in most circumstances to limit our discussions to 
$(3+1)$-dimensional spacetime by integrating out all the effects of extra 
compact dimensions, there is nothing wrong with applying our procedure 
to the full, higher dimensional context.  Such a treatment may be useful 
if we want to postdict the number of our large spacetime dimensions 
to be four (unless the anthropic factor $n(x_i)$ is trivially zero 
except for four large dimensions).  For the universe we live today, 
our knowledge about compact dimensions must come from bounds from 
experimental searches, as well as the structure of the observed physical 
laws if relations between four-dimensional physics and extra dimensional 
geometries are known in the landscape.  The initial space-like 
hypersurface $\Sigma$ may also have arbitrary numbers of non-compact 
and compact spatial dimensions.

In summary, the probabilities are defined through the sampling of past 
light cones that satisfy our prior conditions, by following histories 
of many geodesics $g_i$ emanating from an initial hypersurface $\Sigma$. 
Note that this procedure corresponds precisely to simulating the entire 
multiverse many times as viewed from a single ``observer.''%
\footnote{The procedure may also be viewed as a sort of ``fuzzy'' 
 time cutoff in expanding universes; see Appendix~\ref{app:fuzzy}.}
It is quite satisfactory that this operational way of defining 
probabilities also gives a well-defined prescription for calculating 
probabilities we are interested in:\ Eqs.~(\ref{eq:final-1}) and 
(\ref{eq:final-2}).  Some sample calculations of the probabilities 
are given for toy landscapes in Appendix~\ref{app:calc}.

\subsection{Necessity of a quantum mechanical treatment}
\label{subsec:quantum}

So far, we have been describing our framework using the language of 
classical spacetime.  We know, however, that our world is quantum 
mechanical.  This raises an important question:\ does the prescription 
really make sense in the quantum mechanical world?  In particular, 
can we literally follow the histories of $g_i$'s classically, emanating 
from a small initial region $\Sigma$?

Suppose we want to ``scan'' our current universe finely enough so 
that we can find a past light cone that satisfies our input conditions. 
For example, let us imagine that we need one of the $g_i$'s for every 
$\sim {\rm \mu m}^3$ at the current moment in order to follow the 
procedure described in Section~\ref{subsec:strict}.  In this case, 
the average separation between $g_i$'s at the time when our observable 
inflation, i.e.\ the last $N$-fold of inflation, began is roughly
\begin{equation}
  e^{-N} \frac{T_0}{T_{\rm R}}\, {\rm \mu m}
  \sim 10^{-53}~{\rm m} \left( \frac{10^8~{\rm GeV}}{T_R} \right),
\label{eq:distace}
\end{equation}
where $T_0$ and $T_{\rm R}$ are the current and reheating temperatures, 
respectively.  Here, we have ignored any focusing of geodesics, e.g.\ 
due to structure formation, for simplicity, and used $N = 60$ to obtain 
the final number.  We find that the distance obtained is much shorter 
than the Planck length, $l_P \simeq 1.62 \times 10^{-35}~{\rm m}$; in 
other words, we have a large number of $g_i$'s passing a Planck size 
region at the time when the observable inflation started.  The number 
of $g_i$'s per a Planck size region increases even more if we extrapolate 
the history further back.  This implies that we need to consider a 
huge number of $p_i$'s in a Planck size region on $\Sigma$, from which 
$g_i$'s emanate.

On the other hand, we expect that at the length scale $\sim l_P$ 
gravity becomes strong, so that quantum effects become important 
even for spacetime.%
\footnote{If the theory has a large number of species $n$, the scale 
 where quantum gravity becomes important is actually $\sim l_P 
 \sqrt{n}$ (or the string length $l_s$)~\cite{Dvali:2007hz}.  This, 
 however, does not affect any of our discussions, including the one 
 here.  In the rest of the paper, we will set $n \approx O(1)$ (or 
 $l_s \sim l_P$) for simplicity, but it is straightforward to recover 
 the dependence on $n$, if needed.}
In particular, the holographic principle (or de~Sitter entropy) suggests 
that a Planck size region in de~Sitter space can contain at most of 
$O(1)$ bits of information.  Then, how can we make sense out of histories 
of many $p_i$'s {\it distributed inside a Planck size region}?  This 
naturally suggests the following interpretation of our procedure. 
Starting from a small (even a Planck size) region in the early universe, 
we follow its evolution quantum mechanically.  Because of the quantum 
nature, this probabilistically leads to many cosmic histories; and it 
is these possible histories that correspond to sets of past light cones 
associated with various $g_i$'s in the semi-classical picture.

In the next section, we will see that this interpretation is, in fact, 
{\it forced} by the laws of quantum mechanics.  Interestingly, quantum 
mechanics cannot be viewed as ``small corrections'' to the classical 
picture---{\it quantum mechanics is essential to (correctly) interpret 
the multiverse}.

\section{Multiverse as a Quantum Mechanical Universe}
\label{sec:quantum}

In this section, we show that the quantum mechanical interpretation 
of the multiverse is unavoidable if we assume that the laws of quantum 
mechanics are not violated when physics is described from an observer's 
viewpoint.  In particular, we argue that the {\it entire} multiverse 
can be described from the point of view of a ``single observer.'' 
We develop an explicit quantum mechanical formalism for making 
predictions/postdictions, which corresponds to the framework described 
in Section~\ref{sec:framework} in the semi-classical picture.  We 
first consider the case in which the multiverse is in a pure quantum 
state, and then generalize it to the mixed state case.  We also discuss 
how the quantum-to-classical transition is incorporated in our framework.

\subsection{Quantum mechanics and the ``multiverse''}
\label{subsec:QM}

Recall the laws of quantum mechanics---deterministic, unitary 
evolution of the states and the superposition principle.  (We adopt 
the Schr\"{o}dinger picture throughout.)  These laws say:\ (i) given 
a pure quantum state $\Psi(t_0)$, its history is uniquely determined 
by solving quantum evolution equation, $\Psi(t) = U(t,t_0) \Psi(t_0)$ 
where $U(t,t_0)$ is a unitary operator; and (ii) if $\Psi_1(t)$ and 
$\Psi_2(t)$ are both solutions of the evolution equation, then 
$c_1 \Psi_1(t) + c_2 \Psi_2(t)$ is also a solution for arbitrary, 
complex numbers $c_1$ and $c_2$.

While these properties are manifest in the usual formulation of 
non-relativistic quantum mechanics, they are sometimes obscured 
in quantum field theory---can't an initial $e^+ e^-$ ``state'' 
evolve into $e^+ e^-$, $\mu^+ \mu^-$ or some other ``states''? 
Such an evolution, however, does not contradict the fact that quantum 
evolution is deterministic.  What is happening is simply that the 
state that is initially $e^+ e^-$, $\left| e^+ e^- \right>_{\rm in}$, 
evolves into a {\it unique} final state that has nontrivial overlaps 
with the states approaching $e^+ e^-, \mu^+ \mu^-, \cdots$ in 
the future, $\left| e^+ e^- \right>_{\rm out}, \left| \mu^+ \mu^- 
\right>_{\rm out}, \cdots$.  Expanding in terms of the states in free 
field theories (which is possible at $t \rightarrow \pm \infty$), we 
can write
\begin{equation}
  \Psi(t = -\infty) = \left| e^+ e^- \right>
\quad\rightarrow\quad
  \Psi(t = +\infty) = c_e \left| e^+ e^- \right> 
    + c_\mu \left| \mu^+ \mu^- \right> + \cdots,
\label{eq:QFT-evolution}
\end{equation}
where $c_e = {}_{\rm out}\!\left< e^+ e^- \right. | \left. e^+ e^- 
\right>_{\rm in}$, $c_\mu = {}_{\rm out}\!\left< \mu^+ \mu^- \right. | 
\left. e^+ e^- \right>_{\rm in}$, $\cdots$ are uniquely determined.%
\footnote{Here we have suppressed the momenta and spins of the particles, 
 but including them is straightforward.}
In fact, the evolution of states in quantum field theory is unitary, 
and satisfies the superposition principle.

These properties of quantum mechanics essentially force us to take 
the quantum mechanical view of the multiverse, outlined at the end of 
Section~\ref{subsec:quantum}.  Consider a small initial region $\Sigma$ 
at an early universe, $t = t_0$, which contains an eternally inflating 
point.  If we follow the evolution of this initial (presumably highly 
symmetric) state quantum mechanically, we should find that future 
states are uniquely determined.  On the other hand, the semi-classical 
picture of eternal inflation says that various observers emigrating 
from $\Sigma$ will see different semi-classical histories, or universes. 
Note that, assuming the energy density on $\Sigma$ is (much) smaller 
than the Planck density, we do not have any reason to doubt this 
semi-classical picture.  The only way to reconcile these two facts 
is to consider that various semi-classical histories correspond to 
possible outcomes obtained from a unique quantum state $\Psi(t)$. 
Schematically,
\begin{equation}
  \Psi(t = t_0) = \left| \Sigma \right>
\quad\rightarrow\quad
  \Psi(t) = \sum_i c_i \left| \mbox{cosmic history $i$ at time $t$} \right>.
\label{eq:multiverse-evolution}
\end{equation}
Below, we present more precise arguments on this point and give an 
explicit formalism which makes well-defined predictions/postdictions 
possible in the multiverse.

\subsection{Lessons from black holes}
\label{subsec:BH}

The two expressions in Eqs.~(\ref{eq:QFT-evolution}) and 
(\ref{eq:multiverse-evolution}) look very similar, but there is 
one significant difference.  While the usual ``in-out'' formalism of 
quantum field theory, exemplified in Eq.~(\ref{eq:QFT-evolution}), deals 
only with the states at infinite past or future (which is indeed quite 
sufficient for the purpose of describing experiments involving scattering 
or decay), the cosmological setup of Eq.~(\ref{eq:multiverse-evolution}) 
requires states defined throughout the whole history.  As a relativistic 
version of quantum mechanics, quantum field theory must be able to 
describe such a state, $\Psi(t)$.  However, a naive implementation 
of this under the existence of gravity could cause a problem---the 
laws of quantum mechanics may be violated.

\begin{figure}[t]
\begin{center}
\begin{picture}(200,255)(-25,-10)
  \Line(0,0)(0,160) \Text(-5,80)[r]{$r=0$}
  \Line(0,0)(150,150) \Line(150,150)(70,230) \Line(70,160)(70,230)
  \Line(0,160)(70,160) \Photon(0,160)(70,160){2}{8} \Text(35,168)[b]{$r=0$}
  \DashLine(0,90)(70,160){3}
  \CArc(-116,40)(170,12,40) \CArc(-115.5,40)(170,12,40)
  \Vertex(50.8,74){2}
  \Line(25,132.5)(13,148.5) \Line(27.5,135)(15.5,151)
  \Line(13,152)(13,142) \Line(13,152)(22.6,149.2)
  \Photon(40,120)(70,150){2}{4}
  \DashLine(74,156)(88.5,170.5){2} \DashLine(76,153.5)(91,168.5){2}
  \DashLine(92,172)(88.8,162.5){2} \DashLine(92,172)(82.5,168.8){2}
  \DashLine(0,145)(51.7,145){1}
  \DashCArc(52.7,160.0)(15,270,315){1}
  \DashLine(64.0,150.2)(66.1,152.3){1}
  \DashCArc(95,124.8)(40,77.6,135){1}
  \DashCArc(70,-33.3)(200,66.4,80){1}
\end{picture}
\caption{A Penrose diagram representing a traveler who falls into an 
 evaporating black hole (solid curve) carrying some information.  For 
 the traveler, information appears to be always with him/her (solid 
 arrow), while from a distant observer, the information appears to be 
 sent back from the black hole in Hawking radiation (dashed arrow) 
 An example of ``wrong'' constant time hypersurfaces is depicted with 
 the dotted line.}
\label{fig:BH}
\end{center}
\end{figure}
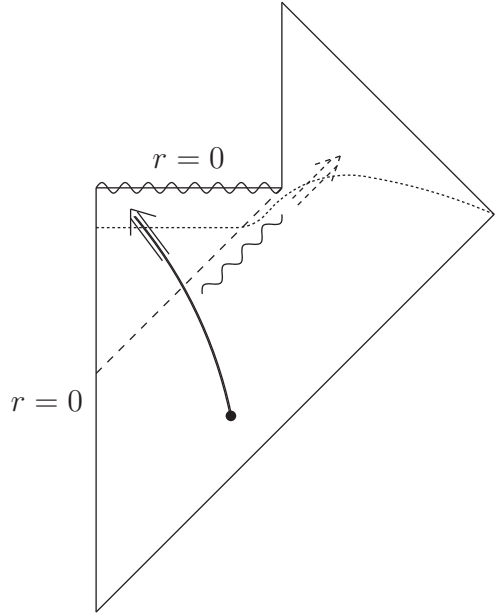

To see what might happen, let us consider a traveler falling into a 
black hole, carrying some information.  For this traveler, the information 
is always with him/her until he/she hits the singularity; in particular, 
it will be inside the black hole at late times.  On the other hand, 
for a distant observer, the information appears to be absorbed into 
the (stretched) horizon, and then sent back in Hawking radiation. 
This apparently indicates that the same information exists in two 
different locations, contradicting the no-cloning theorem of quantum 
mechanics, which says that a faithful duplication of quantum information 
is not possible (see Fig.~\ref{fig:BH}).%
\footnote{The no-cloning theorem can be proved using the laws of quantum 
 mechanics stated in Section~\ref{subsec:QM}.}
This paradox, however, can be resolved by the so-called black hole 
complementarity~\cite{Susskind:1993if}.  Since one cannot be {\it both} 
a distant observer {\it and} the falling traveler at the same time, 
it is perfectly consistent that the two descriptions disagree about 
something that cannot {\it in principle} be compared; in particular, 
they may not agree about where the information exists after the traveler 
crossed the horizon.  When physics is described consistently by a 
distant observer, or by the falling traveler, the violation of quantum 
mechanical principles does not occur.

One can consider this phenomenon to arise as a result of noncommutativity 
of local measurements performed by the two experimenters:\ the traveler 
$T$ and distant observer $D$.  Suppose that $T$ and $D$ have their own 
complete sets of operators ${\cal T}_i$ and ${\cal D}_i$, and that we 
take a Hilbert space basis in which the states that appear local to $T$ 
can be written schematically in the form
\begin{equation}
  \psi_1 = 
    \!\!\!\!\begin{array}{c} \otimes \\ \\ \\ \\ \end{array}\!\!\!\!\!\!
  \left( \begin{array}{c}
    \psi \\ 0_v \\ \vdots
  \end{array} \right),
\quad
  \psi_2 = 
    \!\!\!\!\begin{array}{c} \otimes \\ \\ \\ \\ \end{array}\!\!\!\!\!\!
  \left( \begin{array}{c}
    0_v \\ \psi \\ \vdots
  \end{array} \right),
\quad
  \cdots,
\label{eq:psi-diag}
\end{equation}
where the symbol $\otimes$ means that $\psi_i$ is given by the direct 
product of all the elements in the vector.  Here, each element corresponds 
to a different spatial point viewed from $T$, and $\psi$ and $0_v$ 
represent excited and unexcited states at each point.  The local operators 
for $T$ then take the (block-)diagonal form:
\begin{equation}
  {\cal T}_1 = \left( \begin{array}{ccc}
    \hat{\phi} & 0 & \cdots \\
    0 & {\bf 1} & \cdots \\
    \vdots & \vdots & \ddots
  \end{array} \right),
\quad
  {\cal T}_2 = \left( \begin{array}{ccc}
    {\bf 1} & 0 & \cdots \\
    0 & \hat{\phi} & \cdots \\
    \vdots & \vdots & \ddots
  \end{array} \right),
\quad
  \cdots,
\label{eq:T-diag}
\end{equation}
(ignoring any exponentially damping tails).  In the same Hilbert space 
basis, however, operators local for $D$ are {\it not} necessarily 
diagonal, e.g.\
\begin{equation}
  {\cal D}_1 = \frac{1}{2} \left( \begin{array}{ccc}
    {\bf 1} + \hat{\phi} & -{\bf 1} + \hat{\phi} & \cdots \\
    -{\bf 1} + \hat{\phi} & {\bf 1} + \hat{\phi} & \cdots \\
    \vdots & \vdots & \ddots
  \end{array} \right),
\quad
  {\cal D}_2 = \frac{1}{2} \left( \begin{array}{ccc}
    {\bf 1} + \hat{\phi} & {\bf 1} - \hat{\phi} & \cdots \\
    {\bf 1} - \hat{\phi} & {\bf 1} + \hat{\phi} & \cdots \\
    \vdots & \vdots & \ddots
  \end{array} \right),
\quad
  \cdots.
\label{eq:D-nondiag}
\end{equation}
The states in Eq.~(\ref{eq:psi-diag}) may then look highly delocalized 
for the distant observer $D$.  (This happens at a time after the traveler 
passed the horizon).  In fact, we expect that (local) information 
inside the black hole are stored in long-range correlations in Hawking 
radiation (or in horizon degrees of freedom) from an outside observer 
point of view~\cite{Susskind:1993if}.

Now, if we want to discuss physics as viewed from distant observer 
$D$, it is absurd to take the basis in Eq.~(\ref{eq:D-nondiag})---we 
should go to the basis where local operators for $D$ are diagonalized. 
In the new basis, Eq.~(\ref{eq:D-nondiag}) becomes
\begin{equation}
  {\cal D}_1 = \left( \begin{array}{ccc}
    \hat{\phi} & 0 & \cdots \\
    0 & {\bf 1} & \cdots \\
    \vdots & \vdots & \ddots
  \end{array} \right),
\quad
  {\cal D}_2 = \left( \begin{array}{ccc}
    {\bf 1} & 0 & \cdots \\
    0 & \hat{\phi} & \cdots \\
    \vdots & \vdots & \ddots
  \end{array} \right),
\quad
  \cdots,
\label{eq:D-diag}
\end{equation}
and the states in Eq.~(\ref{eq:psi-diag}) look like
\begin{equation}
  \psi_1 = 
    \!\!\!\!\begin{array}{c} \otimes \\ \\ \\ \\ \end{array}\!\!\!\!\!\!
  \left( \begin{array}{c}
    (0_v + \psi)/\sqrt{2} \\ (0_v - \psi)/\sqrt{2} \\ \vdots
  \end{array} \right),
\quad
  \psi_2 = 
    \!\!\!\!\begin{array}{c} \otimes \\ \\ \\ \\ \end{array}\!\!\!\!\!\!
  \left( \begin{array}{c}
    (0_v + \psi)/\sqrt{2} \\ (-0_v + \psi)/\sqrt{2} \\ \vdots
  \end{array} \right),
\quad
  \cdots.
\label{eq:psi-nondiag}
\end{equation}
Two equations~(\ref{eq:psi-diag}) and (\ref{eq:psi-nondiag}) clearly 
show that when we describe physics from an experimenter's point of 
view, the two experimenters $T$ and $D$ describe the same system very 
differently---one localized and the other delocalized.  (Of course, 
in the Hilbert space, the $\psi_i$'s in the two equations {\it are} 
the same states:\ they simply correspond to different coordinate 
representations.)  Below, when we talk about an observer, we 
always assume that we take the basis such as Eqs.~(\ref{eq:T-diag}) 
and (\ref{eq:D-diag}), i.e.\ the basis where local operators are 
``diagonalized,'' which is the usual basis for local quantum fields.%
\footnote{Later we will quantize the system on null hypersurfaces, in 
 which case Schr\"{o}dinger picture operators at different points do 
 not all commute.  This subtlety, however, does not affect the basic 
 picture; see Section~\ref{subsec:prob}.}
This means, in particular, that when we change the viewpoint, we must 
in general perform the associated basis change in the Hilbert space.

The black hole example discussed here also tells us that the choice 
of constant time slices---a set of hypersurfaces on which quantum states 
are defined---is extremely important in quantum theories with gravity. 
Suppose we want to describe the system from a distant observer's point 
of view, using the local operator basis as described above.  How should 
we define the states?  From Fig.~\ref{fig:BH}, it is evident that 
if we choose a ``wrong'' time slice (often called a ``nice'' slice), 
we would have a state in which quantum information is duplicated, 
contradicting the no-cloning theorem.  In fact, choosing this kind 
of slices, nonlocal effects from quantum gravity do not decouple at 
low energies, and we cannot simultaneously ``diagonalize'' operators 
localized in different spatial points~\cite{Lowe:1995ac}.  Therefore, 
if we want to maintain a local description of physics, such choices 
of time slices should be avoided.  The usual in-out formalism of 
quantum field theory bypasses this issue by considering only states 
at $t = \pm \infty$ in asymptotically Minkowski (or anti de~Sitter) 
spacetime, but it is in general important in the context of the 
cosmology of the multiverse.

\subsection{Quantum observer principle---relativistic quantum 
 mechanics for cosmology}
\label{subsec:rel-QM}

We now present our quantum mechanical framework for describing the 
multiverse.  Our central postulate, which we may call the {\it quantum 
observer principle}, is the following:
\begin{itemize}
\item[]
{\it Physics obeys the laws of quantum mechanics when described from 
 the viewpoint of an ``observer'' (geodesic) traveling the multiverse, 
 although this need not be the case if described in other ways, e.g., 
 using the global spacetime picture with synchronous time slicing. 
 The description involves only spacetime regions inside the (stretched) 
 apparent horizons, as well as the degrees of freedom associated 
 with these horizons.}
\end{itemize}
As we discussed in the previous subsection, we choose a Hilbert space 
basis which ``diagonalizes'' local operators as viewed from the observer; 
in particular, we take {\it the same operator basis} for all observers. 
(We can always do this because, being state independent, complete operator 
sets for observers are all isomorphic.)  To realize the framework 
explicitly, we still need to define quantum states carefully, which 
will be discussed in the next subsection.  However, without such 
a realization, the single principle stated above already leads, together 
with the usual statistical interpretation of horizon entropies, to many 
nontrivial, general consequences for the description of the multiverse.

Let us now discuss implications of each element of the laws of quantum 
mechanics in turn:
\begin{itemize}
\item
{\bf Deterministic evolution} --- Consider a small, eternally inflating 
space-like region $\Sigma$ at some early time.  According to the 
semi-classical picture, observers in this region can see infinitely 
many different universes in the future (i.e.\ the entire multiverse), 
even if they are ``equivalent'' (i.e.\ related by an element of the 
kinematic subgroup of the de~Sitter group).  On the other hand, the 
quantum observer principle says that quantum evolution of states is 
deterministic; and the holographic principle (or de~Sitter entropy) 
says that the amount of information $\Sigma$ can carry, i.e.\ the 
number of different initial conditions one can prepare on $\Sigma$, 
is finite.  This implies that in the quantum picture, the origin of 
various semi-classical universes cannot be attributed to the difference 
of initial conditions on $\Sigma$.%
\footnote{Mathematically, the de~Sitter group cannot be exact 
 at the quantum level, because of the discreteness of energy 
 levels~\cite{Goheer:2002vf}.  However, since typical level spacing 
 is of order the inverse Poincar\'{e} recurrence time, it is plausible 
 that the effect is physically irrelevant, and semi-classically 
 equivalent observers see equivalent quantum systems.  (See 
 also~\cite{Banks:2002wr}.)  Indeed, string theory suggests that 
 lifetimes of de~Sitter vacua are (much) shorter than the recurrence 
 times~\cite{Kachru:2003aw}, so that no energy measurement can 
 physically resolve the discreteness of the levels.}
The only possible interpretation, then, is that these different 
universes arise as possible outcomes obtained from a quantum state 
$\Psi(t)$, which is uniquely determined once an initial condition 
is given at some early moment.%
\footnote{Note that, taking the same operator basis for all observers, 
 the two descriptions in Fig.~\ref{fig:BH}, i.e.\ by the traveler $T$ 
 and distant observer $D$, correspond to having {\it different} (though 
 equivalent) states $\Psi_T$ and $\Psi_D$, subjected to different 
 initial conditions:\ one falling and the other not falling into the 
 black hole.  On the other hand, if the two experimenters $T$ and $D$ 
 arise in the future of an eternally inflating region $\Sigma$, then 
 the two descriptions become parts of a {\it single} multiverse 
 state, $\Psi \sim c_T \left| T\mbox{'s view} \right> + c_D \left| 
 D\mbox{'s view} \right> + \cdots$, in the complete description of 
 the multiverse.}
\item
{\bf Unitarity} --- The quantum state of the multiverse $\Psi(t)$ 
is meaningful if (and, presumably, only if) it is described from the 
point of view of an observer.  In particular, this implies that the 
state $\Psi(t)$ is {\it not} defined outside horizons (of de~Sitter, 
black hole, or any other kind).  On the other hand, in the semi-classical 
picture, there are certainly processes which carry information to outside 
these horizons.  How can unitary evolution of $\Psi(t)$ be ensured 
then?  The answer is that, as in the black hole case, this information 
is stored in the stretched horizons when seen by the observer.  (Indeed, 
there is evidence that any cosmological horizons virtually act as black 
hole horizons; see e.g.~\cite{Unruh:1976db,Jacobson:1995ab,Banks:2001yp}.) 
Namely, assuming a pure initial state, subsequent evolution of the 
multiverse is described by a pure quantum state $\Psi(t)$ if we include 
the microscopic description of the stretched horizons.  For such a 
state, the usual, thermal description of a horizon emerges only after 
considering a suitable statistical ensemble in the multiverse, by 
``coarse-graining'' the horizon degrees of freedom.%
\footnote{For a generic state $\Psi(t)$, the entanglement entropy 
 between the bulk and stretched-horizon degrees of freedom is 
 smaller than the full horizon entropy, which takes into account 
 all {\it possible} microstates associated with the horizon.}
A nontrivial consistency check of the statement made here is provided 
in Appendix~\ref{app:no-cloning}.
\item
{\bf Superposition principle} --- Suppose we want to compute the 
probability of an initial situation $A$ at time $t$ to develop into 
a future situation $C$ at later time $t + \varDelta t$.  We would then 
evolve the state representing $A$ by $\varDelta t$, and take the overlap 
with the state(s) corresponding to $C$: $P(C|A) = |\left< \psi_C \right| 
e^{-i \hat{H} \varDelta t} \left| \psi_A \right>|^2$, where $\hat{H}$ 
is the Hamiltonian describing the dynamics of the system.  On the other 
hand, in the description of the entire multiverse, the situation $A$ 
occurs multiple times at $t = t_i$ ($i = 1,2,\cdots$) as one of the 
possible universes: $\Psi(t_i) = c_i \left| A \right> + \cdots$, where 
$c_i \lll 1$ in general.  Because of the superposition principle, however, 
the evolution of the $c_i \left| A \right>$ term in $\Psi(t_i)$ is 
exactly the same as the state $\left| \psi_A \right>$ (if we use the 
complete multiverse evolution operator for $\hat{H}$, which is most 
accurate though often overkill).  This implies that we may compute the 
probability $P(C|A)$ using the entire multiverse state $\Psi(t)$, instead 
of $\left| \psi_A \right>$; in particular, we need not consider that 
$\Psi(t)$ collapses to $\left| \psi_A \right>$ when we compute $P(C|A)$.
\end{itemize}

An important consequence of the framework described here is that 
the quantum state $\Psi(t)$, defined from a {\it single} observer's 
viewpoint, describes the {\it entire} multiverse.  This is indicated 
by the fact that eternal inflation populates the entire landscape in 
the semi-classical picture, even starting from a measure zero point 
on an initial hypersurface.%
\footnote{Here we have assumed that the multiverse is irreducible, 
 i.e.\ any two points in the landscape are connected by some physical 
 processes.  The case with a reducible landscape will be mentioned 
 in Section~\ref{sec:initial} (in footnote~\ref{ft:reducible}).}
Namely, {\it the state $\Psi(t)$ (which may in general be a pure or 
mixed state) provides a complete description of the multiverse.}

\subsection{The entire multiverse from the viewpoint of a single observer}
\label{subsec:single}

How can the multiverse state $\Psi(t)$ be defined explicitly?  Let us 
suppose, for simplicity, that the multiverse is described by a pure 
state $\left| \Psi(t) \right>$.  (An extension to the mixed state 
case is straightforward and will be discussed later.)  From an observer's 
point of view, the subsequent evolution of this state is uniquely 
determined according to the laws of quantum mechanics.  A question 
is:\ what constant time slices should we choose to describe the system 
``as viewed from an observer''?  In order to maintain locality of the 
description, such slices should not extend to the region that cannot 
be accessed by the observer.

Let us consider how we can define time slices in an operationally 
well-defined manner.  In the multiverse, one obviously cannot carry 
a physical clock through (some of) the cosmic phase transitions occurring 
in the landscape---even low energy degrees of freedom may change across 
bubble walls associated with the transitions.  This implies that we cannot 
use any physical quantities, e.g.\ average CMB temperature, to define 
equal time hypersurfaces {\it throughout the entire multiverse}.  Of 
course, we can adopt any time parametrization {\it along the geodesic 
(observer)}---a natural choice is the proper time, which can be 
assigned a simple, invariant meaning.  But how can we extend it to 
the rest of spacetime without referring to any physical quantities?

We use the causal structure to determine the equal time hypersurfaces, 
i.e.\ the hypersurfaces on which quantum states are defined.  While this 
is not an absolute necessity (see discussion in Section~\ref{subsec:prob}), 
it provides the conceptually simplest formulation of our framework. 
Specifically, consider a spacetime point $p(t)$ along the geodesic 
(observer) corresponding to a proper time $t$.  We then define a quantum 
state $\left| \Psi(t) \right>$ on the past light cone whose tip is at 
$p(t)$.  We assume that $\left| \Psi(t) \right>$ is defined only within 
the stretched horizon, which is located roughly $l_P$ proper distance 
in front of the real, mathematical horizon in the static coordinates. 
The rationale behind this is the following.  In the ``coarse-grained 
picture,'' which considers an ensemble of spacetime regions having the 
same macroscopic properties, the local temperature on the stretched 
horizon is of $O(1/l_P)$, due to the blueshift effect.  Since this thermal 
entropy already saturates the horizon entropy, the number of physical 
degrees of freedom behind the stretched horizon should be at most 
of $O({\cal A}_{\rm horizon}/l_P^2)$, where ${\cal A}_{\rm horizon}$ 
is the horizon area (see, e.g.,~\cite{Susskind:2005js}).  This implies 
that we can describe any physics in that region using degrees of freedom 
``on'' the stretched horizon, which we include in $\left| \Psi(t) \right>$.

\begin{figure}[t]
\begin{center}
\vspace{5mm}
  \includegraphics[scale=0.7]{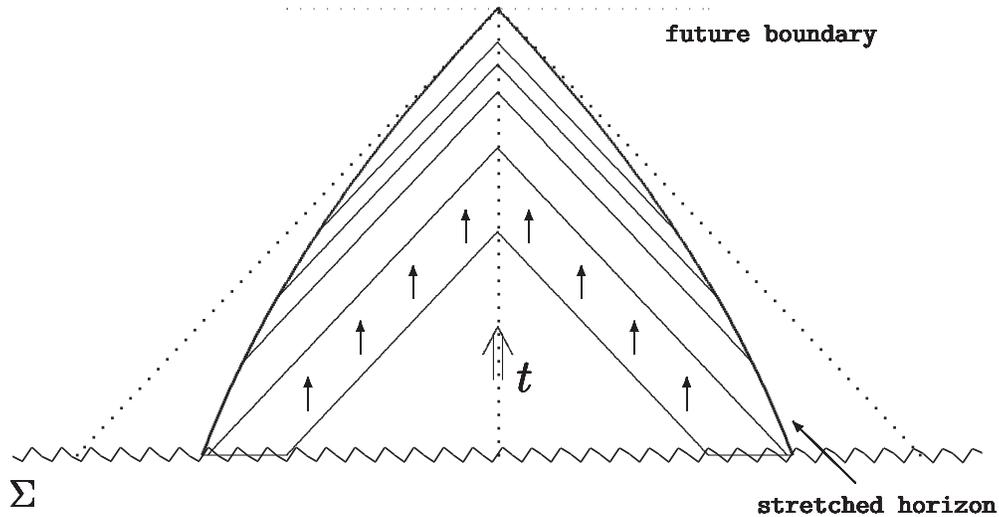}
\vspace{2mm}
\caption{The state $\left| \Psi(t) \right>$ is defined on the past light 
 cones (thin solid lines) bounded either by the initial hypersurface 
 $\Sigma$ (wavy line) or the stretched horizon (thick solid).  The 
 tips of these light cones are on the geodesic corresponding to the 
 ``observer,'' and the time parameter $t$ is chosen such that it agrees 
 with the proper time associated with the observer.  If the initial 
 condition is given on a past light cone and the stretched horizon 
 bounding it, then the introduction of the space-like hypersurface 
 $\Sigma$ is not necessary.}
\label{fig:stretched}
\end{center}
\end{figure}

Suppose that an initial condition for the multiverse is given on a 
space-like hypersurface $\Sigma$ (which need not be the case, as we 
will see shortly).  In this case, the multiverse state $\left| \Psi(t) 
\right>$ is defined at early times on the past light cone within 
$C(t)$ and on $\Sigma$ outside $C(t)$, where $C(t)$ is the intersection 
between the light cone and $\Sigma$ (see Fig.~\ref{fig:states} in 
Section~\ref{sec:intro}).  This situation, however, does not last long. 
After a brief initial period, specifically for $\varDelta t \simgt 
-H^{-1} \ln(l_P H)$ after the initial moment, where $H$ is the Hubble 
parameter, the state $\left| \Psi(t) \right>$ is given entirely on 
the past light cones bounded by the stretched horizon, as depicted 
in Fig.~\ref{fig:stretched}.

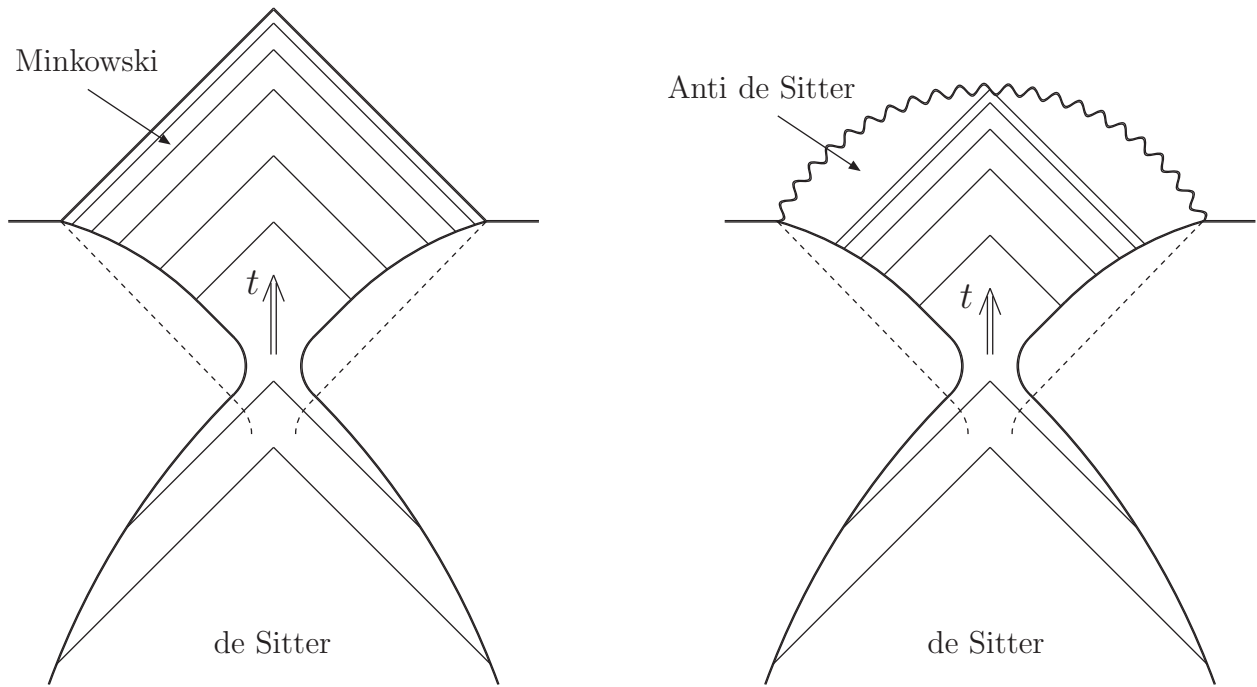
\begin{figure}[t]
\begin{center}
\begin{picture}(470,280)(0,-185)
%
%
  \Line(0,0)(20,0) \Line(0,0.5)(20,0.5)
  \Line(20,0)(100,80) \Line(20,0.7)(100,80.7)
  \Line(100,80)(180,0) \Line(100,80.7)(180,0.7)
  \Line(180,0)(200,0) \Line(180,0.5)(200,0.5)
  \DashLine(20,0)(86.5,-68.5){2} \DashCArc(77,-80)(14.7,0,42){2}
  \DashLine(180,0)(113.5,-68.5){2} \DashCArc(123,-80)(14.7,138,180){2}
  \CArc(297,-277)(300,135,160) \CArc(297.3,-277.3)(300,135,159.95)
  \CArc(74.3,-54.3)(15,-45,45) \CArc(74.8,-54.3)(15,-47,47)
  \Line(84.9,-43.7)(71.6,-30.4) \Line(85.2,-43.4)(71.9,-30.1)
  \CArc(-13.3,-115.4)(120.1,45,74) \CArc(-13.0,-115.1)(120.1,45,74.3)
  \CArc(-97,-277)(300,20,45) \CArc(-97.3,-277.3)(300,20.05,45)
  \CArc(125.7,-54.3)(15,135,225) \CArc(125.2,-54.3)(15,133,227)
  \Line(115.1,-43.7)(128.4,-30.4) \Line(114.8,-43.4)(128.1,-30.1)
  \CArc(213.3,-115.4)(120.1,106,135) \CArc(213.0,-115.1)(120.1,105.7,135)
  \Line(100,75)(24,-1) \Line(100,75)(176,-1)
  \Line(100,65)(31.2,-3.8) \Line(100,65)(168.8,-3.8)
  \Line(100,50)(41.5,-8.5) \Line(100,50)(158.5,-8.5)
  \Line(100,25)(57,-18) \Line(100,25)(143,-18)
  \Line(100,0)(71,-29) \Line(100,0)(129,-29)
  \Line(100,-60)(45,-115) \Line(100,-60)(155,-115)
  \Line(100,-85)(18,-167) \Line(100,-85)(182,-167)
  \Line(99,-50)(99,-23) \Line(101,-50)(101,-23)
  \Line(100,-20)(96,-30) \Line(100,-20)(104,-30)
  \Text(90,-23)[l]{\large $t$}
  \Text(100,-155)[t]{de~Sitter}
  \LongArrow(30,50)(60,30) \Text(30,57)[b]{Minkowski}
%
%
  \Line(270,0)(290,0) \Line(270,0.5)(290,0.5)
  \PhotonArc(370,-40)(89.7,26.5,153.5){2}{22}
  \PhotonArc(370,-40)(90.3,26.5,153.5){2}{22}
  \Line(450,0)(470,0) \Line(450,0.5)(470,0.5)
  \DashLine(290,0)(356.5,-68.5){2} \DashCArc(347,-80)(14.7,0,42){2}
  \DashLine(450,0)(383.5,-68.5){2} \DashCArc(393,-80)(14.7,138,180){2}
  \CArc(567,-277)(300,135,160) \CArc(567.3,-277.3)(300,135,159.95)
  \CArc(344.3,-54.3)(15,-45,45) \CArc(344.8,-54.3)(15,-47,47)
  \Line(354.9,-43.7)(341.6,-30.4) \Line(355.2,-43.4)(341.9,-30.1)
  \CArc(256.7,-115.4)(120.1,45,74) \CArc(257.0,-115.1)(120.1,45,74.3)
  \CArc(173,-277)(300,20,45) \CArc(172.7,-277.3)(300,20.05,45)
  \CArc(395.7,-54.3)(15,135,225) \CArc(395.2,-54.3)(15,133,227)
  \Line(385.1,-43.7)(398.4,-30.4) \Line(384.8,-43.4)(398.1,-30.1)
  \CArc(483.3,-115.4)(120.1,106,135) \CArc(483.0,-115.1)(120.1,105.7,135)
  \Line(370,50)(311.5,-8.5) \Line(370,50)(428.5,-8.5)
  \Line(370,45)(315,-10) \Line(370,45)(425,-10)
  \Line(370,35)(321,-14) \Line(370,35)(419,-14)
  \Line(370,20)(330,-20) \Line(370,20)(410,-20)
  \Line(370,-5)(343,-32) \Line(370,-5)(397,-32)
  \Line(370,-60)(315,-115) \Line(370,-60)(425,-115)
  \Line(370,-85)(288,-167) \Line(370,-85)(452,-167)
  \Line(369,-50)(369,-28) \Line(371,-50)(371,-28)
  \Line(370,-25)(366,-35) \Line(370,-25)(374,-35)
  \Text(360,-28)[l]{\large $t$}
  \Text(370,-155)[t]{de~Sitter}
  \LongArrow(290,40)(320,20) \Text(285,47)[b]{Anti de~Sitter}
\end{picture}
\caption{A quantum state for the multiverse, $\left| \Psi(t) \right>$, 
 is defined on past light cones ($45^\circ$ lines) bounded by apparent 
 horizons (thick solid lines).  The left (right) diagram represents 
 a nucleation of a Minkowski (anti de~Sitter) bubble in a meta-stable 
 de~Sitter vacuum.  The bubble walls are depicted by dashed lines.}
\label{fig:apparent}
\end{center}
\end{figure}

One may wonder what happens if the observer enters into a Minkowski 
or anti de~Sitter bubble at a late stage in the multiverse evolution. 
In this case, the (stretched) event horizon disappears, which would 
be bounding the region where $\left| \Psi(t) \right>$ is defined. 
Does this mean that we need to know the entire (future) cosmic history 
to determine the region where $\left| \Psi(t) \right>$ is defined? 
Or should we suddenly change the defining region of $\left| \Psi(t) 
\right>$ to the entire past light cone {\it all the way} back to 
$\Sigma$, when the observer undergoes a transition from a de~Sitter 
to a Minkowski/anti de~Sitter phase?  Both of these would be highly 
unreasonable.  Considerations along these lines lead us to the following 
picture:\ the defining region for $\left| \Psi(t) \right>$ is determined 
not by the event horizon but by the {\it apparent horizon}, a surface 
on which the (local) expansion of the cross sectional area of the past 
light cone turns from positive to negative.%
\footnote{For a use of apparent horizons in the context of a geometric 
 cutoff, see~\cite{Bousso:2010im}.}
With this definition, the state $\left| \Psi(t) \right>$ lives on a 
smaller portion of the light cone, as sketched in Fig.~\ref{fig:apparent} 
for nucleations of Minkowski and anti de~Sitter bubbles.
Note that since the apparent horizon agrees with the Hubble horizon in 
the limit of small curvature, the picture of Fig.~\ref{fig:stretched} 
is essentially unchanged.  In particular, the horizon should still be 
stretched.

In fact, there are several reasons to choose the apparent horizon as 
the boundary of the defining region for the multiverse state:
\begin{itemize}
\item
The concept of apparent horizon does not depend on low energy physics, 
such as the gauge group or matter content, so that the definition can be 
extended to the entire multiverse.
\item
While the apparent horizon does not exist for all spacetimes, it does 
exists in {\it any} Friedmann-Robertson-Walker (FRW) universe that starts 
from a big-bang or bubble nucleation in a higher energy meta-stable vacuum, 
i.e., spacetime relevant for cosmology in the multiverse.
\item
The location of an apparent horizon is determined ``locally'' within 
the bubble universe.  In particular, unlike a particle or event horizon, 
it does {\it not} depend on what happens in the infinite past or future. 
This is very comfortable.  In particular, it allows us to describe 
physics without knowing the future.
\item
The apparent horizon plays a role of the preferred screen in the holographic 
principle~\cite{Bousso:1999xy}, i.e., its area bounds the amount of 
entropy on a light sheet associated with it.  Here, a light sheet is 
defined as a congruence of light rays orthogonal to the screen whose 
cross sectional areas are non-increasing in the direction away from it.
\end{itemize}

The last point is especially interesting, since our constant time slice 
is given by a past light cone, which is precisely one of the light sheets 
associated with the horizon.  This allows us to state the holographic 
principle in a very simple form:\ {\it at a given time $t$, the maximal 
number of degrees of freedom in the bulk is bounded by the area of the 
boundary in Planck units}.  Namely,
\begin{equation}
  \ln \left\{ {\rm max}\,{\rm dim} 
    \left( \left| \Psi_{\rm bulk}(t) \right> \right) \right\} 
  \leq \frac{1}{4 l_P^2} {\cal A}_{\rm horizon}(t),
\label{eq:holographic}
\end{equation}
where $\left| \Psi_{\rm bulk}(t) \right>$ is the bulk quantum state, and 
${\cal A}_{\rm horizon}(t)$ the area of the horizon bounding the ``space'' 
(i.e.\ the light cone).  Another important point is that the mandatory 
existence of the apparent horizon (the second item above) allows us to 
specify the initial condition on a stretched apparent horizon and the 
past light cone bounded by it.  This has an advantage that the introduction 
of a space-like hypersurface $\Sigma$ is not necessary, which considerably 
simplifies the formalism.  In fact, by specifying initial conditions 
this way, the state $\left| \Psi(t) \right>$ is defined {\it consistently} 
on the past light cone bounded by the stretched apparent horizon.

We now have the multiverse state $\left| \Psi(t) \right>$ which is 
consistent with the principles of quantum mechanics.  In general, $\left| 
\Psi(t) \right>$ consists of a superposition of different semi-classical 
spacetime configurations, and the above analysis applies to each of these 
components.  The Hilbert space for the multiverse state is, therefore, 
given by
\begin{equation}
  {\cal H} = \oplus_{\cal M} \bigl( {\cal H}_{{\cal M}, {\rm bulk}} 
    \otimes {\cal H}_{{\cal M}, {\rm horizon}} \bigr),
\label{eq:H-decomp}
\end{equation}
where ${\cal H}_{{\cal M}, {\rm bulk}}$ and ${\cal H}_{{\cal M}, {\rm 
horizon}}$ represent Hilbert spaces for the degrees of freedom on the 
past light cones {\it inside} and {\it on} the stretched apparent horizon, 
respectively, for a fixed semi-classical configuration ${\cal M}$. 
The representation of ${\cal H}$ in Eq.~(\ref{eq:H-decomp}) is analogous 
to the Fock space in usual quantum field theories.

How does the multiverse state $\left| \Psi(t) \right>$ evolve?  The 
quantum observer principle states that the evolution is deterministic 
and unitary:
\begin{equation}
  \left| \Psi(t_1) \right> = U(t_1,t_2) \left| \Psi(t_2) \right>,
\label{eq:evolution}
\end{equation}
where $U(t_1,t_2)$ is a unitary operator acting on the multiverse state 
$\left| \Psi(t) \right>$.  How can we determine the form of $U(t_1,t_2)$? 
Since the definition of the state involves horizon degrees of freedom, 
an explicit expression of $U(t_1,t_2)$ is obtained only with knowledge 
of quantum gravity.%
\footnote{In particular, quantum gravity must provide a picture of 
 how unitarity is ensured despite the fact that some of the semi-classical 
 histories represented in $\left| \Psi(t) \right>$ end up with big crunch 
 or black hole singularities; a conjecture on this will be given in 
 Section~\ref{subsec:fate}.  Also, since the apparent horizon can 
 have space-like world volumes in some regions, its ``time evolution'' 
 should be into space-like directions in those regions.  This is 
 not a problem.  Since the holographic principle allows the apparent 
 horizon to encode all the information on the portion of the observer's 
 past light cone that lies {\it in the past of} the horizon, such a 
 ``space-like evolution'' can be equivalent to the standard time evolution 
 of the system {\it outside} the horizon.}
Therefore, while the description in terms of $\left| \Psi(t) \right>$
is essential for the understanding of the conceptual issues, it is not 
very ``practical'' in making predictions for low energy physics (with 
current theoretical technology).  This difficulty, however, can be 
circumvented if we adopt the following ``approximation.'' Suppose we 
are maximally ignorant about the state of the horizon degrees of freedom, 
so the multiverse is described by a {\it bulk density matrix}
\begin{equation}
  \rho_{\rm bulk}(t) = {\rm Tr}_{\rm horizon} 
    \left| \Psi(t) \right> \left< \Psi(t) \right|,
\label{eq:rho_bulk}
\end{equation}
where ${\rm Tr}_{\rm horizon}$ means the partial trace over the horizon 
degrees of freedom.  This corresponds to the usual statistical 
description of the horizons, so that the evolution of $\rho_{\rm bulk}(t)$ 
can be determined by semi-classical calculations in low energy quantum 
field theories (for example, through thermal treatments for future 
horizons~\cite{Bekenstein:1973ur,Unruh:1976db} and stochastic approaches 
for past horizons~\cite{Mukhanov:1981xt}).  Note that the unitarity 
of the evolution is {\it not} preserved in this description, so that 
information {\it appears} to be lost in some processes, as was found 
in the classic black hole analysis by Hawking~\cite{Hawking:1976ra}. 
The unitarity, however, is recovered once we include the horizon degrees 
of freedom, as in Eq.~(\ref{eq:evolution}).

We finally mention the possibility that the multiverse itself is in 
a mixed state, rather than a pure state (which occurs, e.g., if the 
theory of initial conditions requires it or if our knowledge of the 
system is incomplete).  In this case, the multiverse is described by 
a density matrix
\begin{equation}
  \rho(t) = \sum_i \lambda_i 
    \left| \Psi_i(t) \right> \left< \Psi_i(t) \right|,
\label{eq:rho}
\end{equation}
whose evolution is given by
\begin{equation}
  \rho(t_1) = U(t_1,t_2)\, \rho(t_2)\, U(t_2,t_1),
\label{eq:rho-evolve}
\end{equation}
where we take $\sum_i \lambda_i = 1$ following the convention.  Again, 
quantum gravity is needed to obtain the complete evolution.  The 
description corresponding to Eq.~(\ref{eq:rho_bulk}) can be obtained 
by considering a reduced density matrix
\begin{equation}
  \rho_{\rm bulk}(t) = {\rm Tr}_{\rm horizon}\, \rho(t),
\label{eq:rho-red_bulk}
\end{equation}
whose evolution can be determined using semi-classical calculations.

\subsection{Probabilities in the quantum universe}
\label{subsec:prob}

Having defined states, we now consider operators.  As discussed in 
Section~\ref{subsec:BH}, we take the Hilbert space basis in which local 
operators are ``diagonalized.''  Strictly speaking, with quantization 
on past light cones, operators at different points do not all commute, 
as in the case of usual light front quantization~\cite{Dirac:1949cp}. 
For example, in an ordinary Minkowski space, operators in the same 
angular direction do not necessarily commute due to possible causal 
connections.  This subtlety, however, is not essential.  In fact, we 
can avoid it if we adopt quantization on space-like hypersurfaces, 
instead of null hypersurfaces.  The only crucial thing for our framework 
is to restrict the spacetime region to inside the causal patch bounded 
by apparent horizons---with this restriction, the problem discussed 
in Section~\ref{subsec:BH} does not arise, and our previous formulae 
all persist.  Here, however, we keep using past light cones as our 
``equal time'' hypersurfaces.  This has an advantage that observational 
conditions are easier to impose, since our ``direct'' knowledge (without 
using the evolution equation) is intrinsically limited to the spacetime 
region inside our past light cone.

The probabilities are defined through operators projecting Hilbert 
space onto subspaces which satisfy specified (observational) conditions. 
Let $\left| \Psi_{A,i} \right>$ be a set of orthonormal states that 
satisfy condition $A$.  Then the corresponding projection operator 
is given by
\begin{equation}
  {\cal O}_A = \sum_i \left| \Psi_{A,i} \right> \left< \Psi_{A,i} \right|.
\label{eq:O_A-psi}
\end{equation}
The probability that a past light cone satisfying $A$ also has a property 
$B$ is then given by
\begin{equation}
  P(B|A) = \frac{\int\!dt \left< \Psi(t) \right| 
    {\cal O}_{A \cap B} \left| \Psi(t) \right>}
    {\int\!dt \left< \Psi(t) \right| {\cal O}_A \left| \Psi(t) \right>},
\label{eq:probability-AB}
\end{equation}
where we have assumed that the multiverse is in a pure state $\left| 
\Psi(t) \right>$.  This probability corresponds to the semi-classical 
probability given in Eq.~(\ref{eq:final-1}).  We can similarly define 
the probability of a past light cone $A$ to evolve into a particular 
future situation $C$:
\begin{equation}
  P(C|A) = \frac{\int\!dt \left< \Psi(t) \right| 
    {\cal O}_{A \rightarrow C} \left| \Psi(t) \right>}
    {\int\!dt \left< \Psi(t) \right| {\cal O}_A \left| \Psi(t) \right>},
\label{eq:probability-AC}
\end{equation}
where ${\cal O}_{A \rightarrow C}$ is the operator projecting onto 
states that satisfy $A$ and evolve into $C$.  This corresponds to 
Eq.~(\ref{eq:final-2}) in the semi-classical picture.  Note that the 
comment in footnote~\ref{ft:AB-AC} still applies for the probabilities 
defined here, since the quantum evolution of $\left| \Psi(t) \right>$ 
is deterministic.

The probabilities in Eqs.~(\ref{eq:probability-AB}) and 
(\ref{eq:probability-AC}) are well-defined.  In particular, the 
numerators and denominators in these expressions are separately finite.%
\footnote{Strictly speaking, they are finite only if $\left| \Psi(t) 
 \right>$ is normalized.  If not, they can be infinite, but these 
 infinities cancel between the numerators and denominators, giving 
 well-defined, finite probabilities.}
This finiteness can be understood as follows.  Let us expand $\left| 
\Psi(t) \right>$ as
\begin{equation}
  \left| \Psi(t) \right> = a_1 \left| \Psi_1(t) \right> 
    + a_2 \left| \Psi_2(t) \right> + \cdots,
\label{eq:a_n-exp}
\end{equation}
where $\left| \Psi_n(t) \right>$ ($n=1,2,\cdots$) corresponds 
to a component which encounters $A$-satisfying past light cones 
$n$~times in the cosmic history.  The quantity in the denominators 
in Eqs.~(\ref{eq:probability-AB}) and (\ref{eq:probability-AC}) 
is then given by
\begin{equation}
  \int\!dt \left< \Psi(t) \right| {\cal O}_A \left| \Psi(t) \right> 
  = \sum_{n=1}^{\infty} n\, |a_n|^2.
\label{eq:n-sum}
\end{equation}
Now, starting from any generic initial conditions, the probability of 
finding a past light cone that satisfies $A$ is exponentially small 
(since it requires the geodesic to tunnel into a particular vacuum, 
or vacua, in which $A$ can be satisfied).  This implies that $a_n$ 
scales as
\begin{equation}
  a_n \sim e^{-c n}
\quad
  (c > 0),
\end{equation}
so that the sum in Eq.~(\ref{eq:n-sum}) converges.  The expression 
in Eq.~(\ref{eq:n-sum}) is thus finite, as long as the multiverse state 
is appropriately normalized, e.g.\ $\left< \Psi(t) | \Psi(t) 
\right> = 1$.  The finiteness of other quantities can also be understood 
in a similar way.

The probabilities defined here are also ``gauge invariant,'' i.e.\ they 
do not depend on time parametrization.  This is fairly obvious from the 
expressions in Eqs.~(\ref{eq:probability-AB}) and (\ref{eq:probability-AC}), 
but it can also be understood along the lines above.  Basically, our 
probabilities count the number of times each history, $\left| \Psi_n(t) 
\right>$, encounters a particular past light cone(s).  Since the 
coefficients $a_n$ do not depend on time parametrization, the resulting 
probabilities do not depend on it either.

As discussed in Section~\ref{subsec:single}, the complete evolution 
of $\left| \Psi(t) \right>$ requires the knowledge of quantum gravity, 
so that the probabilities in Eqs.~(\ref{eq:probability-AB}) and 
(\ref{eq:probability-AC}) can only serve the role of defining the 
framework.  To do a ``practical'' calculation, we need to focus on bulk 
physics, i.e.\ $\rho_{\rm bulk}(t)$ introduced in Eq.~(\ref{eq:rho_bulk}) 
(at the cost of unitarity in processes involving horizons).  The 
observational conditions, such as $A$, should then be imposed only 
on the bulk part of the state.  This is not a strong restriction, 
since our observational data are always on bulk physics in practice. 
The relevant projection operator is given by
\begin{equation}
  {\cal O}_{{\rm bulk},A} = \sum_i \left| \Psi_{{\rm bulk},A,i} \right> 
    \left< \Psi_{{\rm bulk},A,i} \right|,
\label{eq:O_A-rho}
\end{equation}
where $\left| \Psi_{{\rm bulk},A,i} \right>$ is an orthonormal 
set of the bulk part of the states satisfying condition $A$; see 
Eq.~(\ref{eq:H-decomp}).  The probability $P(B|A)$ is then given by
\begin{equation}
  P(B|A) = \frac{\int\!dt\, {\rm Tr}\left[ \rho_{\rm bulk}(t)\, 
    {\cal O}_{{\rm bulk}, A \cap B} \right]}
    {\int\!dt\, {\rm Tr}\left[ \rho_{\rm bulk}(t)\, 
    {\cal O}_{{\rm bulk}, A} \right]},
\label{eq:bulk-probab-AB}
\end{equation}
where the trace is over the bulk part of Hilbert space.  The 
probability $P(C|A)$ is defined similarly, with the replacement 
${\cal O}_{{\rm bulk}, A \cap B} \rightarrow {\cal O}_{{\rm bulk}, 
A \rightarrow C}$.

The definition of Eq.~(\ref{eq:bulk-probab-AB}) allows us to calculate 
probabilities without knowing quantum gravity.  In principle it 
allows us to answer any questions, except for the ones regarding 
information stored on a horizon at some time in the history.  In 
Appendix~\ref{app:calc-quantum}, we give sample calculations in toy 
landscapes, where it is shown that the results agree with those obtained 
using the semi-classical definition.  Of course, to have definite 
numbers, an initial condition for $\rho_{\rm bulk}(t)$ needs to be 
specified.  The issue of initial conditions will be discussed in 
Sections~\ref{sec:initial} and \ref{sec:discuss}.

We finally mention the case where the multiverse is in a mixed state 
$\rho(t)$.  In this case, the probability is given by
\begin{equation}
  P(B|A) = \frac{\int\!dt\, 
    {\rm Tr}\left[ \rho(t)\, {\cal O}_{A \cap B} \right]} 
    {\int\!dt\, {\rm Tr}\left[ \rho(t)\, {\cal O}_A \right]},
\label{eq:rho-probab-AB}
\end{equation}
where the trace and projection operators act on the entire Hilbert space. 
The probability $P(C|A)$ can be defined similarly.

\subsection{Quantum-to-classical transition}
\label{subsec:Q-to-C}

As we have seen, our framework is (necessarily) quantum mechanical. 
On the other hand, our daily experience is certainly (almost) classical. 
How does this dichotomy arise?  Why do we not observe, e.g., a state 
which is a superposition of different macroscopic configurations?

To be specific, let us suppose that the multiverse was in some highly 
symmetric state $\left| \Psi(t_0) \right>$ at an early moment.  In 
particular, we consider that $\left| \Psi(t_0) \right>$ respects 
rotational symmetry.  This leads to a question:\ why do we not observe 
a chair next to us in a rotationally invariant, $s$-wave state, i.e., 
a superposition of chairs with all different orientations?  One might 
think that various physical processes, such as bubble nucleations, 
spontaneously break rotational symmetry.  This is, however, too naive. 
The resulting state will still be a linear combination of states with 
various bubbles nucleating in all different locations such that rotational 
invariance is respected.  Note that this problem is particularly acute 
in our context.  With $\left| \Psi \right>$ being the quantum state 
for the entire multiverse, there is no ``environment'' with which 
$\left| \Psi \right>$ interacts to feel violation of rotational 
invariance.

Let us focus on a particular macroscopic object, such as a chair, desk, 
or human.  As discussed above, we expect it to be in a rotationally 
invariant state:
\begin{equation}
  \left| \Psi \right> \sim \Bigl( \bigl| \chairup \bigr> 
    + \bigl| \chairdown \bigr> + \cdots \Bigr),
\label{eq:chair}
\end{equation}
where we have used a chair as an example and displayed explicitly only 
two configurations (upward and downward).%
\footnote{Since the chair is a macroscopic object, $\bigl| \ftchairup 
 \bigr>$ ($\bigl| \ftchairdown \bigr>$) itself can be an arbitrary 
 superposition of microscopic states that macroscopically look like 
 an upward (downward) chair.}
Now, consider a second object, say a human, next to the chair.  Is the 
quantum state a direct product of the rotationally invariant chair state 
and the rotationally invariant human state?  Namely,
\begin{equation}
  \left| \Psi \right> \stackrel{?}{\sim} 
    \Bigl( \bigl| \chairup \bigr> + \bigl| \chairdown \bigr> + \cdots \Bigr) 
  \otimes \Bigl( \bigl| \manup \bigr> + \bigl| \mandown \bigr> + \cdots \Bigr).
\label{eq:combined-1}
\end{equation}

The answer is no.  To see this explicitly, consider the Hamiltonian for 
the chair
\begin{equation}
  \left( \begin{array}{cc}
    \bigl< \chairup \bigr| & \bigl< \chairdown \bigr|
  \end{array} \right)\,
  \mbox{\large $\hat{H}$}\,
  \left( \begin{array}{c}
    \bigl| \chairup \bigr> \\ \bigl| \chairdown \bigr>
  \end{array} \right)
  = \left( \begin{array}{cc}
    A & B \\ B & A
  \end{array} \right),
\label{eq:H-chair}
\end{equation}
where we have reduced the system to two states for presentation 
purposes, and the particular form of the matrix in the right-hand side 
is dictated by ``rotational'' symmetry ($Z_2$ in the reduced system). 
We find that the two eigenstates of the Hamiltonian are $\bigl( \bigl| 
\chairup \bigr> \pm \bigl| \chairdown \bigr> \bigr)/\sqrt{2}$ with 
eigenvalues $A \pm B$, which are also eigenstates of ``rotational'' 
symmetry.  However, for a macroscopic object, $B$ is exponentially small, 
since it involves quantum tunneling between the upward and downward 
configurations.  Consequently, the two states are nearly degenerate, 
and the preferred basis for the chair is determined by any small 
``rotational''-symmetry violating perturbations~\cite{Weinberg:QFT-2}. 
In the above chair-human system, the perturbation is the existence 
of the human, so that the state of the entire system is {\it not} 
given by Eq.~(\ref{eq:combined-1}), but by
\begin{equation}
  \left| \Psi \right> \sim 
    \Bigl( \bigl| \chairup \bigr> \otimes \bigl| \manup \bigr> 
    + \bigl| \chairdown \bigr> \otimes \bigl| \mandown \bigr> + \cdots \Bigr),
\label{eq:combined-2}
\end{equation}
where we have arbitrarily assumed that the dynamics prefers to align 
the orientations of the chair and human, rather than anti-align.  (Our 
conclusion does not depend on this particular choice.)  The process 
transforming a (macroscopic) composite system to an entangled form, e.g.\ 
as in Eq.~(\ref{eq:combined-2}), is called decoherence, which typically 
occurs with extremely short timescales~\cite{Schlosshauer}.

We now see why we do not observe a superposition of different macroscopic 
configurations in our daily life.  Equation~(\ref{eq:combined-2}) 
says that the chair always has a definite orientation {\it with 
respect to} the human, which we may identify with ourselves.  We do 
not observe a superposition of chairs, which would have been possible 
if the state contained a term such as $\bigl( \bigl| \chairup \bigr> 
+ \bigl| \chairdown \bigr> \bigr) \otimes \bigl| \manup \bigr>$ as 
in Eq.~(\ref{eq:combined-1}).  Note that quantum interferences between 
different terms in Eq.~(\ref{eq:combined-2}) are extremely small, since 
overlaps between macroscopically different configurations, such as 
$\bigl| \manup \bigr>$ and $\bigl| \mandown \bigr>$, are suppressed 
by the huge dimensionality of the corresponding Hilbert space.  In fact, 
for any observables constructed out of local operators, matrix elements 
between macroscopically distinct states are highly suppressed, e.g.\ 
$\bigl< \manup \big| {\cal O}_A \big| \mandown \bigr> \lll \bigl< 
\manup \big| {\cal O}_A \big| \manup \bigr>, \bigl< \mandown \big| 
{\cal O}_A \big| \mandown \bigr>$.  This, therefore, provides preferred 
bases for any macroscopic systems.

Of course, the {\it entire} state in Eq.~(\ref{eq:combined-2}) is still 
rotationally invariant.  This, however, does not matter.  Since we 
ourselves are a part of the state, we never observe the rotationally 
invariant state $\left| \Psi \right>$.  In fact, rotational invariance 
of the state $\left| \Psi \right>$ implies that for any component $\left| 
\psi \right>$ in $\left| \Psi \right>$, there are terms whose {\it entire 
histories} are related to $\left| \psi \right>$ via rotational symmetry 
(see e.g.\ the first two terms in Eq.~(\ref{eq:combined-2})).  For 
a macroscopic system, the evolution of these terms are mutually independent 
for all practical purposes.  Thus, we may simply use $\left| \psi \right>$ 
(or any other term related to it by the symmetry) when making predictions. 
Of course, we may still use the entire state $\left| \Psi \right>$ if 
we want---the two procedures give identical results.

The consideration here addresses various questions associated with the 
quantum-to-classical transition, for example, why spontaneous symmetry 
breaking occurs at all~\cite{Weinberg:QFT-2}, why density perturbation 
in inflation becomes classical~\cite{Weinberg:cosmo}, and why we do 
not observe a superposition of bubble universes.  In our context, the 
multiverse state is indeed a superposition of various macroscopically 
different configurations.  We see (almost) classical physics because 
we---who are {\it a part of} the state---are correlated (entangled) with 
the rest of the multiverse.  Technically, our framework incorporates 
quantum measurement processes in the form of the imposition of observational 
conditions, such as $A$, $B$, and $C$ in Eqs.~(\ref{eq:probability-AB}) 
and (\ref{eq:probability-AC}).  This procedure exactly extracts 
information encoded in the correlations between ourselves and the 
rest of the world.

\subsection{The possibility of a reduced Hilbert space}
\label{subsec:reduced-H}

We finally mention the possibility that the Hilbert space of the theory 
is actually smaller than that in Eq.~(\ref{eq:H-decomp}).

Consider an arbitrary semi-classical configuration of spacetime ${\cal M}$. 
In the multiverse, such a configuration arises as components of $\left| 
\Psi(t) \right>$ at various times $t_a$ ($a=1,2,\cdots$).  Let us define 
an ensemble of all these states:
\begin{equation}
  {\cal E} = \{ {\cal O}_{\cal M} \left| \Psi(t_a) \right> \},
\label{eq:holo-E}
\end{equation}
where the states are at fixed times $t_a$, and ${\cal O}_{\cal M}$ is 
the projection operator.  Since a component of a pure state $\left| 
\Psi(t) \right>$ is also a pure state, elements of ${\cal E}$ are all 
pure states, which take the form
\begin{equation}
  \left| \Psi_i \right> = \left| \Psi_{\rm bulk,i} \right> 
    \otimes \left| \Psi_{\rm horizon,i} \right>,
\label{eq:holo-decomp}
\end{equation}
where $i = 1,\cdots,{\rm dim}({\cal E})$.

Now, the holographic principle says that (the logarithm of) the number 
of {\it all possible} bulk configurations is bounded by the horizon area 
in Planck units:
\begin{equation}
  S_{{\cal M},{\rm bulk}} \equiv \ln\left\{ {\rm dim}\left( 
    \cup_i \left| \Psi_{{\rm bulk},i} \right> \right) \right\} 
  \leq \frac{{\cal A}_{\rm horizon}}{4 l_P^2}.
\label{eq:holo}
\end{equation}
We expect that this inequality is saturated, since the multiverse 
realizes all possible states throughout the history.  It is also 
reasonable to assume that the horizon can contain only $O(1)$ bits of 
information per Planck area.  Suppose that this number takes a special 
value of $1/4$, i.e.\ $S_{{\cal M},{\rm horizon}} \equiv \ln\left\{ 
{\rm dim}\left( \cup_i \left| \Psi_{{\rm horizon},i} \right> \right) 
\right\} = {\cal A}_{\rm horizon}/4 l_P^2$.  Then we find
\begin{equation}
  S_{{\cal M},{\rm bulk}} = S_{{\cal M},{\rm horizon}},
\label{eq:S=S}
\end{equation}
i.e.\ the numbers of bulk and horizon degrees of freedom are the same 
for a fixed ${\cal M}$.

If Eq.~(\ref{eq:S=S}) is true for arbitrary ${\cal M}$, then it allows 
for an interesting possibility that the Hilbert space of the theory 
is actually the square root of Eq.~(\ref{eq:H-decomp}), namely, there 
are one-to-one correspondences between elements of ${\cal H}_{{\cal M}, 
{\rm bulk}}$ and ${\cal H}_{{\cal M}, {\rm horizon}}$ for all ${\cal M}$. 
If this is the case, then we can provide a {\it complete} description 
of the multiverse in terms of {\it either} $\left| \Psi_{\rm bulk}(t) 
\right>$ {\it or} $\left| \Psi_{\rm horizon}(t) \right>$.

\section{Initial Conditions}
\label{sec:initial}

The framework developed so far allows us to make predictions/postdictions 
once initial conditions are given.  On the other hand, the framework 
itself does not provide a unique initial condition.  This is, in fact, 
as it should be.  Remember that the measure problem of eternal inflation 
has {\it a priori} nothing to do with the ``beginning'' of spacetime. 
Once eternal inflation occurs in the regime where semi-classical analyses 
are valid, then it already leads to the predictivity crisis---any event 
that can happen will happen infinitely many times.  Our general framework 
should be able to regulate these infinities, starting from {\it any} 
eternally inflating vacuum.  Namely, the framework should be sufficiently 
modular such that any of such vacua can be used as initial conditions, 
e.g., on $\Sigma$.

On the other hand, the framework also provides a useful tool to probe 
the real beginning of spacetime, i.e., the initial condition of the 
multiverse.  In this section, we discuss the issue of initial conditions 
from both these two perspectives.  We first consider what initial 
conditions we need to use to calculate probabilities starting from 
an arbitrary eternally inflating state.  We then briefly discuss 
the beginning of spacetime, deferring full discussions to 
Section~\ref{sec:discuss}.

\subsection{Semi-classical predictions from eternal inflation}
\label{subsec:eternal-init}

Suppose we consider a spacetime region that is eternally inflating, 
with the fields taking values collectively denoted as $\phi$.  We want 
to derive predictions for the future, e.g.\ $P(B|A)$ and $P(C|A)$, 
using our framework.  What initial condition should we impose on 
the multiverse state?

We first note that the quantum state of the universe is {\it not} 
uniquely determined by saying that the universe is in a (approximately) 
de~Sitter state with field values $\phi$.  There are two issues associated 
with this.  The first comes from the fact that the de~Sitter space has 
a nontrivial entropy $S = 3/8 G_N^2 V(\phi)$, which implies that this 
space, in fact, represents a statistical ensemble of many quantum states 
with $S$ degrees of freedom.  Here, $G_N$ is Newton's constant and 
$V(\phi)$ is the potential energy density.  The only thing we know 
is that the system is in a thermal state with de~Sitter temperature 
$T = \sqrt{2 G_N V(\phi)/3\pi}$ at the level of semi-classical 
approximation.  In our framework, such a thermal picture arises 
(only) after integrating out horizon degrees of freedom, assuming 
complete ignorance about these degrees of freedom.  This implies 
that we can (only) use the formalism based on the bulk density matrix, 
$\rho_{\rm bulk}(t)$, when we address the questions discussed here.

The second issue is that eternally inflating spacetime is in 
fact not pure de~Sitter---it has a nonzero decay width.  This 
implies that we need to specify an initial hypersurface on which 
the universe was still in an eternally inflating phase.  This 
specification affects physical predictions at late times, albeit 
weakly~\cite{Garriga:2006hw,Freivogel:2009it}; namely, different 
choices for the hypersurface correspond to different physical setups. 
In our framework, a canonical choice for equal time hypersurfaces is to 
take them to be the observer's past light cones bounded by the stretched 
horizon.  Therefore, it is natural to define an initial condition on 
one of these hypersurfaces.  Indeed, this choice is physically well 
motivated, since the spacetime region the observer can actually see 
is limited to that within his/her past light cone.  With this choice, 
we are specifying that the entire universe was in an eternally inflating 
phase at some point in the history, {\it as seen by} the observer.

\begin{figure}[t]
\begin{center}
\vspace{2mm}
  \includegraphics[scale=0.7]{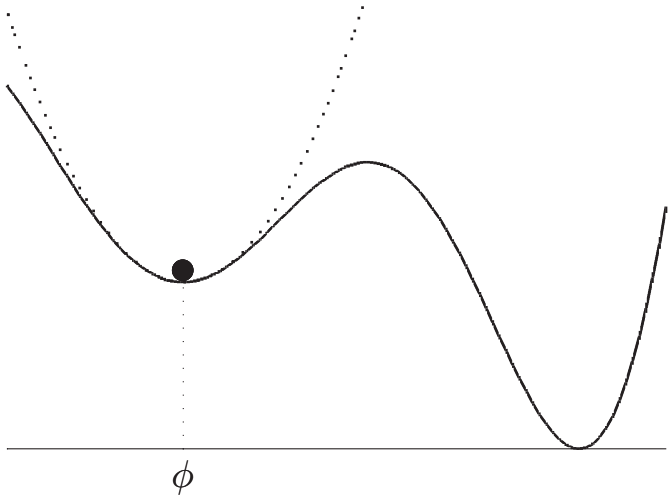}
\vspace{1mm}
\caption{The Hamiltonian $\hat{H}_\phi$ is that of a theory with the 
 potential having the global minimum at $\phi$ and mimicking the original 
 theory around $\phi$.  The potential of this theory and that of the 
 original theory are drawn by the dotted and solid lines, respectively.}
\label{fig:false}
\end{center}
\end{figure}

The initial condition we need to use, then, is that the universe is in 
a (approximately) de~Sitter state at some initial moment, say at $t=t_0$:
\begin{equation}
  \rho_{\rm bulk}(t_0) \simeq \frac{1}{{\rm Tr}\,e^{-\beta \hat{H}_\phi}}
    e^{-\beta \hat{H}_\phi},
\label{eq:rho-init}
\end{equation}
where $\beta = 1/T = \sqrt{3\pi/2 G_N V(\phi)}$, which depends on 
$\phi$, and $\hat{H}_\phi$ is the Hamiltonian operator defined for 
a theory which mimics the original theory around $\phi$ but has the 
stable vacuum at $\phi$ (see Fig.~\ref{fig:false}).  (The trace in 
the denominator is taken in the Hilbert space of this theory.) 
There is, of course, an ambiguity in defining such a theory, and thus 
$\hat{H}_\phi$, since there is no guideline for choosing a potential 
of the new theory above the potential barrier of the original theory. 
This ambiguity, however, leads to only exponentially suppressed 
corrections in the final results, as long as the de~Sitter temperature 
$T$ is much smaller than the barrier height, which is necessary anyway 
for the minimum $\phi$ to be meta-stable.  (And we do not expect 
semi-classical calculations to correctly reproduce exponentially 
suppressed contributions anyway.)

With the initial condition of Eq.~(\ref{eq:rho-init}), we can now derive 
any predictions of eternal inflation using Eq.~(\ref{eq:bulk-probab-AB}). 
As discussed before, this can be done without knowing quantum gravity.

\subsection{Issues with the initial condition of the multiverse}
\label{subsec:initial}

Let us now consider the beginning of spacetime.  This will require 
a theory of initial conditions beyond what we have developed so far. 
Here we only discuss several issues associated with it, leaving more 
complete discussions to Section~\ref{sec:discuss}.

One possibility for the beginning is that it is purely determined 
by semi-classical considerations.  Suppose that spacetime starts from 
some highly symmetric state, with the fields sitting at an extremum of 
the potential.  Let $\phi_a$ ($a = 1,\cdots,n$) denote $n$ such extrema 
that have positive energy densities.  We may then consider the initial 
bulk density matrix
\begin{equation}
  \rho_{\rm bulk}(t_0) \simeq 
    \sum_{a=1}^{n} \lambda_a \frac{1}{{\rm Tr}\,e^{-\beta_a \hat{H}_{\phi_a}}}
    e^{-\beta_a \hat{H}_{\phi_a}},
\label{eq:init-multiverse}
\end{equation}
where $\sum_{a=1}^n \lambda_a = 1$, $\beta_a = \sqrt{3\pi/2 G_N V(\phi_a)}$, 
and $\hat{H}_{\phi_a}$ is the Hamiltonian defined around the extremum 
at $\phi_a$.%
\footnote{This equation may also apply if the landscape is reducible, 
 i.e., if there are multiple sectors in the potential that are not 
 mutually connected by any physical processes. \label{ft:reducible}}
For
\begin{equation}
  \lambda_a \propto \exp\left( \pm \frac{3}{8 G_N^2 V(\phi_a)} \right),
\label{eq:HH-tunneling}
\end{equation}
this corresponds to Hartle-Hawking~\cite{Hartle:1983ai} (the upper 
sign) and tunneling~\cite{Vilenkin:1984wp} (the lower sign) proposals.%
\footnote{The Hartle-Hawking probability can be understood not as a 
 probability of creating universes, but as a thermal distribution of 
 universes in a theory with a de~Sitter ground state~\cite{Linde:1998gs}. 
 In addition, this probability has the serious problem of overwhelming 
 Boltzmann brain observers~\cite{Page:2006hr}.  Therefore, the tunneling 
 probability seems (relatively) more suitable in the present context.}
There are some unsatisfactory features in these ``semi-classical 
beginning'' pictures (although they may, in some sense, be unavoidable; 
see discussion in Section~\ref{subsec:mega-multiverse}).  First, the 
fact that the initial condition specifies only $\rho_{\rm bulk}$, 
and not $\left| \Psi \right>$, means that there is an {\it intrinsic 
uncertainty} which we cannot hope to reduce; in particular, we in 
principle cannot predict exact states for certain (horizon) degrees 
of freedom.  Second, the expression of Eq.~(\ref{eq:init-multiverse}) 
is only approximate, as the definition of $\hat{H}_{\phi_a}$ has an 
ambiguity, and we do not see any obvious way to make it exact.

An alternative possibility for the beginning is that the theory of initial 
conditions determines the multiverse to be in a specific pure state $\left| 
\Psi(t_0) \right>$ at the earliest moment.  According to our hypothesis, 
this implies that the multiverse is in a pure state for arbitrary time 
$t$, $\left| \Psi(t) \right>$.

In either of these cases, if we admit the existence of {\it the} initial 
state---which is the case in any scenario along the lines of creation 
from ``nothing''~\cite{Zeldovich:1981}---then the quantum observer 
principle is necessarily violated there, since we cannot evolve the 
state further back.  While this is possible, in Section~\ref{sec:discuss} 
we will explore the possibility that the principle is in fact not 
violated throughout the whole history, which will suggest that our 
multiverse is a ``fluctuation'' in some larger structure.

\section{Quantum Measurements and Global Spacetime}
\label{sec:measure-spacetime}

In this section, we see that our framework allows for a unified treatment 
of quantum measurement processes and the eternally inflating multiverse. 
We conclude that the eternally inflating multiverse is the same as many 
worlds in quantum mechanics.  We also discuss the relation of our single 
observer picture to the conventional, global spacetime picture.%
\footnote{A similar relation is discussed independently by Raphael 
 Bousso in the context of geometric cutoff measures~\cite{Bousso}.}

\subsection{Unification of the multiverse and many worlds in 
 quantum mechanics}
\label{subsec:unification}

So far, we have mainly focused on issues at very large scales, e.g.\ 
bubble universes in eternal inflation, in applying our formalism.  This, 
however, need not be the case.  In fact, the formalism applies equally 
to any (even microscopic) quantum processes without modification.

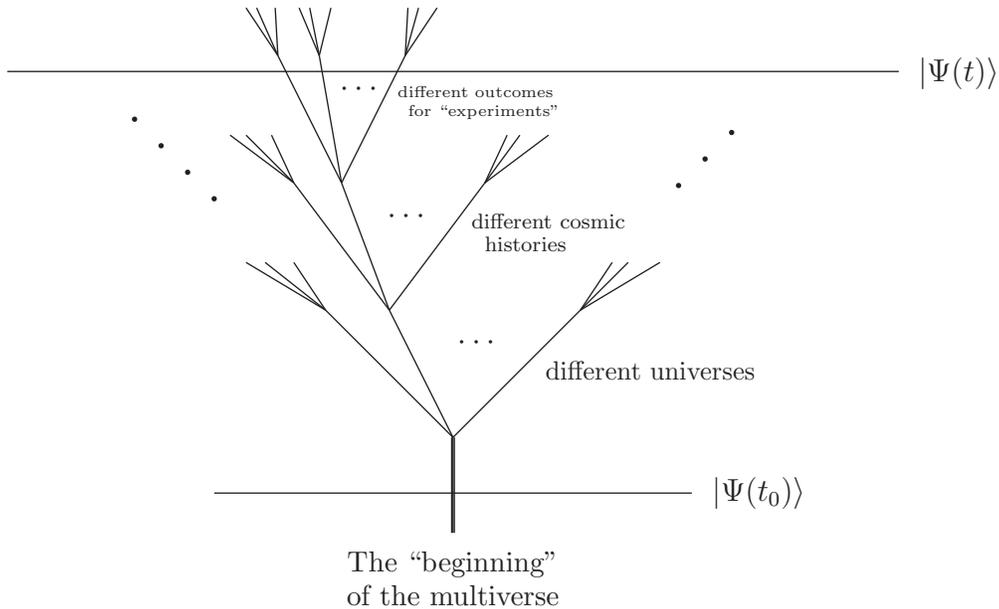
\begin{figure}[t]
\begin{center}
\begin{picture}(375,250)(-170,-35)
  \Line(0,0)(0,36) \Line(0.5,0)(0.5,36) \Line(-0.5,0)(-0.5,36)
  \Text(0,-7)[t]{\small The ``beginning''}
  \Text(0,-20)[t]{\small of the multiverse}
  \Line(0,36)(-48,84)
  \Line(-48,84)(-78,102) \Line(-48,84)(-70.8,102) \Line(-48,84)(-60,102)
  \Vertex(-90,126){1} \Vertex(-100,136){1}
  \Vertex(-110,146){1} \Vertex(-120,156){1}
  \Line(0,36)(-24,84) \Text(9.6,72)[]{$\cdots$}
  \Line(-24,84)(-60,132)
  \Line(-60,132)(-84,150) \Line(-60,132)(-78,150) \Line(-60,132)(-68.4,150)
  \Line(-24,84)(-42,132) \Text(-16.8,120)[]{$\cdots$}
  \Line(-24,84)(12,132)
  \Line(12,132)(20.4,150) \Line(12,132)(25.2,150) \Line(12,132)(36,150)
  \Text(7,121)[lt]{\scriptsize different cosmic}
  \Text(12,112)[lt]{\scriptsize histories}
  \Line(-42,132)(-66,180)
  \Line(-66,180)(-78,198) \Line(-66,180)(-74,198) \Line(-66,180)(-67,198) 
  \Line(-42,132)(-50.4,180) \Text(-34.8,168)[]{$\cdots$}
  \Line(-50.4,180)(-58,198) \Line(-50.4,180)(-54,198) \Line(-50.4,180)(-46,198) 
  \Line(-42,132)(-18,180)
  \Line(-18,180)(-17,198) \Line(-18,180)(-13,198) \Line(-18,180)(-6,198) 
  \Text(-21,169)[lt]{\tiny different outcomes}
  \Text(-17,162)[lt]{\tiny for ``experiments''}
  \Line(0,36)(48,84)
  \Line(48,84)(60,102) \Line(48,84)(66,102) \Line(48,84)(78,102)
  \Vertex(85,131){1} \Vertex(95,141){1} \Vertex(105,151){1}
  \Text(35,65)[lt]{\footnotesize different universes}
  \Line(-168,174)(168,174) \Text(176,174)[l]{$\left| \Psi(t) \right>$}
  \Line(-90,15)(90,15) \Text(98,15)[l]{$\left| \Psi(t_0) \right>$}
\end{picture}
\caption{A schematic picture for the evolution of the multiverse state 
 $\left| \Psi(t) \right>$.  The state is described from the viewpoint of 
 a single ``observer'' (geodesic) traveling the multiverse.}
\label{fig:branching}
\end{center}
\end{figure}

Suppose the multiverse starts from a highly symmetric state $\left| 
\Psi(t_0) \right>$.  This state evolves into a superposition of states 
in which various bubble universes nucleate in various spacetime locations. 
As time passes, a state representing each universe further evolves 
into a superposition of states representing various possible cosmic 
histories, including different outcomes of ``experiments'' performed 
within that universe.  (These ``experiments'' may, but need not, be 
scientific experiments---they can be any physical processes.)  At late 
times, the multiverse state $\left| \Psi(t) \right>$ will thus contain 
an enormous number of terms, each of which represents a possible world 
that may arise from $\left| \Psi(t_0) \right>$ consistently with the 
laws of physics.  A schematic picture of these ``branching'' processes 
is given in Fig.~\ref{fig:branching}.

The resulting picture is remarkably simple.  From the initial state 
$\left| \Psi(t_0) \right>$, the multiverse simply evolves deterministically 
according to the quantum mechanical evolution law.  This evolution, 
however, is not along an axis in Hilbert space that is determined by 
operators local in spacetime.  Therefore, at late times, the multiverse 
state is misaligned with (an enormous number of) axes determined by 
the local operators.  (These axes are analogues of Fock states in usual 
quantum field theories.)  This makes the multiverse state a superposition 
of a huge number of terms corresponding to different cosmic histories:
\begin{equation}
  \left| \Psi(t) \right> \approx 
    \sum_i \left| \mbox{possible world $i$ at time $t$} \right>,
\label{eq:multiverse-t}
\end{equation}
when expanded in the basis determined by local operators.  As discussed 
before, all these histories are described from the viewpoint of a single 
``observer'' (geodesic).

The picture given above is precisely that of the many-worlds interpretation 
of quantum mechanics~\cite{Everett:1957hd}.  Therefore, we conclude 
that {\it the multiverse is the same as (or a specific manifestation 
of) many worlds in quantum mechanics}.  In fact, when we ask physical 
questions, we ``inject'' these questions to the theory by imposing 
conditions such as $A$, $B$, and $C$ in Eqs.~(\ref{eq:probability-AB}) 
and (\ref{eq:probability-AC}).  The questions may be about global 
properties of the universe, or about outcomes of a specific experiment 
being performed.  Our framework, therefore, provides a fully unified 
treatment of (even microscopic) experiments and the multiverse.  Indeed, 
it applies to any physical processes from the smallest (the Planck 
length) to the largest (the apparent horizon) scales.

Incidentally, if we are interested only in the future of a particular 
macroscopic configuration at time $t$, then we may drop all the terms 
in Eq.~(\ref{eq:multiverse-t}) except for ones corresponding to the 
specified configuration.  This is because the superposition principle 
guarantees that the evolution of the retained terms is independent 
of the dropped terms, and matrix elements of macroscopic observables 
between states with different macroscopic configurations are highly 
suppressed (see Section~\ref{subsec:Q-to-C}) so the dropped terms do 
not affect future predictions.  {\it This operation of truncating 
the state is precisely what is called wavefunction collapse}, which 
is indeed an extremely good approximation when we ask questions about 
a system with large degrees of freedom.%
\footnote{The approximation may even be ``exact'' if the system is large 
 enough that the effect from the dropped terms is smaller than what is 
 measurable within the limitation of the uncertainty principle.}
(Note that a system here includes an experimental apparatus, in addition 
to the object being measured.)  It is, however, not an exact procedure 
in general, nor necessary for making predictions.

\subsection{Connection to (``reconstruction'' of) the global picture}
\label{subsec:global}

What is the relation of our picture based on a single quantum observer 
to the picture based on global spacetime?  One (extreme) attitude is 
to consider that ``physical reality'' is simply the multiverse quantum 
state, so that no other picture is needed.  This indeed makes sense 
because all physical questions (regarding predictions/postdictions) 
can be answered using that state, following the prescription in 
Section~\ref{sec:quantum}.  On the other hand, it is useful to 
understand a connection between the present picture and more 
conventional, global spacetime picture.  Here we study this issue.

One way of developing intuition about the connection is to consider 
a large (infinite) number of identical multiverse states.  As shown 
in conventional analyses of frequency operators~\cite{Caves-Schack}, 
the resulting product state can be reorganized as
\begin{equation}
  \left| \Psi(t) \right>^{\otimes N} 
  = \left( \sum_i c_i \left| \alpha_i \right> \right)^{\otimes N} 
  \,\,\stackrel{N \rightarrow \infty}{\longrightarrow}\quad
  \propto \left( \left| \alpha_1 \right>^{|c_1|^2 N} \otimes 
    \left| \alpha_2 \right>^{|c_2|^2 N} \otimes \cdots \right) 
    + {\rm permutations},
\label{eq:frequency}
\end{equation}
up to a zero-norm part of the state.  Along the lines of 
Ref.~\cite{Aguirre:2010rw}, we may interpret the rightmost expression 
to represent each outcome $\left| \alpha_i \right>$ spreading over 
(global) spacetime.  In particular, any experimental results may be 
viewed as distributed over the multiverse, allowing us to interpret 
outcomes of a quantum measurement in the frequentist's sense over 
spacetime.%
\footnote{In contrast to Ref.~\cite{Aguirre:2010rw}, in which the 
 assumption of statistical uniformity played an important role, our 
 argument here requires only Eq.~(\ref{eq:frequency}), since we do not 
 have the measure problem.  In fact, $N$ $\left| \Psi(t) \right>$'s in 
 Eq.~(\ref{eq:frequency}) are {\it identical} copies by construction.}

\begin{figure}[t]
\begin{center}
\begin{picture}(500,140)(-80,-5)
%
%
  \Text(-71,100)[l]{{\large $\Psi(t) \,\sim\, \cdots \,+\, 
    \Biggl($}$\cdots \,\,+$}
  \DashLine(61,80)(101,120){1} \DashLine(101,120)(141,80){1}
  \Text(90,92)[]{\Large $\star$} \Text(150,100)[]{$+$}
  \Line(156,80)(196,120) \Line(196,120)(236,80)
  \Text(197,92)[]{\Large $\star$} \Text(245,100)[]{$+$}
  \DashLine(251,80)(293,120){2} \DashLine(291,120)(333,80){2}
  \Text(304,92)[]{\Large $\star$}
  \Text(333,100)[l]{$+\,\, \cdots${\large $\Biggr) \,+\, \cdots$}}
%
%
  \Line(65,33)(85,33) \Line(80,36)(90,30) \Line(70,36)(60,30)
  \Line(65,27)(85,27) \Line(80,24)(90,30) \Line(70,24)(60,30)
  \Line(62.5,31.5)(60.625,29.625) \Line(66,33)(61.875,28.875)
  \Line(68,33)(63.125,28.125) \Line(70,33)(64.375,27.375)
  \Line(72,33)(66,27) \Line(74,33)(68,27) \Line(76,33)(70,27)
  \Line(78,33)(72,27) \Line(80,33)(74,27) \Line(82,33)(76,27)
  \Line(84,33)(78,27) \Line(85.7,32.7)(80,27) \Line(87.0,32.0)(82,27)
  \Line(88.2,31.2)(84,27) \Line(89.4,30.4)(87.6,28.6)
%
%
  \DashLine(150,0)(150,65){2}
  \DashLine(150,65)(147,60){2} \DashLine(150,65)(153,60){2}
  \DashLine(100,0)(150,50){2} \DashLine(150,50)(200,0){2}
  \Line(165,0)(165,65)
  \Line(165,65)(162,60) \Line(165,65)(168,60)
  \Line(115,0)(165,50) \Line(165,50)(215,0)
  \DashLine(180,0)(180,65){1}
  \DashLine(180,65)(177,60){1} \DashLine(180,65)(183,60){1}
  \DashLine(130,0)(180,50){1} \DashLine(180,50)(230,0){1}
  \Text(165.7,15)[]{\LARGE $\star$}
\end{picture}
\caption{A schematic depiction of the relation between the single quantum 
 observer picture (the upper line) and the global spacetime picture 
 (the lower line).  The location of an event, represented by stars, 
 is discretized for illustrative purposes.}
\label{fig:corresp-1}
\end{center}
\end{figure}
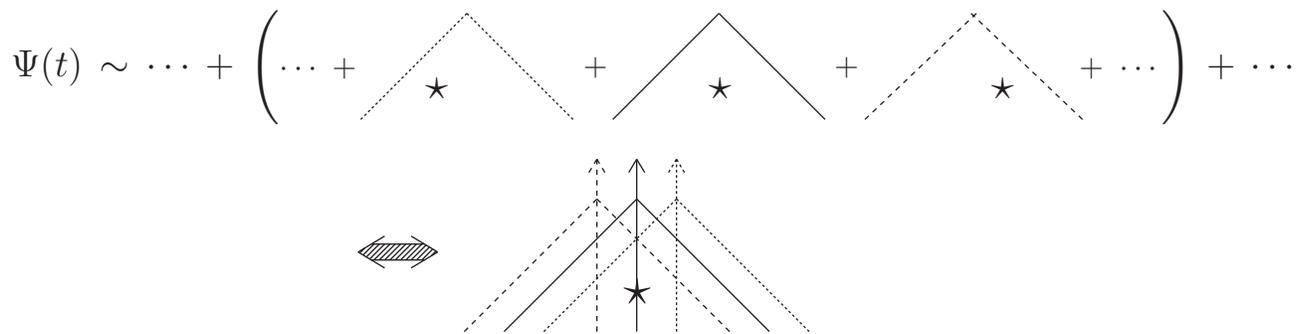

Let us now study the connection between the two pictures in more detail. 
(The following analysis does not require Eq.~(\ref{eq:frequency}), 
which was presented simply to help developing intuition.)  Remember 
that the multiverse state $\left| \Psi(t) \right>$ contains all 
possible worlds that can consistently arise in the single observer's 
viewpoint.  For example, in a spatially homogeneous region, an event 
that can occur may occur anywhere in space, so the multiverse state 
contains a superposition of terms in which the event occurs in all 
different locations, as schematically illustrated in the upper line 
of Fig.~\ref{fig:corresp-1}.
Now, in the global spacetime picture, these terms can be interpreted 
to correspond to the {\it same} spacetime region as viewed from 
{\it different observers (geodesics)}, as illustrated in the lower 
line of the figure.

In general, space is not homogeneous, i.e., the values of the 
coefficients are not identical for terms that have the event occurring 
in different locations relative to some other object (since the 
existence of the object breaks the symmetry).  In the global spacetime 
picture, this means that the ``density'' of observers (geodesics) are 
not uniform in such regions.  Note that the picture that spacetime is 
scanned/penetrated by a large number of geodesics is precisely what 
we had in Section~\ref{sec:framework}.  In fact, {\it (semi-classical) 
global spacetime can always be ``reconstructed'' from the single quantum 
observer picture} in the manner described here.

\section{(No) Problems associated with Geometric Cutoffs}
\label{sec:no-problem}

In a conventional treatment of the multiverse, one introduces a geometric 
cutoff to regularize infinite spacetime.  The (relative) probabilities 
are then defined by counting the number of events within the cutoff, 
which is eliminated at the end of the calculation either by sending it 
to infinity or through a certain averaging procedure~\cite{Guth:2000ka}. 
This treatment, however, introduces an arbitrariness associated with 
the choice of the cutoff, and also leads to peculiar conclusions such 
as the ``end'' of time~\cite{Bousso:2010yn}.  Here we show that our 
framework does not suffer from these difficulties.

\subsection{Youngness paradox, Boltzmann brains, and other problems}
\label{subsec:paradoxes}

Arguably, the earliest, and numerically most severe, problem 
encountered in the context of eternal inflation is the youngness 
paradox~\cite{Guth:2000ka-3}.  Suppose we take FRW time while the 
universe is still in the eternally inflating phase, and then extend 
this time coordinate to future keeping a synchronous gauge condition. 
This defines a global time coordinate over the entire (semi-classical) 
multiverse.  We can now define the relative probability of two classes 
of events by the ratio of the numbers of events occurring before 
a fixed time $t_c$, taking $t_c \rightarrow \infty$ at the end of 
the calculation~\cite{Linde:1993nz}.

This most naive---and seemingly innocuous---procedure, however, leads 
to violent contradictions with observations.  Since the spatial volume 
of the eternally inflating region is growing exponentially with time, 
the rate at which bubble universes form is also increasing exponentially, 
proportional to the volume.  This implies that, at any given time, there 
are enormously more younger universes than older ones.  For instance, 
the number of universes with $T_{\rm CMB} \simeq 3~{\rm K}$ is a factor 
of $\sim 10^{10^{59}}$ larger than that with $T_{\rm CMB} = T_0 \simeq 
2.725~{\rm K}$.  It is hard to imagine that such an immense bias 
towards younger universes is compensated by any possible anthropic 
weight factor.

Our framework does not suffer from this problem.  This is because the 
increase of spatial volume is not rewarded in calculating probabilities. 
(This is also true in classes of geometric cutoff measures proposed 
in Refs.~\cite{DeSimone:2008bq,Bousso:2006ev}.)  To see this, let us 
consider the quantum probabilities defined in Section~\ref{sec:quantum}. 
In this picture, the probabilities are calculated from the viewpoint 
of a single observer, so it is rather evident that the volume increase 
is not rewarded.  For instance, there is obviously no gain in probabilities 
by staying longer in an eternally inflating phase.  The same can also 
be seen in the semi-classical picture---spacetime expansion increases 
spatial volume, but only at the cost of diluting observers (geodesics). 
Since there is no reward for volume increase, any problems associated 
with volume weighting (such as $Q$-catastrophe~\cite{Feldstein:2005bm}) 
do not arise in our framework.

The present framework is also free from ambiguities associated with 
the question ``what is an observer?'';%
\footnote{So far, we have been using the word ``observer'' to 
 simply mean a geodesic in the multiverse.  In contrast, the 
 observer here means a ``real observer'' (called an experimenter in 
 Section~\ref{subsec:BH}), who performs experiments/observations 
 and collects data.  I hope there is no confusion.}
e.g., does an observer mean a civilization, an individual, or some sort 
of consciousness?  (Even what about a dog seeing a tree?)  In fact, 
the answer to it should be specified {\it already in a question we ask}, 
i.e.\ in prior condition $A$.  For example, we can ask the probability 
of an individual (defined carefully) to see a certain event, a 
civilization (again, carefully defined) to obtain a certain result 
from a certain experiment, and so on---{\it as long as the question 
is well defined, we will obtain a well defined answer}.  We also 
mention that our framework does not suffer from the ambiguity for 
defining probabilities which arises if an observer can condition only 
a part of a wavefunction, as occurs in a global (quantum) description 
of the universe~\cite{Page:2009qe}.  Our framework avoids this problem 
because the state is defined as viewed from a single observer, so 
that the conditioning (such as $A$, $B$, and $C$) is on the {\it 
entire} multiverse state, through Eqs.~(\ref{eq:probability-AB}) and 
(\ref{eq:probability-AC}).  In fact, these equations are nothing but 
the standard Born rule, which plays an essential role in our framework.

The Boltzmann brain problem is the statement that if (a part of) 
the universe stays in a (approximately) de~Sitter phase too long, 
then ordinary observations made by ordinary observers are completely 
overwhelmed by disordered ``observations'' made by vacuum thermal 
fluctuations~\cite{Page:2006dt}.%
\footnote{If the multiverse is in a pure state, then a particular 
 de~Sitter vacuum does not have thermal fluctuations because it 
 corresponds to a single quantum state.  We can, however, consider 
 an ensemble of de~Sitter vacua throughout the multiverse which 
 macroscopically look the same.  We then find that the standard argument 
 for Boltzmann brains still applies to this ensemble; namely, the 
 problem may exist even in the case where the multiverse is in 
 a pure state.}
For example, if the naive synchronous time cutoff is adopted (as 
in the first paragraph of this subsection), then this leads to the 
peculiar conclusion that our universe must decay within $\approx 20$ 
billion years.  In fact, this conclusion does not even require the 
existence of other universes.  The problem is generically worse if 
we take into account other universes.

Our framework avoids this problem under rather mild assumptions about 
the vacuum structure of the landscape.  The situation is similar 
to those in classes of geometric cutoffs that do not reward volume 
increase~\cite{Bousso:2008hz}.  Let us first formulate the problem 
precisely in the current framework.  Suppose that some ``observation'' 
is made.  This requires us to take prior condition $A$ to be classes 
of information processing occurring in spacetime in some physical 
form (which includes firings of neural signals in a human brain). 
We then take condition $B$ to be that these ``observations'' find 
some ordered pattern (e.g.\ the sight ``seen'' by this process is 
an ordered world obeying regular rules).  The Boltzmann brain problem 
arises if we obtain $P(B|A) \lll 1$.  Since the world we (or I/you) 
see is ordered, such a result would contradict observations.

The probability of a random thermal fluctuation to compose an ``observation'' 
is suppressed by a huge Boltzmann factor associated with a (macroscopic) 
configuration corresponding to the observation---estimates for this factor 
span the range $\approx \exp(10^{O(10\mbox{--}100)})$~\cite{Bousso:2008hz}. 
Since our framework does not reward volume increase, the problem is 
avoided if de~Sitter vacua decay before Boltzmann brains start dominating. 
This gives only weak constraints on lifetimes of de~Sitter vacua: 
$\tau_{\rm dS} \simlt \exp(10^{O(10\mbox{--}100)})$, where the exponent 
depends on a vacuum.%
\footnote{Units of time do not matter here, since the right-hand side 
 is so large.  Also, if a vacuum does not support Boltzmann brains, then 
 that vacuum need not obey this bound.}
Indeed, the fact that de~Sitter vacua decay within such timescales is 
consistent with what we know about string theory~\cite{Freivogel:2008wm}. 
The resulting lifetimes are also sufficiently short to avoid Poincar\'{e} 
recurrences~\cite{Dyson:2002pf}.

\subsection{The ``end'' of time}
\label{subsec:end-time}

We now discuss the issue of the ``end'' of time.  It has recently 
been pointed out that {\it any} simple geometric cutoffs lead to the 
peculiar conclusion that time should ``end''~\cite{Bousso:2010yn}. 
Suppose we consider two classes of events:\ an experimenter sees (i) 
1~o'clock and (ii) 2~o'clock on his/her watch.  Since there are always 
experimenters for whom the observation of 1~o'clock occurs before the 
cutoff while that of 2~o'clock after, the numbers of observations of 
1 and 2~o'clocks, $N_{1,2}$, satisfy $N_1 > N_2$.  An important point 
is that, since events in eternally inflating spacetime are dominated 
by a late-time attractor regime, the effect of the cutoff does {\it not} 
decouple when it is sent to infinity, i.e., $P_{\rm end} \equiv 1-N_2/N_1 
\nrightarrow 0$.  This implies that there is a nonzero probability that 
an experimenter who saw 1~o'clock never sees 2~o'clock, {\it even if} 
the watch or experimenter does not break or die in between---for some 
experimenters, time simply ``ends.''

Our framework does not lead to this conclusion.  Consider the probability 
that an experimenter who saw 1~o'clock will see 2~o'clock.  This is 
given by Eq.~(\ref{eq:probability-AB}) (or Eq.~(\ref{eq:final-1})), 
taking $A$ to be an experimenter seeing 1~o'clock, while $B$ the same 
experimenter seeing 2~o'clock.  It is then obvious that, since the time 
evolution of the system follows standard physical laws, we should find
\begin{equation}
  P(\mbox{2~o'clock}|\mbox{1~o'clock}) = 1,
\label{eq:no-end-time}
\end{equation}
as long as the watch or experimenter does not break/die, which we are 
assuming here.  Namely, time does {\it not} end in a way discussed in 
Ref.~\cite{Bousso:2010yn}.  (Of course, it can still end in other ways, 
e.g.\ at spacetime singularities.)  The ultimate reason behind this 
can be traced in a sentence in Ref.~\cite{Bousso:2010yn}:\ ``In eternal 
inflation, however, one first picks a causal patch; then one looks for 
observers in it.''  Our framework does not follow this approach.  We 
instead pick an observer first, and then construct the relevant spacetime 
regions associated with it.

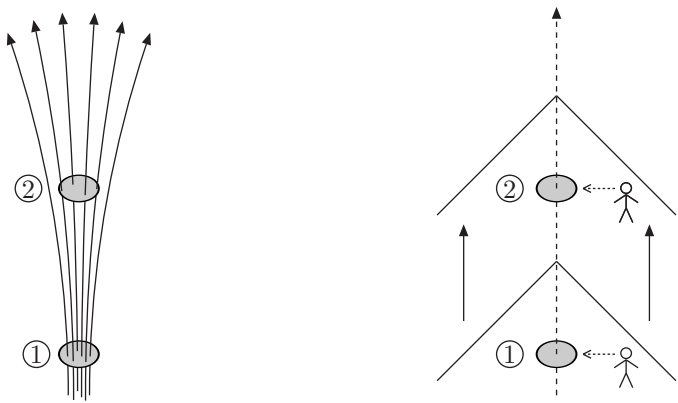
\begin{figure}[t]
\begin{center}
\begin{picture}(300,175)(90,10)
%
%
  \CArc(-274,22)(420,0,1.6)
  \CArc(-522,22)(670,-0.15,0.9)
  \CArc(-1650.5,22)(1800,0.05,0.4)
  \CArc(2251,22)(2100,179.7,180.02)
  \CArc(872.5,22)(720,179.1,180.16)
  \CArc(574,22)(420,178.4,180)
  \GOval(150,37.5)(5,7.5)(0){0.8}
  \Text(137,37.5)[r]{\footnotesize 1} \CArc(134,37.5)(5,0,360)
  \LongArrowArc(-274,22)(420,2,19)
  \CArc(-522,22)(670,1.1,6.5)
  \CArc(-1650.5,22)(1800,0.53,2.4)
  \CArc(2251,22)(2100,178.0,179.6)
  \CArc(872.5,22)(720,174.0,179)
  \LongArrowArcn(574,22)(420,178,161)
  \GOval(150,100)(5,7.5)(0){0.8}
  \Text(134,100)[r]{\footnotesize 2} \CArc(131,100)(5,0,360)
  \LongArrowArc(-522,22)(670,6.6,12.2)
  \LongArrowArc(-1650.5,22)(1800,2.53,4.6)
  \LongArrowArcn(2251,22)(2100,177.94,176.06)
  \LongArrowArcn(872.5,22)(720,173.8,168.7)
%
%
  \DashLine(330,22)(330,32.5){2}
  \GOval(330,37.5)(5,7.5)(0){0.8}
  \Text(316,37.5)[r]{\footnotesize 1} \CArc(312.4,37.5)(5,0,360)
  \Text(350,37.5)[l]{$\manup$} \DashLine(352,37.5)(340,37.5){1}
  \Line(340,37.5)(343,39) \Line(340,37.5)(343,36)
  \DashLine(330,37.5)(330,95){2}
  \GOval(330,100)(5,7.5)(0){0.8}
  \Text(316,100)[r]{\footnotesize 2} \CArc(312.4,100)(5,0,360)
  \Text(350,100)[l]{$\manup$}
  \Text(350,100)[l]{$\manup$} \DashLine(352,100)(340,100){1}
  \Line(340,100)(343,101.5) \Line(340,100)(343,98.5)
  \DashLine(330,100)(330,167){2}
  \LongArrow(330,166)(330,167)
  \Line(330,72.5)(285,27.5) \Line(330,72.5)(375,27.5)
  \Line(330,135)(285,90) \Line(330,135)(375,90)
  \LongArrow(295,50)(295,85) \LongArrow(365,50)(365,85)
\end{picture}
\caption{Some of the ``observers'' (geodesics) passing the watch at 
 1~o'clock do {\it not} pass it at 2~o'clock (left).  On the other hand, 
 an experimenter who saw the watch at 1~o'clock {\it will} see it at 
 2~o'clock (right).  Here, the hats in the right panel represent past 
 light cones.}
\label{fig:watch}
\end{center}
\end{figure}

Let us now ask:\ what is {\it the} relative probability of the watch 
showing 1~o'clock and 2~o'clock?  A natural way of defining the question 
is to take conditions $A$ and $B$ as:\ (A) A watch is located at (or 
has an overlap with) the tip of the past light cone of an observer 
(geodesic), and its long hand is on the twelve; (B) The short hand of 
the watch is on the (i) one or (ii) two.  Then it is easy to see that, 
in an expanding universe, we obtain
\begin{equation}
  \frac{P(\mbox{``2~o'clock''})}{P(\mbox{``1~o'clock''})} < 1,
\label{eq:Prel-1-2}
\end{equation}
as depicted in the left panel of Fig.~\ref{fig:watch}.
(This is related to the fact that our procedure can be viewed as a sort 
of time cutoff in expanding universes; see Appendix~\ref{app:fuzzy}.) 
How can this be consistent with Eq.~(\ref{eq:no-end-time})?%
\footnote{I thank Raphael Bousso for bringing my attention to this issue.}

The answer is that the probabilities $P(\mbox{``1~o'clock''})$ 
in Eq.~(\ref{eq:Prel-1-2}) and $P(\mbox{1~o'clock})$ in 
Eq.~(\ref{eq:no-end-time}) (and similarly for 2~o'clock) are 
different quantities.  Specifically, Eq.~(\ref{eq:no-end-time}) 
should better be written more explicitly as
\begin{equation}
  P(\mbox{The same experimenter sees 2~o'clock.}|\mbox{An experimenter 
    sees 1~o'clock.}) = 1,
\label{eq:no-end-time-2}
\end{equation}
which can certainly be consistent with Eq.~(\ref{eq:Prel-1-2}); see 
the right panel of Fig.~\ref{fig:watch}.%
\footnote{Strictly speaking, there can be a (small) possibility that the 
 experimenter or watch leaves the causal horizon of the observer (geodesic), 
 making the probability of Eq.~(\ref{eq:no-end-time-2}) smaller than $1$. 
 This is, however, a usual physical process already existing in general 
 relativity, and has nothing to do with the end of time.}
The probability is ``lost'' in Eq.~(\ref{eq:Prel-1-2}) not because time 
ended in between, but because some of the ``observers'' (geodesics) who 
``saw'' 1~o'clock simply miss the watch at 2~o'clock because of spacetime 
expansion (Fig.~\ref{fig:watch}, left).

\section{Discussions}
\label{sec:discuss}

A predictivity crisis in eternal inflation has long been a major 
problem in cosmology.  An important aspect of the problem comes 
from its robustness.  Once we have a sufficiently meta-stable de~Sitter 
vacuum, it produces an infinite number of events in an infinite 
number of lower energy vacua, making predictions impossible without 
an appropriate regularization.  The high sensitivity of predictions 
on the regularization prescription has plagued many physicists over 
the last two decades.  In fact, the problem has, a priori, nothing 
to do with the landscape or quantum gravity---it already exists in 
classical general relativity.

Like black hole physics, this predictivity problem has told us a lot 
about fundamental aspects of gravity and spacetime.  The youngness 
paradox says that the measure based on a synchronous time cutoff in 
global spacetime does not work, despite the fact that it apparently 
seems most natural and innocuous.  The Boltzmann brain problem implies 
that de~Sitter vacua (at least ones that can support complexities 
required for ``observations'') must be unstable, with lifetimes 
much shorter than their Poincar\'{e} recurrence times.  Measures 
avoiding these problems, especially ones based on local 
pictures~\cite{Bousso:2006ev,Bousso:2006ge}, were put forward, 
but they still suffer from a peculiar conclusion that time should 
``end'' even if there is no corresponding singularity in general 
relativity.

In this paper, we presented a framework which gives well-defined 
predictions and yet does not suffer from these difficulties.  The 
framework is formulated consistently within a fully quantum mechanical 
treatment of the multiverse.  (The semi-classical picture can be derived 
by ``integrating out'' physics associated with quantum gravity.)  We 
argued that the {\it entire} multiverse is described {\it purely} from 
the viewpoint of a single ``observer.''  A complete description of the 
physics is obtained in spacetime regions that the observer can causally 
access and are bounded by his/her apparent horizons.  In conventional 
geometric cutoff measures, one first picks a spacetime region and then 
looks for observers in it.  We do not follow this approach.  We instead 
pick an observer first, and then construct the relevant spacetime 
regions associated with it.

The resulting picture is quite satisfactory.  As viewed from a single 
observer, probabilities keep being ``diluted'' because of continuous 
branching of the state into different semi-classical possibilities, 
which is caused by the fact that the evolution of the multiverse state 
is not along an axis in Hilbert space determined by operators local 
in spacetime.  This makes it possible to obtain well-defined predictions 
according to the standard Born rule.  We may say that {\it it is quantum 
mechanics that solves the measure problem in eternal inflation}.  Indeed, 
our framework allows for a completely unified treatment of quantum 
measurement processes and the multiverse.  We conclude that {\it the 
eternally inflating multiverse and many worlds in quantum mechanics 
are the same}.

The multiverse state $\left| \Psi(t) \right>$, or $\rho(t)$, is 
literally ``everything'' for making predictions.  For example, even 
our own existence appears in components of $\left| \Psi(t) \right>$ 
at some time(s) $t_j$.  Physical predictions can then be made by 
extracting correlations between ourselves (or our experimental 
apparatus) and the surroundings, using Eq.~(\ref{eq:probability-AB}) 
or (\ref{eq:probability-AC}).  There is no need to introduce anything 
beyond these basic elements---in particular, it is not necessary to 
introduce wavefunction collapse, environmental decoherence, or anything 
like those (although these concepts will still be useful when applied 
to understanding structures {\it inside} $\left| \Psi(t) \right>$). 
Indeed, there is no ``external observer'' that performs measurements 
on $\left| \Psi(t) \right>$, and there is no ``environment'' with which 
$\left| \Psi(t) \right>$ interacts.  The only task left for us is 
to find the ``Hamiltonian'' for quantum gravity, $\hat{H}$, and the 
boundary conditions determining the multiverse state, which, hopefully, 
a complete understanding of string theory would give us.

The picture described above provides a satisfactory answer to the 
question raised at the beginning of this paper:\ what is the meaning 
of the phrase ``In an eternally inflating universe, anything that 
can happen will happen; in fact, it will happen an infinite number 
of times''?  We can now say that anything that can happen will happen 
with a nonzero quantum mechanical probability.  The same event can 
occur many times during the cosmic history, although the probability 
for that to happen is small.  The volume increase in the semi-classical 
picture of eternal inflation ensures that all these events can 
look identical, but it does not mean that probabilities of multiple 
occurrences are larger.  In fact, the probabilities are {\it not} 
obtained by simply {\it counting} the number of events in (regulated) 
semi-classical spacetime, which already assign (implicitly) equal 
probability weights for all these events.  Infinities appear only 
if one asks a question an infinite number of times, which obviously 
does not affect the probabilities.  Indeed, one can view that the 
probabilities are more fundamental, which exist regardless of whether 
one asks a question or not.

Our framework has a number of implications beyond what have been 
discussed so far.  Below, we consider some of them.

\subsection{The ultimate fate of the multiverse}
\label{subsec:fate}

The evolution of the multiverse state $\left| \Psi(t) \right>$ is supposed 
to be unitary.  On the other hand, some of the components in $\left| 
\Psi(t) \right>$ hit black hole or big crunch singularities at finite 
proper times.  What happens to these components?

\begin{figure}[t]
\begin{center}
\begin{picture}(120,160)(0,-10)
  \Line(0,0)(0,120) \Text(-6,60)[r]{$r=0$}
  \Line(0,120)(120,120) \Photon(0,120)(120,120){2}{14} \Text(50,129)[b]{$r=0$}
  \DashLine(0,0)(120,120){3}
  \Text(16,68)[t]{\scriptsize not} \Text(16,61)[t]{\scriptsize counted}
  \DashLine(10,70)(10,115){2} \LongArrow(10,115)(10,116)
  \DashLine(20,70)(25,115){2} \LongArrow(25,115)(25.111,116)
  \Text(45,68)[t]{\scriptsize counted}
  \Line(40,70)(65,115) \LongArrow(65,115)(65.556,116)
  \Line(50,70)(85,115) \LongArrow(85,115)(85.778,116)
  \Line(40,120)(80,80) \Line(40.5,120)(80,80.5) \Line(39.5,120)(79.5,80)
  \Vertex(80,80){3} \Text(84,76)[lt]{\small horizon}
\end{picture}
\caption{A Penrose diagram representing inside a black hole.  The light 
 sheet (thick solid line) associated with a horizon (dot) counts only 
 degrees of freedom that pass the sheet before hitting the singularity.}
\label{fig:singularity}
\end{center}
\end{figure}
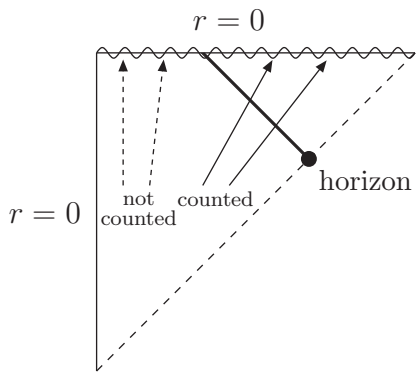

We conjecture that these components simply ``disappear'' from $\left| 
\Psi(t) \right>$.  This conjecture is motivated by the holographic 
principle, especially the covariant entropy bound~\cite{Bousso:1999xy}. 
Let us consider a black hole and count the entropy inside the horizon. 
The covariant entropy bound implies that degrees of freedom are counted 
only if they pass the light sheet before hitting the singularity; 
see Fig.~\ref{fig:singularity}.
One may interpret this such that the light sheet is extended all the way 
to the center, but the degrees of freedom that hit the singularity---where 
the curvature is of order the Planck scale---``disappear'' from the 
theory.  In fact, this interpretation is consistent with what is suggested 
by black hole physics:\ there are essentially no degrees of freedom 
behind the stretched horizon where the local temperature is above 
the Planck scale.  In our context, this implies that a component that 
hits the singularity should be eliminated from $\left| \Psi(t) \right>$ 
(i.e.\ the coefficient of the term should be set to zero) at the time 
of the hitting.  The information about the disappearing component remains 
in other components in $\left| \Psi(t) \right>$, e.g.\ in the form of 
Hawking radiation, so that no contradiction with reversibility of quantum 
evolution arises.  Note that this argument would fail if there were exact 
global symmetries that prohibit degrees of freedom to disappear; but 
string theory suggests that any global symmetries are inevitably broken 
by Planck scale physics~\cite{Banks:2010zn}.%
\footnote{Unbroken gauge symmetries do not cause a problem, since at 
 any given time, components of $\left| \Psi(t) \right>$ have vanishing 
 gauge charges because of the charge neutrality.}
The same argument as the one here also applies to big crunch 
singularities.

The conjecture given above has an important consequence for the ultimate 
fate of the multiverse.  Starting from any initial state, components 
that hit singularities disappear from $\left| \Psi(t) \right>$.  Since 
observers (geodesics) inside anti de~Sitter bubbles hit singularities 
within finite times, they do not remain in $\left| \Psi(t) \right>$ 
far in the future.%
\footnote{There is a (small) possibility that an anti de~Sitter 
 vacuum transitions into a de~Sitter vacuum due to a bubble 
 collision~\cite{Johnson:2010bn}.  This, however, does not affect 
 our conclusion.}
In addition, any observers in de~Sitter bubbles or non-supersymmetric 
Minkowski bubbles eventually experience decays into lower energy vacua,%
\footnote{With a (presumably huge) number of anti de~Sitter vacua, 
 non-supersymmetric Minkowski vacua will not be absolutely stable, 
 even taking into account gravitational effects~\cite{Coleman:1980aw}. 
 It is, however, a logical possibility that there are stable 
 such vacua, in which case they should be added to the right-hand 
 side of Eq.~(\ref{eq:SUSY}).  Such an addition would not affect 
 discussions below.}
so they either disappear into singularities or end up in supersymmetric 
Minkowski vacua, which are stable due to the positive energy 
theorem~\cite{Weinberg:1982id}.  Therefore, in the infinite future, 
the multiverse state becomes
\begin{equation}
  \left| \Psi(t) \right>
  \quad \stackrel{t \rightarrow \infty}{\longrightarrow} \quad 
    \sum_i \left| \mbox{Supersymmetric Minkowski world $i$} \right>,
\label{eq:SUSY}  
\end{equation}
where the sum runs over states in vacua with varying low energy 
physics, e.g., matter content, spatial dimensions, and the amount 
of supersymmetries.  This result has an important implication on 
a possible theory describing the entire history of the multiverse, 
as discussed in Section~\ref{subsec:transient}.

One interesting question is which vacua can exist in the sum in the 
right-hand side of Eq.~(\ref{eq:SUSY}).  In the string landscape, 
not all possible supersymmetric theories may be realized with vanishing 
cosmological constant.  For example, moduli might not all be stabilized 
supersymmetrically with vanishing superpotential, and the unfixed 
moduli might be pushed away by nonperturbative effects, excluding 
four dimensional ${\cal N} = 1$ worlds from the final state.  (The 
existence of stable such vacua, however, is suggested in string 
theory~\cite{Micu:2007rd}.)  In any case, it is possible that the 
sum in Eq.~(\ref{eq:SUSY}) consists of only states in certain limited 
vacua, although the states themselves may contain complex objects, 
such as a fractal of stable black holes filling fixed portions of 
the sky~\cite{Freivogel:2007fx}.  A detailed study of this issue 
is warranted, especially in the context of full string theory.

\subsection{The multiverse as a transient phenomenon}
\label{subsec:transient}

What is the ``beginning'' of the multiverse?  From the observed second 
law of thermodynamics (in our ``vicinity'' in the multiverse), we 
expect that the initial state of the multiverse is a (extremely) 
low entropy state.  (In fact, this is almost the only way to 
understand the arrow of time from the statistical point of view.) 
Let $\left| \alpha_{\rm beginning} \right>$ be such a state.  (We 
treat it as a pure state, for simplicity, but an extension to the 
mixed state case is straightforward.)  What do we know about the 
evolution of the multiverse afterwards?

Consider a complete set of states $\left| \alpha_a \right>$ defined 
at infinite Minkowski future, namely all possible past light cones 
of future time-like infinity of Minkowski spacetime, available in the 
landscape.  In view of the result in the previous subsection, the 
evolution of the multiverse is given by
\begin{equation}
  \left| \alpha_{\rm beginning} \right>
\quad\rightarrow\quad
  \left| \Psi(t = +\infty) \right> = \sum_a c_a \left| \alpha_a \right>.
\label{eq:Psi-infty}
\end{equation}
Here, the coefficients $c_a$ are given by $c_a = \left< \alpha_a 
\right| e^{-i \hat{H} t} \left| \alpha_{\rm beginning} \right>$ with 
$t \rightarrow \infty$, where $\hat{H}$ is the ``time evolution operator'' 
of full quantum gravity in our parametrization of spacetime.

An interesting feature of Eq.~(\ref{eq:Psi-infty}) is that the final 
state is given at infinite Minkowski future, where the basis of local 
operators is particularly simple, and asymptotically free particles are 
well defined.  (In fact, it is these properties that allow us to take 
the limit $t \rightarrow +\infty$ without difficulty.)  This might be 
helpful when we try to calculate $c_a$ in quantum gravity, e.g., by 
summing up all possible ``string world sheets'' for a fixed $\left| 
\alpha_a \right>$.  Such a calculation, however, is far beyond the current 
technology; it requires, at least, incorporation of all the (including 
nonperturbative) effects in time-dependent (not necessarily asymptotically 
Minkowski or anti de~Sitter) geometries.  Therefore, we focus here 
on what we can say about Eq.~(\ref{eq:Psi-infty}) without having 
explicit information from quantum gravity.

Is there anything we can say about $\left| \Psi(t = +\infty) \right>$ 
without input from quantum gravity?  We expect it to have a fractal 
structure when expanded in the basis $\left| \alpha_a \right>$, 
determined by local operators.  We first note that unitarity of 
the evolution implies that there are nonzero probabilities of 
forming $\left| \alpha_{\rm beginning} \right>$ (whatever it is) 
from other states.  For example, if $\left| \alpha_{\rm beginning} 
\right>$ is a vacuum with a Planckian energy density, it will be 
formed in lower energy de~Sitter vacua through upward transitions. 
This implies that $\left| \alpha_{\rm beginning} \right>$ will appear 
{\it as a component} in $\left| \Psi(t) \right>$ at some (late) time 
$t_I$, which will start the entire multiverse as a branch of $\left| 
\Psi(t) \right>$ for $t > t_I$.  In fact, this process will be 
repeated an infinite number of times, making $\left| \Psi(t = +\infty) 
\right>$ fractal; more precisely, $c_a(t_\Lambda)$ in $\left| 
\Psi(t_\Lambda) \right>$ have a fractal structure as $t_\Lambda 
\rightarrow \infty$.  Note that this does not affect the 
well-definedness of the probabilities, since the occurrence 
of $\left| \alpha_{\rm beginning} \right>$ in $\left| \Psi(t) \right>$ 
is exponentially suppressed, because of the difference of the density 
of states between $\left| \alpha_{\rm beginning} \right>$ and vacua 
in which $\left| \alpha_{\rm beginning} \right>$ is generated.

The picture presented here indicates that our entire multiverse 
is a transient phenomenon while a low entropy state relaxes into 
a supersymmetric Minkowski state that has a fractal structure. 
A natural question is:\ what is the origin of $\left| \alpha_{\rm 
beginning} \right>$?  As discussed in Section~\ref{subsec:initial}, 
if $\left| \alpha_{\rm beginning} \right>$ is {\it the} real initial 
state, then the quantum observer principle is violated there since 
we cannot evolve the state further back.  Is there any way of avoiding 
this conclusion?  In Section~\ref{subsec:mega-multiverse} we will study 
this issue, and speculate that $\left| \alpha_{\rm beginning} \right>$ 
may not be the real beginning, and that the quantum observer principle 
may be respected throughout the entire history of spacetime.

\subsection{General covariance, the arrow of time, and a holographic 
 quantum multiverse}
\label{subsec:holo}

A general covariant theory of gravity based on the global spacetime 
picture has huge redundancies in its description of physics.  There 
are (at least) three kinds of redundancies:
\begin{itemize}
\item
{\bf General covariance} --- The theory is formulated in such a way 
that the form of physical laws is invariant under arbitrary coordinate 
transformations.  This introduces large redundancies.  While quantities 
appearing in the theory may depend on (arbitrarily chosen) coordinates, 
only the ones that are invariant under coordinate transformations are 
physically observable.
\item
{\bf Global spacetime} --- The theory describes (global) spacetime in 
such a way that some portions of it are redundant.  This occurs when 
there are spacetime regions that cannot have causal contact later.  A 
classic example of this is given by a black hole~\cite{Susskind:1993if}:\ 
having both the interior of the horizon {\it and} Hawking radiation 
within a single description is double counting.
\item
{\bf Local spacetime} --- While a naive picture of spacetime suggests 
that an $O(1)$ amount of information can be stored per each Planck 
size region, the actual number of physical degrees of freedom is 
much smaller~\cite{'tHooft:1993gx}.  In fact, the maximal number of 
degrees of freedom is bounded by the area of the ``holographic screen'' 
in Planck units~\cite{Bousso:1999xy}.
\end{itemize}
The first of these three already appears at the level of classical 
general relativity, while the last two at the quantum (semi-classical) 
level.  It is important to treat these redundancies appropriately when 
we apply quantum mechanics to a system with gravity.

Our framework explicitly addresses the first two.  Recall that we 
describe a system from the viewpoint of a single observer, parametrizing 
spacetime {\it using (past) light cones}.  Since causal relations 
between events are invariant under coordinate transformations, this 
extracts invariant information about the system.  Of course, our time 
parametrization (in terms of the proper time along the observer) should 
also be a redundant gauge choice.  Indeed, there is no concept of absolute 
time in quantum gravity---``time evolution'' we perceive is simply 
correlations between physical quantities~\cite{DeWitt:1967yk}.  Our 
framework incorporates this idea by formulating physical questions in 
the form of Eqs.~(\ref{eq:probability-AB},~\ref{eq:probability-AC}) 
(or Eqs.~(\ref{eq:final-1},~\ref{eq:final-2})), in which time $t$ is 
nothing but an auxiliary parameter relating different physical events 
or configurations.  For example, when we describe a location of a 
ball as a function of time, ${\bf x}(\tau)$, what we really mean is 
the value of ${\bf x}$ that provides a nonzero support of
\begin{equation}
    P(\,\mbox{The ball is at ${\bf x}$.} \,|\,
    \mbox{The hands of a clock show $\tau$ ``in our universe''.})
\label{eq:ball-move}
\end{equation}
for each $\tau$, where the phrase ``in our universe'' represents the 
conditions needed to specify ``a clock'' in the multiverse, e.g.\ 
the model of a clock, the configuration of other neighboring objects 
(including ourselves), the fact that the entire system is located on 
a planet called the earth, which is in a universe whose low energy 
effective theory is given by an $SU(3)_C \times SU(2)_L \times U(1)_Y$ 
gauge theory, etc.\ etc.%
\footnote{Here, we have assumed, for simplicity, that $P({\bf x}|\tau)$ 
 is nonzero only at some value of ${\bf x}$ for a fixed $\tau$; 
 otherwise, ${\bf x}(\tau)$ should be given by a certain average, 
 e.g., $\int\! {\bf x}\, P({\bf x}|\tau)\, d{\bf x}/\int\! 
 P({\bf x}|\tau)\, d{\bf x}$.}
The rest of general covariance can also be satisfied straightforwardly 
by making the ``time evolution'' operator $\hat{H}$, as well as 
projection operators ${\cal O}_A$, invariant under ``spatial'' 
coordinatizations.  The redundancy of global spacetime (the second 
item in the above list) does not exist either, since we limit our 
description to the regions that can be physically observed.

The resulting description is, as we discussed throughout this paper, 
built on the Hilbert space for bulk and horizon degrees of freedom, 
Eq.~(\ref{eq:H-decomp}), as well as the ``time evolution'' operator 
$\hat{H}$ encoding dynamics of full quantum gravity.  What form does 
$\hat{H}$ take?  We have not discussed it explicitly, but it is likely 
to be rather complicated.  Indeed, even in standard QED, the gauge-fixed 
Hamiltonian (in Coulomb gauge) contains an apparent, Lorentz-violating 
instantaneous force, whose effects are canceled only after performing 
full quantum calculations~\cite{Weinberg:QFT-1}.  The situation in 
a gravitational theory is expected to be worse, especially because 
the holographic principle implies that the number of degrees of freedom 
in the bulk is (much) smaller than that indicated by local quantum field 
theory (as described as the third class of redundancies in the list).

It is possible, however, that a description of the bulk exists in 
which all the redundancies are fixed, and that such a ``holographic 
description'' has a simple(r) form of ``time evolution'' operator. 
In this respect, it is encouraging that such simple holographic 
descriptions do seem to exist (at least) in certain limited 
cases~\cite{Maldacena:2010un,Freivogel:2006xu}.  Since the holographic 
principle indicates the existence of a holographic description at the 
horizon $\partial{\cal M}$ for each semi-classical spacetime ${\cal M}$, 
the Hilbert space for the holographic description of the multiverse 
would take the form
\begin{equation}
  {\cal H} = \oplus_{\cal M} \bigl( \tilde{\cal H}_{{\cal M}, {\rm bulk}} 
    \otimes {\cal H}_{{\cal M}, {\rm horizon}} \bigr).
\label{eq:Hilbert}
\end{equation}
Here, $\tilde{\cal H}_{{\cal M},{\rm bulk}}$ represents the holographic 
bulk Hilbert space, whose size is manifestly
\begin{equation}
  \ln\Bigl( {\rm dim}\,\tilde{\cal H}_{{\cal M},{\rm bulk}} \Bigr) 
  = \frac{{\cal A}_{\partial{\cal M}}}{4 l_P^2},
\label{eq:Hilbert-bulk-M}
\end{equation}
where ${\cal A}_{\partial{\cal M}}$ is the area of the horizon 
$\partial{\cal M}$.  Since the dimension of ${\cal H}_{{\cal M},{\rm 
horizon}}$ is expected to be similar to that of $\tilde{\cal H}_{{\cal 
M},{\rm bulk}}$, it is possible that these two subspaces are actually 
the same, having a one-to-one correspondence between the elements (as 
discussed in Section~\ref{subsec:reduced-H}).  If this is the case, 
then the complete Hilbert space is actually the square root of 
Eq.~(\ref{eq:Hilbert}).

The evolution of the multiverse state starts from some initial ``point'' 
in ${\cal H}$, $\left| \alpha_{\rm beginning} \right>$, and ends up 
with a point corresponding to a supersymmetric Minkowski world.%
\footnote{Our picture shares some features in common with the 
 proposal of Refs.~\cite{Freivogel:2006xu,Susskind:2007pv}, which 
 claims that the entire multiverse is described by a two-dimensional 
 Euclidean Liouville theory on the boundary of a stable Minkowski 
 bubble.  A relation between the two pictures, if any, is not clear.}
One may consider a course-grained entropy of this system defined 
by Eq.~(\ref{eq:Hilbert-bulk-M}), i.e.\ $S \equiv \ln( {\rm dim}\, 
\tilde{\cal H}_{{\cal M},{\rm bulk}} )$.  The evolution of the 
multiverse is then
\begin{equation}
  S_{\rm beginning} \sim 0 \quad\rightarrow\quad
  S_{\rm Minkowski} = \infty,
\label{eq:S-evolve}
\end{equation}
where $S_{\rm beginning} \sim 0$ simply means that the entropy of 
$\left| \alpha_{\rm beginning} \right>$ is low, e.g.\ somewhere in 
the range between $0$ and a few orders of magnitude.  This evolution 
of $S$ is ultimately the origin of the ``global arrow of time'' in 
the multiverse, which also dictates our local arrow of time.

\subsection{The fractal ``mega-multiverse''}
\label{subsec:mega-multiverse}

What is the origin of $\left| \alpha_{\rm beginning} \right>$ at the 
beginning of our multiverse?  Does it arise somehow from outside our 
framework?  Or can we understand it consistently within the framework?

Recall that no matter what $\left| \alpha_{\rm beginning} \right>$ is, 
it also appears at late times as components in $\left| \Psi(t) \right>$, 
reproducing the entire multiverse from there.  Then, why can't we 
identify ``the initial state'' of $\left| \Psi(t) \right>$, from 
which we started, to arise as a component in a ``larger'' state $\left| 
\Phi(t) \right>$?  In some sense, this is unavoidable.  Since the same 
$\left| \alpha_{\rm beginning} \right>$ appears multiple (infinitely 
many) times during the evolution, we will not know which branch ``our 
multiverse'' corresponds to.  In this picture, the initial state of 
our multiverse arises as a statistical fluctuation in a larger structure 
$\left| \Phi(t) \right>$.  It implies that we only know the initial 
state statistically, i.e., our multiverse begins as a mixed state
\begin{equation}
  \rho_{\rm beginning} \simeq \sum_i \lambda_i 
    \left| \alpha_{{\rm beginning},i} \right> 
    \left< \alpha_{{\rm beginning},i} \right|,
\label{eq:rho-beginning}
\end{equation}
where $\lambda_i$ is the probability of producing $\left| 
\alpha_{{\rm beginning},i} \right>$ as a component in $\left| 
\Phi(t) \right>$.  Grouping together the states having the 
same semi-classical geometry $a$, Eq.~(\ref{eq:rho-beginning}) 
can be written as 
\begin{equation}
  \rho_{\rm beginning} \simeq \sum_a \lambda_a \rho_{{\rm beginning},a},
\qquad
  \rho_{{\rm beginning},a} \equiv \frac{1}{\sum_{i \in a} \lambda_i}
    \sum_{i \in a} \lambda_i \left| \alpha_{{\rm beginning},i} \right> 
    \left< \alpha_{{\rm beginning},i} \right|.
\label{eq:rho-beginning-2}
\end{equation}

What about the beginning of $\left| \Phi(t) \right>$?  Given a 
landscape potential, we can explore dynamics of a structure larger 
than our multiverse, defined as a particular branch evolving as 
Eq.~(\ref{eq:S-evolve}).  The relevant machinery was developed in 
Ref.~\cite{Garriga:2005av} in the context of geometric cutoff measures. 
Let us introduce an object that captures coarse-grained dynamics of 
the large(st) structure at the semi-classical level:
\begin{equation}
  \aleph(t) = \sum_a f_a(t)\, \rho_a,
\label{eq:rho-tilde}
\end{equation}
where $\rho_a$ is the density matrix representing the states in 
vacuum $a$.  The rate equation governing the evolution of $\aleph(t)$ 
is given by
\begin{equation}
  \frac{d f_a}{dt}(t) = M_{ab}\, f_b(t),
\label{eq:rate-eq}
\end{equation}
where $M_{ab}$ is the transition matrix between different vacua.

The solutions of Eq.~(\ref{eq:rate-eq}) can be written in terms 
of (generalized) eigenvectors $v_a^X$ of $M_{ab}$, which form a 
complete basis, i.e., $X$ runs from $1$ to the number of vacua $n$. 
The eigenvalues and eigenvectors of $M_{ab}$ have the following 
structure~\cite{Garriga:2005av}.  Each terminal vacuum is an eigenvector 
with eigenvalue zero, which we denote by $v_a^I$ ($I = 1,\cdots,n_T$), 
where $n_T$ is the number of terminal vacua.  All the other eigenvalues 
have negative real parts.  In particular, the eigenvalue that 
has the smallest real part in magnitude, $\alpha_D$, is pure real 
and has a nondegenerate eigenvector $v_a^D$ (called the dominant 
eigenvalue/eigenvector).  The rest of the eigenvalues $\alpha_L$ may 
or may not be degenerate, having (generalized) eigenvectors $v_a^L$, 
where $L = 1,\cdots,n_S$ with $n_S + n_T + 1 = n$.  The general 
solution to Eq.~(\ref{eq:rate-eq}) is then given by
\begin{equation}
  f_a(t) = \sum_{I=1}^{n_T} c_I v_a^I + c_D e^{-|\alpha_D|\, t} v_a^D 
    + \sum_{L=1}^{n_S} c_L(t) e^{-|{\rm Re}\,\alpha_L| t} v_a^L,
\label{eq:f_a}
\end{equation}
where $c_{I,D}$ are constants, and $c_L(t)$ are composed of polynomial 
and trigonometric functions.  

There are two scenarios one can imagine here:

\underline{\bf The multiverse with a beginning} --- There is some 
theory beyond our framework, e.g.\ creation from ``nothing'' or quantum 
gravity, that provides the initial condition for the {\it largest 
possible} structure.  In this case, we can simply call this structure 
the multiverse, and the situation is reduced to that discussed in 
Section~\ref{subsec:initial}.  The quantum observer principle is 
violated at the earliest moment, since we cannot evolve the state 
further back.  In general, the phenomenological predictions do depend 
on the initial condition.

\underline{\bf The stationary, fractal ``mega-multiverse''} --- Instead 
of admitting the existence of the ``beginning,'' we may require that 
the quantum observer principle is respected for the {\it whole history} 
of spacetime.  In this case, the beginning of our multiverse is a 
fluctuation of a larger structure, whose beginning is also a fluctuation 
of an even larger structure, and this series goes on forever.  This 
leads to the picture that our multiverse arises as a fluctuation in a 
huge, {\it stationary} ``mega-multiverse,'' which has a fractal structure. 
Here the ``stationary'' means that the same predictions are obtained 
by Eq.~(\ref{eq:probability-AB}) (or Eq.~(\ref{eq:probability-AC})) 
regardless of the limits of the $t$ integrals, as long as the interval 
is taken sufficiently long.  For this to be true, inflating regions 
of the mega-multiverse state must be in one of the eigenvectors 
of $M_{ab}$.  Assuming that we are interested only in ``transient 
phenomena'' in the multiverse, the initial condition {\it for our 
multiverse} can be taken to be
\begin{equation}
  \lambda_a \simprop \lambda_{ab} H_b^{-3} v_b^D 
    \quad\mbox{or}\quad \lambda_{ab} H_b^{-3} v_b^L,
\label{eq:lambda_a-2}
\end{equation}
where $\lambda_{ab}$ is the rate of the transition $b \rightarrow a$ 
per unit physical spacetime, $H_b$ the Hubble parameter in vacuum $b$, 
and $L$ can be any of the $n_S$ possibilities.  In this scenario, 
no ``real'' initial condition needs to be imposed---there is simply 
no beginning or end for the mega-multiverse.  On the other hand, we 
need to choose which of the stationary states the mega-multiverse is 
in.  This is, in some sense, a choice of ``theories'' (rather than 
``vacua'' or ``initial conditions'') because there is no physical 
process allowing transitions between different choices.

In either of the scenarios described above, our multiverse is one 
of the infinite series of multiverses created in a larger structure 
as statistical fluctuations.  The global arrow of time in 
Eq.~(\ref{eq:S-evolve})---a part of which is our arrow of time---is 
simply a manifestation of the fluctuations relaxing into the 
equilibrium state of a supersymmetric Minkowski fractal.

Can we explore the larger structure $\left| \Phi(t) \right>$ 
experimentally by evolving our current multiverse state back in time?%
\footnote{Note that, because of the deterministic nature of quantum 
 evolution, we do not need to regard that time flows from smaller 
 to larger $t$---we could equally view that time evolves the other 
 way.  Such a picture is obviously highly unintuitive---for example, 
 our brains ``evolve in time'' in such a way that we keep losing our 
 ``memories of the future''---but this is not a real problem.  The 
 real problem is that such a description is highly sensitive to small 
 perturbations; namely, the evolution of the system is unstable against 
 small errors in the initial data.  Here, we assume that we have a 
 perfect knowledge about the current state of the multiverse.}
Unfortunately, we cannot.  If we evolve the multiverse state to the 
past, we would at some point reach one of the ``initial states,'' 
$\left| \alpha_{{\rm beginning},i} \right>$.  To go back further, 
however, we need to know other components in $\left| \Phi(t) \right>$, 
which are ``outside'' our multiverse.  In the conventional language, 
our multiverse state $\left| \Psi(t) \right>$ is a state after 
``wavefunction collapse,'' i.e.\ a state obtained after throwing 
away components that are irrelevant for {\it future} measurements 
(namely future of $\left| \alpha_{{\rm beginning},i} \right>$). 
But to evolve the state further back, it is not enough to know the 
wavefunction ``already collapsed''; we need to know the state ``before 
the collapse.''  This may make one worry that the considerations here 
might just be ``meta-physics.''  However, the scenarios described here 
{\it do} affect our multiverse through their implications on the initial 
condition, e.g.\ through Eq.~(\ref{eq:lambda_a-2}), whose consequences 
can, in principle, be worked out and compared with the observations. 
At the very least, different choices of the scenario lead to different 
fractal patterns in the final Minkowski bubbles, which can be 
observed by civilizations (if any) living in these bubbles.  Since 
the coarse-grained entropy of Minkowski space is infinite, there 
is enough room in these bubbles to store all the information about 
the ``beginning.''

The scenarios presented here are speculative, but attractive.  In 
particular, the fractal mega-multiverse finally eliminates the necessity 
of imposing initial conditions from cosmology (though at the cost of 
introducing a choice of the state), and it seems to be an inevitable 
consequence of the ultimate extrapolation of the quantum observer 
principle.  Maybe, quantum mechanics already tells us the entire history 
of the multiverse, and even about an infinite series of multiverses, 
which are created in stationary, fractal spacetime.

\section*{Acknowledgments}

I am grateful for stimulating discussions with Raphael Bousso, Clifford 
Cheung, Ben Freivogel, Alan Guth, and Vladimir Rosenhaus.  This work 
was supported in part by the Director, Office of Science, Office of 
High Energy and Nuclear Physics, of the US Department of Energy under 
Contract DE-AC02-05CH11231, and in part by the National Science Foundation 
under grants PHY-0855653.

\appendix

\section{Interpretation as a ``Fuzzy'' Time Cutoff}
\label{app:fuzzy}

In expanding universes, our method in Section~\ref{subsec:classical} 
can be viewed as a sort of ``fuzzy'' time cutoff.  Suppose that prior 
conditions $A$ can be satisfied in a class of vacua $X$ if they are 
nucleated inside any of vacua $Y$.  We assume that the conditions are 
met in a fraction $r_X$ of the spatial volume at time $\tau_{X,{\rm obs}}$ 
after the nucleation, where $\tau_X$ is the Friedmann-Robertson-Walker 
time inside $X$ bubbles.  Now, the comoving volume of a bubble $X$ 
nucleated in $Y$ is
\begin{equation}
  V_{Y \rightarrow X} 
  = \frac{4\pi}{3} H_Y^{-3} e^{-3\, t_{\rm nuc}},
\label{eq:V_X}
\end{equation}
where $t_{\rm nuc}$ is the scale factor time at the bubble nucleation, 
which depends on $X$ and $Y$, and $H_Y$ is the Hubble expansion rate 
in vacuum $Y$.

Now, consider a set of geodesics each occupying comoving volume 
$V_\epsilon = \epsilon^3$.  On average, a past light cone satisfying 
$A$ intersects with one of the geodesics if $r_X V_{Y \rightarrow X} 
\simgt V_\epsilon$, i.e.
\begin{equation}
  t_{\rm nuc} \simlt \ln\frac{(\frac{4\pi}{3} r_X)^{1/3}}{\epsilon H_Y},
\label{eq:t_nuc}
\end{equation}
where we have adopted the square bubble approximation, which is appropriate 
in the present context~\cite{Bousso:2007nd}.  This implies that light 
cones are ``counted'' only if bubbles to which they belong form early 
enough, and therefore provides an effective time cutoff which becomes 
infinity for $\epsilon \rightarrow 0$.  This cutoff, however, is 
``fuzzy'' and depends on properties of $X$ and $Y$.

Incidentally, the well-definedness of quantum probabilities, introduced 
in Section~\ref{sec:quantum}, can also be understood similarly.  While 
a particular event may happen multiple times in the history of the 
multiverse, the probabilities of that to occur decrease exponentially 
with the number of times.  Thus, events that happen later make smaller 
contributions in calculation of the probabilities (e.g.\ to the 
numerators and denominators of Eqs.~(\ref{eq:probability-AB}) and 
(\ref{eq:probability-AC})).  This provides an effective time cutoff 
in the quantum probabilities.

\section{Sample Calculations in Toy Landscapes}
\label{app:calc}

\subsection{Semi-classical picture}
\label{app:calc-classical}

Here we present explicit calculations of probabilities in our framework, 
using simplified toy landscape models.  We assume that all the vacua 
lead to physical laws that reduce to the standard model of particle 
physics at low energies; namely, they differ only in properties that 
are not yet measured experimentally, e.g.\ the Higgs boson mass or 
TeV-scale physics, which we assume not to affect our own evolution 
in these vacua.  We also suppose that certain transitions between 
vacua lead to the standard model of cosmology; for example, they 
provide sufficiently high reheating temperature that later histories 
are consistent with the current observations.  We see that under these 
simplifying assumptions, the calculations reduce essentially to those 
for comoving probabilities~\cite{Garriga:1997ef}.

Let us consider the landscape consisting of several discrete vacua. 
The fraction of comoving volume occupied by vacuum $X$ at time $t$, 
$f_X(t)$, then obeys the following rate equation:
\begin{equation}
  \frac{d f_X}{dt} 
  = \sum_Y \Bigl( \kappa_{XY} - \delta_{XY} \sum_Z \kappa_{ZX} \Bigr) f_Y,
\label{eq:evol-eq}
\end{equation}
where $\kappa_{XY} \equiv (4\pi/3) \lambda_{XY} H_Y^{-4}$, and 
$\lambda_{XY}$ and $H_Y$ are the bubble nucleation rate per unit 
physical spacetime for the $Y \rightarrow X$ transition and the Hubble 
expansion rate in vacuum $Y$, respectively.  Here, we have taken $t$ 
to be the scale factor time, i.e.\ $t = \ln a(t)$ where $a(t)$ is the 
scale factor, but the final results do not depend on the choice of 
the time coordinate.

Now, we want to find the past light cones that are consistent with 
our prior conditions and encountered by geodesics emanating from an 
initial space-like hypersurface $\Sigma$.  Assuming that transitions 
$Y \rightarrow X$ lead to standard cosmology, the number of such past 
light cones whose tips are in vacuum $X$ can be estimated as
\begin{equation}
  {\cal N}_X 
  \propto \sum_Y \int_0^{t_c}\! \frac{V_{XY}(t)}{V_\epsilon}\, dt 
  \propto \sum_Y \int_0^{t_c}\! \kappa_{XY} f_Y(t)\, dt,
\qquad
  t_c \approx \ln\left[ \frac{(4\pi/3)^{1/3}}{\epsilon H_Y} \right],
\label{eq:N_X}
\end{equation}
where $V_{XY}(t)\, dt$ and $V_\epsilon = \epsilon^3$ are the comoving 
volume of $X$ created by the transition $Y \rightarrow X$ between time 
$t$ and $t + dt$ and the average comoving volume occupied by a single 
geodesic, respectively.  The function $f_Y$ is obtained by solving 
Eq.~(\ref{eq:evol-eq}) with a given initial condition $f_X(0)$, and 
the ``cutoff time'' $t_c$ is determined from the fact that the comoving 
volume of a bubble formed after $t_c$ is smaller than $\epsilon^3$, 
so that the geodesics typically do not intersect with these bubbles. 
The relative probability of finding ourselves in vacua $X_1$ and $X_2$ 
is then given by
\begin{equation}
  \frac{P_{X_1}}{P_{X_2}} = \lim_{\epsilon \rightarrow 0} 
    \frac{{\cal N}_{X_1}}{{\cal N}_{X_2}}.
\label{eq:rel-P}
\end{equation}
This probability agrees with the comoving probability.

The calculation can be simplified significantly if the landscape possesses 
(at least one) terminal vacua, as suggested by string theory.  In this 
case, we can use the integrated version of Eq.~(\ref{eq:evol-eq}):
\begin{equation}
  f_X(\infty) - f_X(0) 
  = \sum_Y \Bigl( \kappa_{XY} - \delta_{XY} \sum_Z \kappa_{ZX} \Bigr) F_X,
\label{eq:evol-eq-integ}
\end{equation}
where $F_X \equiv \int_0^\infty\! f(t)\, dt$.  Since $f_X(\infty) = 0$ for 
{\it non-}terminal vacua (as all the comoving volume decays into terminal 
vacua at $t \rightarrow \infty$), we can obtain $F_X$ for these vacua by 
simply solving the set of linear equations (\ref{eq:evol-eq-integ}) 
for a given initial condition $f_X(0)$.  The probabilities are then 
obtained using Eq.~(\ref{eq:N_X}), which now takes the form
\begin{equation}
  \lim_{\epsilon \rightarrow 0} {\cal N}_X \propto \sum_Y \kappa_{XY} F_Y.
\label{eq:N_X-lim}
\end{equation}
Note that the summation in the right-hand side runs only over non-terminal 
vacua, so that we need to solve Eq.~(\ref{eq:evol-eq-integ}) only for 
these vacua.

We now apply the above formulae to some simple examples.

\subsubsection*{Example~1---A system with three recyclable vacua}

\begin{figure}[t]
\begin{center}
\vspace{2mm}
  \includegraphics[scale=0.7]{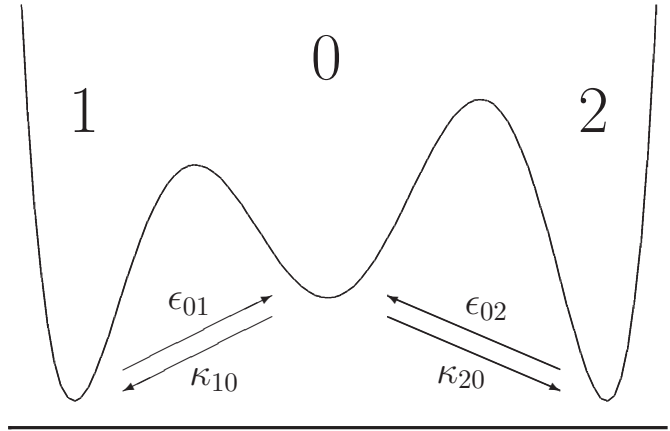}
\vspace{2mm}
\caption{Toy landscape (I)---A system with three recyclable vacua.}
\label{fig:ex-1}
\end{center}
\end{figure}
Let us consider a system with three de~Sitter vacua between which there 
are several allowed transitions as depicted in Fig.~\ref{fig:ex-1}.  The 
rate equation (\ref{eq:evol-eq}) is then given by
\begin{equation}
  \frac{d}{dt} \left(\begin{array}{c} f_0 \\ f_1 \\ f_2 \end{array}\right) 
  = \left(\begin{array}{ccc} 
      -(\kappa_{10}+\kappa_{20}) &  \epsilon_{01} &  \epsilon_{02} \\
                     \kappa_{10} & -\epsilon_{01} &              0 \\
                     \kappa_{20} &              0 & -\epsilon_{02} 
    \end{array}\right) 
    \left(\begin{array}{c} f_0 \\ f_1 \\ f_2 \end{array}\right),
\label{eq:recy-eq}
\end{equation}
where we have written the rates for upward transitions (transitions from 
lower energy minima to higher energy ones) as $\epsilon_{XY}$, instead 
of $\kappa_{XY}$, to emphasize that they are generically much smaller 
than the rates for usual, downward transitions~\cite{Lee:1987qc}. 
We assume that the standard cosmology is obtained when the universe 
experiences a transition $0 \rightarrow 1$ or $0 \rightarrow 2$, so 
that we can be living either vacuum $1$ or $2$.%
\footnote{A landscape without a terminal vacuum will be 
 unrealistic, as it leads to the problem of Boltzmann brains; 
 see Section~\ref{subsec:paradoxes}.  Here we consider such 
 a model simply for illustrative purposes.}

Suppose the initial condition of spacetime is given by
\begin{equation}
  f_0(0) = 1,
\qquad
  f_1(0) = f_2(0) = 0,
\label{eq:recy-init-1}
\end{equation}
i.e.\ the multiverse begins from the highest energy minimum.  Then, 
Eq.~(\ref{eq:recy-eq}) gives
\begin{eqnarray}
  f_0(t) &=& e^{-(\kappa_{10} + \kappa_{20})t} 
    + \frac{\epsilon_{01}\kappa_{10} + \epsilon_{02}\kappa_{20}}
      {(\kappa_{10} + \kappa_{20})^2} 
    \left\{ 1 - e^{-(\kappa_{10} + \kappa_{20})t} 
      - t(\kappa_{10} + \kappa_{20}) e^{-(\kappa_{10} + \kappa_{20})t} 
    \right\} + O(\epsilon^2),
\\
  f_1(t) &=& \frac{\kappa_{10}}{\kappa_{10} + \kappa_{20}} 
    \left\{ 1 - e^{-(\kappa_{10} + \kappa_{20})t} \right\} + O(\epsilon),
\\
  f_2(t) &=& \frac{\kappa_{20}}{\kappa_{10} + \kappa_{20}} 
    \left\{ 1 - e^{-(\kappa_{10} + \kappa_{20})t} \right\} + O(\epsilon),
\label{eq:recy-sol-1}
\end{eqnarray}
so that $\int_0^T\! f_0(t)\, dt = ((\epsilon_{01} \kappa_{10} + \epsilon_{02} 
\kappa_{20})/(\kappa_{10} + \kappa_{20})^2) T$ for $T \rightarrow \infty$. 
The relative probability for finding ourselves in vacua $1$ and $2$ is thus
\begin{equation}
  \frac{P_1}{P_2} 
  = \lim_{T \rightarrow \infty} \frac{\int_0^T\! \kappa_{10} f_0(t)\, dt} 
    {\int_0^T\! \kappa_{20} f_0(t)\, dt}
  = \frac{\kappa_{10}}{\kappa_{20}}.
\label{eq:recy-prob-1}
\end{equation}

Similarly, if the initial condition is given by
\begin{equation}
  f_1(0) = 1,
\qquad
  f_0(0) = f_2(0) = 0,
\label{eq:recy-init-2}
\end{equation}
then
\begin{eqnarray}
  f_0(t) &=& \frac{\epsilon_{01}}{\kappa_{10} + \kappa_{20}} 
    \left\{ 1 - e^{-(\kappa_{10} + \kappa_{20})t} \right\} + O(\epsilon^2),
\\
  f_1(t) &=& 1 + O(\epsilon),
\\
  f_2(t) &=& O(\epsilon),
\label{eq:recy-sol-2}
\end{eqnarray}
leading to $\lim_{T \rightarrow \infty} \int_0^T\! f_0(t)\, dt = 
(\epsilon_{01}/(\kappa_{10}+\kappa_{20})) T$.  The probability is, 
thus, again given by
\begin{equation}
  \frac{P_1}{P_2} = \frac{\kappa_{10}}{\kappa_{20}}.
\label{eq:recy-prob-2}
\end{equation}

We find that the relative probability does not depend on the initial 
condition in this particular example.  This can be understood from 
the fact that ``our universe'' arises only through a transition from 
vacuum~$0$ to $1$ or $2$, and that the branching ratio for these transitions 
is given by ${\rm Br}(0 \rightarrow 1)/{\rm Br}(0 \rightarrow 2) = 
\kappa_{10}/\kappa_{20}$.

\subsubsection*{Example~2---A system with two non-terminal and 
 one terminal vacua}

\begin{figure}[t]
\begin{center}
\vspace{2mm}
  \includegraphics[scale=0.7]{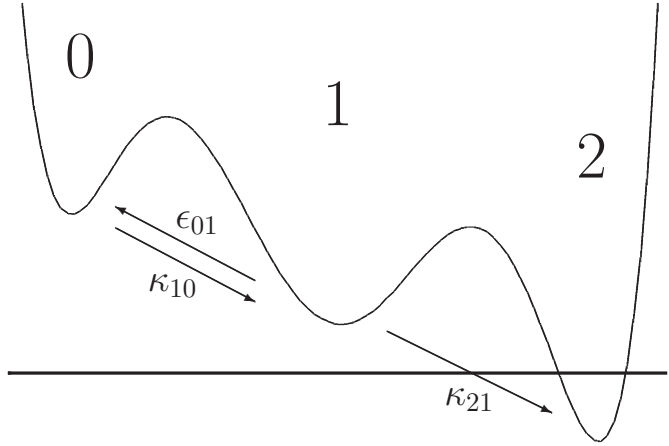}
\vspace{2mm}
\caption{Toy landscape (II)---A system with two non-terminal and one 
 terminal vacua.}
\label{fig:ex-2}
\end{center}
\end{figure}
We now consider a system with two non-terminal and one terminal vacua, 
illustrated in Fig.~\ref{fig:ex-2}.  The rate equation for this system 
is given by
\begin{equation}
  \frac{d}{dt} \left(\begin{array}{c} f_0 \\ f_1 \\ f_2 \end{array}\right) 
  = \left(\begin{array}{ccc} 
      -\kappa_{10} &                \epsilon_{01} & 0 \\
       \kappa_{10} & -(\kappa_{21}+\epsilon_{01}) & 0 \\
                 0 &                  \kappa_{21} & 0 
    \end{array}\right) 
    \left(\begin{array}{c} f_0 \\ f_1 \\ f_2 \end{array}\right).
\label{eq:term-eq}
\end{equation}
As discussed before, we can focus on the integrated version of this 
equation for non-terminal vacua:
\begin{equation}
  - \left(\begin{array}{c} f_0(0) \\ f_1(0) \end{array}\right) 
  = \left(\begin{array}{cc} 
      -\kappa_{10} &                \epsilon_{01} \\
       \kappa_{10} & -(\kappa_{21}+\epsilon_{01}) 
    \end{array}\right) 
    \left(\begin{array}{c} F_0 \\ F_1 \end{array}\right),
\label{eq:term-eq-integ}
\end{equation}
where we have used $f_0(\infty) = f_1(\infty) = 0$.  We assume that 
the standard cosmology arises after the transition $0 \rightarrow 1$ 
or $1 \rightarrow 2$.%
\footnote{Here we ignore the fact that we have already measured a positive 
 vacuum energy in our universe.}

If the initial condition is given by
\begin{equation}
  f_0(0) = 1,
\qquad
  f_1(0) = 0,
\label{eq:term-init-1}
\end{equation}
i.e.\ the multiverse starts from the higher de~Sitter minimum, then 
Eq.~(\ref{eq:term-eq-integ}) gives
\begin{equation}
  F_0 = \frac{\kappa_{21} + \epsilon_{01}}{\kappa_{10} \kappa_{21}},
\qquad
  F_1 = \frac{1}{\kappa_{21}},
\label{eq:term-sol-1}
\end{equation}
and the relative probability of finding ourselves in vacua $1$ and $2$ is
\begin{equation}
  \frac{P_1}{P_2} 
  = \frac{\kappa_{10} F_0}{\kappa_{21} F_1} 
  = \frac{\kappa_{21} + \epsilon_{01}}{\kappa_{21}}.
\label{eq:term-prob-1}
\end{equation}
On the other hand, if the multiverse starts from the lower de~Sitter 
minimum
\begin{equation}
  f_0(0) = 0,
\qquad
  f_1(0) = 1,
\label{eq:term-init-2}
\end{equation}
then
\begin{equation}
  F_0 = \frac{\epsilon_{01}}{\kappa_{10} \kappa_{21}},
\qquad
  F_1 = \frac{1}{\kappa_{21}},
\label{eq:term-sol-2}
\end{equation}
so that
\begin{equation}
  \frac{P_1}{P_2} 
  = \frac{\epsilon_{01}}{\kappa_{21}}.
\label{eq:term-prob-2}
\end{equation}

The results in Eqs.~(\ref{eq:term-prob-1}) and (\ref{eq:term-prob-2}) 
show that the probability does depend on the initial condition.  The 
physical picture in each case is clear.  Let us first consider the former 
case, where the initial condition is given by Eq.~(\ref{eq:term-init-1}). 
In this case, if the multiverse starting from vacuum~$0$ simply decays 
into $2$, i.e.\ if $\epsilon_{01} = 0$, then we would obtain $P_1/P_2 = 1$ 
since the decay must always occur through vacuum~$1$.  Now, turning on 
a recycling process $\epsilon_{01} \neq 0$ makes $P_1/P_2$ slightly larger 
than $1$, since there is then a small possibility of the universe experiencing 
the transition $0 \rightarrow 1$ more than once, although $1 \rightarrow 2$ 
occurs only once.  This explains the form of Eq.~(\ref{eq:term-prob-1}). 
On the other hand, in the latter case where the initial condition is 
given by Eq.~(\ref{eq:term-init-2}), the multiverse starts from vacuum~$1$ 
which, in the limit of $\epsilon_{01} = 0$, simply decays into vacuum~$2$, 
giving $P_1/P_2 = 0$.  However, a recycling process provides a small 
possibility that the transition $1 \rightarrow 0$ happens before 
the $1 \rightarrow 2$ decay, making $P_1/P_2$ slightly nonzero.  In 
fact, the expression of Eq.~(\ref{eq:term-prob-2}) is nothing but the 
branching ratio ${\rm Br}(1 \rightarrow 0)/{\rm Br}(1 \rightarrow 2) 
= \epsilon_{01}/\kappa_{21}$.

\subsection{Quantum picture}
\label{app:calc-quantum}

Here we compute probabilities in toy landscapes, using the quantum 
mechanical definition given in Section~\ref{sec:quantum}.  We consider 
the same setup as in Section~\ref{app:calc-classical}:\ all the vacua 
lead to the standard model of particle physics at low energies, and 
measurements are performed (i.e.\ civilizations exist) right after 
certain cosmic phase transitions.  Note that while we adopt the 
quantum mechanical definition of the probabilities, the computation 
is (necessarily) semi-classical, as we do not know the theory of 
quantum gravity.

Following Section~\ref{subsec:single}, we describe the multiverse in 
terms of bulk density matrices.  With the level of approximation we need, 
the complete set for these density matrices ${\cal S}$ can be taken as 
all possible past light cones whose tips are in vacuum $X$ at proper 
time $\tau$ after the last bubble nucleation:
\begin{equation}
  {\cal S} = \left\{ \rho_{X,\tau} \right\}.
\label{eq:toy-Hilbert}
\end{equation}
The general multiverse state can then be written as
\begin{equation}
  \rho_{\rm bulk}(t)
  = \sum_X \int\!d\tau\, C_{X,\tau}(t)\, \rho_{X,\tau},
\label{eq:toy-Psi}
\end{equation}
where $t$ represents proper time along the observer (geodesic). 
The evolution of the coefficients $C_{X,\tau}(t)$ can be calculated 
semi-classically, and is governed by the usual rate equations:
\begin{eqnarray}
  |C_{X,0}(t+\varDelta t)|^2 &=& \sum_Y \lambda_{XY} 
    \frac{4\pi}{3 H_Y^3} \int\!d\tau\, |C_{Y,\tau}(t)|^2,
\label{eq:C-evolve-1}\\
  |C_{X,\tau+\varDelta \tau}(t+\varDelta t)|^2 
  &=& |C_{X,\tau}(t)|^2 - \sum_Z \lambda_{ZX} \frac{4\pi}{3 H_X^3} 
    \varDelta \tau\, |C_{X,\tau}(t)|^2,
\label{eq:C-evolve-2}
\end{eqnarray}
where $\varDelta\tau = \varDelta t$ by construction, and $\lambda_{XY}$ 
and $H_Y$ are as defined in Section~\ref{app:calc-classical}.  Defining 
\begin{equation}
  f_X(t) = \int\!d\tau\, |C_{X,\tau}(t)|^2,
\label{eq:f-def}
\end{equation}
i.e.\ the probability of the tip of the light cone being in vacuum 
$X$ at time $t$, the evolution equation for $f_X$ is obtained from 
Eqs.~(\ref{eq:C-evolve-1}) and (\ref{eq:C-evolve-2}) as
\begin{equation}
  \frac{d f_X(t)}{dt} = \sum_Y \lambda_{XY} \frac{4\pi}{3 H_Y^3} f_Y(t) 
    - \sum_Z \lambda_{ZX} \frac{4\pi}{3 H_X^3} f_X(t).
\label{eq:toy-f-evolve}
\end{equation}

Now, suppose that the transitions to vacua $X_1$ and $X_2$ lead to 
intelligent life just after the transitions.  An important point is 
that the anthropic factor $n_X$, i.e.\ the probability of finding an 
experimenter in vacuum $X$, is the same for {\it all bubbles} with the 
{\it same vacuum} $X$, because all these bubbles look identical to the 
observer (geodesic) traveling the multiverse.  In particular, bubbles 
formed at later times are {\it not} rewarded by the volume increase 
during eternal inflation.  Assuming that $X_1$ and $X_2$ have equal 
anthropic factors (as in Section~\ref{app:calc-classical} and also 
implicit in Eq.~(\ref{eq:toy-Psi})), the relative probability of 
finding these vacua is given through the definition of 
Eq.~(\ref{eq:bulk-probab-AB}) as
\begin{equation}
  \frac{P_{X_1}}{P_{X_2}} = 
    \frac{\int\!dt\, {\rm Tr}\{ \rho_{\rm bulk}(t)\, 
    {\cal O}_{{\rm bulk},X_1} \}}
    {\int\!dt\, {\rm Tr}\{ \rho_{\rm bulk}(t)\, 
    {\cal O}_{{\rm bulk},X_2} \}}
  = \frac{\int\!dt\, |C_{X_1,0}(t)|^2}{\int\!dt\, |C_{X_2,0}(t)|^2},
\label{eq:toy-Qprob}
\end{equation}
where ${\cal O}_{{\rm bulk},X_i} = \rho_{X_i,0}$.  The factors 
appearing in the rightmost expression can be calculated from 
Eq.~(\ref{eq:C-evolve-1}):
\begin{equation}
  \int\!dt\, |C_{X,0}(t)|^2 = \sum_Y \lambda_{XY} 
    \frac{4\pi}{3 H_Y^3} \int\! f_Y(t)\, dt,
\label{eq:C0-det}
\end{equation}
where $f_Y(t)$ is obtained as a solution to Eq.~(\ref{eq:toy-f-evolve}), 
once the initial conditions are given.  

Incidentally, equations~(\ref{eq:toy-f-evolve}) and (\ref{eq:C0-det}) 
may be rewritten in terms of scale factor time $t_{\rm SF}$, 
$dt_{\rm SF} = H\, dt$, as
\begin{equation}
  \frac{d f_X(t_{\rm SF})}{dt_{\rm SF}} 
  = \sum_Y \kappa_{XY} f_Y(t_{\rm SF}) - \sum_Z \kappa_{ZX} f_X(t_{\rm SF}),
\label{eq:toy-f-evolve-2}
\end{equation}
\begin{equation}
  \int\!dt\, |C_{X,0}(t)|^2 = \sum_Y \kappa_{XY} 
    \int\! f_Y(t_{\rm SF})\, dt_{\rm SF},
\label{eq:C0-det-2}
\end{equation}
where $\kappa_{XY} = (4\pi/3) \lambda_{XY} H_Y^{-4}$, as in 
Section~\ref{app:calc-classical}.  We then find, comparing 
Eqs.~(\ref{eq:evol-eq},~\ref{eq:N_X},~\ref{eq:rel-P}) and 
(\ref{eq:toy-f-evolve-2},~\ref{eq:C0-det-2},~\ref{eq:toy-Qprob}), 
that relative probabilities defined using the quantum picture (in 
Section~\ref{sec:quantum}) precisely agree with those defined using 
the semi-classical picture (in Section~\ref{sec:framework}).

\section{No Quantum Cloning in Bubble Universes}
\label{app:no-cloning}

Our framework postulates that all the information behind apparent 
horizons (as viewed from the observer) is encoded on the (stretched) 
apparent horizons.  For past horizons, this provides ``initial 
conditions'' for the subsequent evolution in the bulk of spacetime; 
and for future horizons, the information stored can be sent back 
as Hawking radiation.  In this Appendix, we provide a nontrivial 
consistency check of this picture in the eternally inflating multiverse. 
Our analysis closely follows that of Ref.~\cite{Susskind:1993mu}, 
performed in the context of black hole physics.

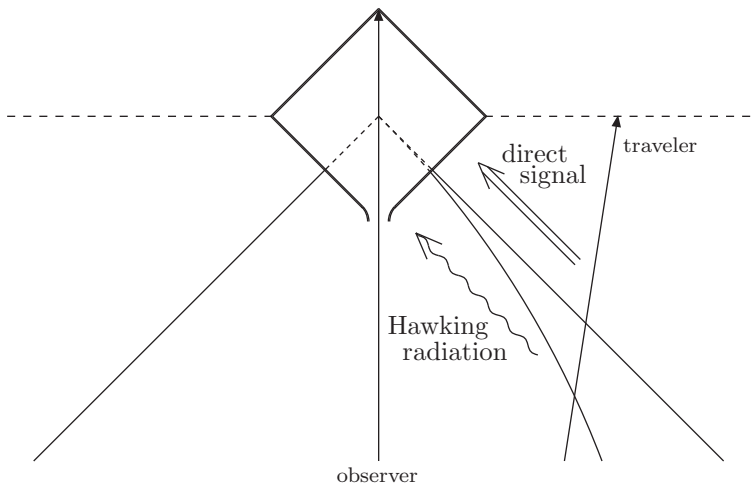
\begin{figure}[t]
\begin{center}
\begin{picture}(280,205)(0,-150)
  \DashLine(0,0)(100,0){3}
  \Line(100,0)(140,40) \Line(99.3,0)(140,40.7)
  \Line(140,40)(180,0) \Line(140,40.7)(180.7,0)
  \DashLine(180,0)(280,0){3}
  \Line(100,0)(134,-34) \Line(99.3,0)(134,-34.7)
  \CArc(128.34,-39.66)(8,0,45) \CArc(127.84,-39.66)(8,0,45)
  \Line(180,0)(146,-34) \Line(180.7,0)(146,-34.7)
  \CArc(151.66,-39.66)(8,135,180) \CArc(152.16,-39.66)(8,135,180)
  \DashLine(140,0)(120,-20){2} \Line(120,-20)(10,-130)
  \DashLine(140,0)(160,-20){2} \Line(160,-20)(270,-130)
  \DashCArc(-121.63,-261.63)(370,40.5,45){2}
  \CArc(-121.63,-261.63)(370,20.84,40.5)
  \LongArrow(140,-130)(140,39) \Text(140,-133)[t]{\scriptsize observer}
  \LongArrow(210,-130)(230,-1) \Text(233,-8)[lt]{\scriptsize traveler}
  \Photon(155,-45)(200,-90){1}{6} \Line(154,-44)(155,-45)
  \Line(154,-44)(163.90,-48.24) \Line(154,-44)(158.24,-53.90)
  \Text(182,-80)[r]{\footnotesize Hawking}
  \Text(189,-89)[r]{\footnotesize radiation}
  \Line(179,-21)(214,-56) \Line(181,-19)(216,-54)
  \Line(177.5,-17.5)(187.40,-21.74) \Line(177.5,-17.5)(181.74,-27.40)
  \Text(212,-14)[r]{\footnotesize direct}
  \Text(219,-23)[r]{\footnotesize signal}
\end{picture}
\caption{An observer traveling the multiverse may apparently receive 
 the same quantum information both from Hawking radiation and from 
 a direct signal sent by the traveler, which would violate the no-cloning 
 theorem of quantum mechanics.  A careful consideration, however, reveals 
 that this cannot happen.}
\label{fig:gedanken}
\end{center}
\end{figure}

Consider an observer in an inflationary phase.  We describe the 
spacetime using the flat coordinates:
\begin{equation}
  ds^2 = -dt^2 + \frac{1}{H^2} e^{2Ht} (dr^2 + r^2 d\Omega_2^2),
\label{eq:dS-metric}
\end{equation}
where $H$ is the Hubble parameter, and we set the observer at $r = 0$. 
Suppose there is a traveler falling behind the observer's horizon, as 
depicted in Fig.~\ref{fig:gedanken}.
Then, all the information he/she carries will be stored on the stretched 
horizon, from the observer's viewpoint.  Let $t_{\rm esc}$ be the time 
when the traveler crosses the horizon.  The traveler's location at 
$t = t_{\rm esc}$ is then given by
\begin{equation}
  r_{\rm traveler} = e^{-H t_{\rm esc}}.
\label{eq:r_esc}
\end{equation}
If the traveler follows a comoving path, $r_{\rm traveler}$ stays 
constant throughout the future history.

According to the present picture, the observer can retrieve the information 
carried away by the traveler, from Hawking radiation at late times (the 
wavy arrow in Fig.~\ref{fig:gedanken}).  On the other hand, the traveler 
may also try to communicate the information to the observer by sending 
some signal after he/she crosses the horizon (the solid arrow).  Now, 
suppose a Minkowski bubble forms in the future of the observer.  Then, 
the observer may receive the signal sent by the traveler, as shown in 
Fig.~\ref{fig:gedanken}.  This is a dangerous situation.  If the observer 
could obtain the information {\it both} from Hawking radiation {\it and} 
from the signal, then the observer would have duplicate quantum 
information, contradicting the basic principles of quantum mechanics:\ 
the no-cloning theorem.

Let us examine the condition under which this inconsistency might occur. 
Suppose, for simplicity, that the Minkowski bubble nucleates at time 
$t_{\rm nuc}$ on the observer's trajectory, $r=0$.  The observer can 
then collect the information before entering into the bubble, if
\begin{equation}
  t_{\rm nuc} \simgt t_{\rm esc} + t_{\rm ret}.
\label{eq:t-rel}
\end{equation}
Here, $t_{\rm ret}$ is the information retrieval time, i.e.\ the minimal 
time needed to retrieve any information from the horizon, which for a 
de~Sitter space is given by~\cite{Sekino:2008he}
\begin{equation}
  t_{\rm ret} = \frac{1}{H} \left\{ \ln\left(\frac{1}{l_P H}\right) 
    + O(1) \right\}.
\label{eq:t_ret}
\end{equation}
(Note that the coefficient is fixed to be $1/2\pi T = 1/H$ where $T$ 
is the de~Sitter temperature.)  In de~Sitter space, the bubble nucleated 
at $t_{\rm nuc}$ grows to the size $r \approx e^{-H t_{\rm nuc}}$ at 
future infinity.%
\footnote{Strictly speaking, the metric after the information gathering 
 must change from de~Sitter to Schwarzschild-de~Sitter, but the correction 
 from this is negligible.}
Therefore, the largest distance to the bubble wall from the observer 
who collected the information from Hawking radiation is
\begin{equation}
  r_{\rm wall} \simlt r_{\rm wall,\, max} \approx 
    e^{-H t_{\rm esc}} e^{-H t_{\rm ret}}.
\label{eq:r_wall}
\end{equation}
Here, we have used Eq.~(\ref{eq:t-rel}).

We now consider the signal sent by the traveler.  Suppose the traveler 
sent it to the observer at time $\varDelta t$ after he/she crossed the 
horizon, i.e.\ at $t = t_{\rm esc} + \varDelta t$.  The signal then reaches
\begin{equation}
  r_{\rm signal} \simgt r_{\rm signal,\, min} 
    \approx e^{-H t_{\rm esc}} (1 - e^{-H \varDelta t}),
\label{eq:r_signal}
\end{equation}
at future infinity.  Here, we have assumed that the traveler follows a 
comoving trajectory, which is a good approximation for small $\varDelta t$. 
The contradiction would occur if
\begin{equation}
  r_{\rm signal,\, min} < r_{\rm wall,\, max},
\label{eq:contradiction}
\end{equation}
since then the observer may receive the same information both from 
Hawking radiation and the signal.  Using Eqs.~(\ref{eq:t_ret}), 
(\ref{eq:r_wall}), and (\ref{eq:r_signal}), this is translated into
\begin{equation}
  \varDelta t \simlt l_P.
\label{eq:dt-bound}
\end{equation}
The same inequality is also obtained by considering another (extreme) 
setup where the bubble nucleates on the stretched horizon in the 
direction antipodal to the traveler.

The above analysis indicates that the contradiction may occur only if 
the traveler can send the information fast enough (in a super-Planckian 
time) that the condition of Eq.~(\ref{eq:dt-bound}) is satisfied.  Is 
it possible to do that?  The holographic principle implies that the 
amount of information the traveler can emit during time interval 
$\varDelta t$ is bounded by
\begin{equation}
  I \leq \frac{\pi}{4 l_P^2} \varDelta t^2,
\end{equation}
so that sending even one bit of information, $I \approx \ln 2$, requires 
the Planck time:
\begin{equation}
  \varDelta t \simgt l_P.
\label{eq:info-send}
\end{equation}
Therefore, we find that the violation of the principles of quantum 
mechanics actually does {\it not} occur, as viewed from the observer. 
In fact, it is quite convincing that this ``quantum censorship'' 
works but only barely.

\end{document}